\documentclass[journal,twoside,web]{ieeecolor}
\usepackage{generic}
\usepackage{cite}
\usepackage{amsmath,amssymb,amsfonts}
\usepackage{subfigure}
\usepackage{graphicx}
\usepackage{textcomp}
\usepackage{xfrac}
\usepackage{algorithm}
\usepackage[end]{algpseudocode}
\usepackage{lipsum}
\newcommand\itema{\item[\textbf{{}}]}
\newcommand\itemb{\item[\textbf{{}}]}
\newcommand\itemc{\item[\textbf{{}}]}
\newcommand\itemd{\item[\textbf{{}}]}
\usepackage[utf8]{inputenc}
\def\BibTeX{{\rm B\kern-.05em{\sc i\kern-.025em b}\kern-.08em
		T\kern-.1667em\lower.7ex\hbox{E}\kern-.125emX}}
\begin{document}
	\title{Robust Consensus of Higher-Order Multi-Agent Systems With Attrition and Inclusion of Agents and Switching Topologies } 
	\author{Jinraj V Pushpangathan, Harikumar Kandath \IEEEmembership{Member, IEEE}, Rajdeep Dutta, Rajarshi Bardhan, and J. Senthilnath \IEEEmembership{Senior Member, IEEE}
		\thanks{Freelance aerospace consultant @ Kerala-683512,
			India, e-mail: (jinrajaero@gmail.com).}	
		\thanks{Assistant professor @ Robotics Research Center, International Institute of Information Technology, Hyderabad-32, India, e-mail:(harikumar.k@iiit.ac.in)}
	\thanks{Scientist @Institute for Infocomm Research, Agency for Science, Technology
		and Research (A*STAR), Singapore 138632, Singapore,
 e-mail:(Rajdeep$\_$Dutta@i2r.a-star.edu.sg) }
		\thanks{Scientist @SIMTech, Agency for Science, Technology and Research (A*STAR), Singapore, e-mail:(rajarshi$\_$bardhan@simtech.a-star.edu.s) }
\thanks{Research Scientist @Institute for Infocomm Research, Agency for Science, Technology
	and Research (A*STAR), Singapore-138632, Singapore
, e-mail:(J$\_$Senthilnath@i2r.a-star.edu.sg) }
	}
	
	\maketitle
	
	\begin{abstract}
	Some of the issues associated with the practical applications of consensus of  multi-agent systems (MAS) include switching topologies, attrition and inclusion of agents from an existing network, and model uncertainties of agents. In this paper, a single distributed dynamic state-feedback protocol referred to as the Robust Attrition-Inclusion Distributed Dynamic (RAIDD) consensus protocol, is synthesized for achieving the consensus of MAS with attrition and inclusion of linear time-invariant higher-order uncertain homogeneous agents and switching topologies.  A state consensus problem termed as the Robust Attrition-Inclusion (RAI) consensus problem is formulated to find this RAIDD consensus protocol. To solve this RAI consensus problem, first,   the sufficient condition for the existence of the RAIDD protocol is obtained using the $\nu$-gap metric-based simultaneous stabilization approach.  Next,  the RAIDD consensus protocol is attained using the  Glover-McFarlane robust stabilization method if the sufficient condition is satisfied. The performance of this RAIDD protocol is validated by numerical simulations.
	\end{abstract}
	
	
	\section{Introduction}
	\label{sec:intro}
	\raggedbottom
A multi-agent system (MAS) is made up of multiple independently operated autonomous agents that can work together as a group through communication. The state consensus problem is a fundamental issue in the cooperative control of  MASs.  This problem is concerned with the synthesize of a distributed consensus protocol that drives the desired states of all the agents to a  common value.  Now, to reach consensus with the distributed protocol,  each agent must be able to access the states of its neighboring agents via a communication network or sensing devices. This type of    MAS with communicating agents is modeled using the model of the dynamics of each agent,  a communication protocol that describes the interaction among the agents, and a graph that represents the interconnection topologies between agents. The potential applications of consensus are in flocking, formation control, oscillation synchronization, firefighting, multi-agent rendezvous, and satellite reconfiguration. It is worth noting that some of the practical applications of consensus have issues. For example, in the consensus of mobile agents  the communication topologies between agents need to switch between several fixed topologies from time to time due to finite communication radius as well as due to the presence of an obstacle between two agents.  Furthermore, the number of agents in some applications, such as firefighting, can change over time as some agents are relieved or new agents join the existing network depending upon the workload. In this paper, the removal and addition of agents from an existing network of agents are referred to as \textit{attrition} and \textit{inclusion} of agents, respectively. Moreover, the model of each agent is of higher-order and could possess parametric as well as unmodeled dynamics uncertainties. Considering all the aforementioned issues, some  practical applications require  a single distributed robust consensus protocol that can handle a varying number of higher-order agents, switching topologies, and model uncertainties.
\par
Concerning the cooperative control of MAS with attrition of agents, a cooperative relay tracking strategy is developed in \cite{dong} to ensure successful tracking even when a second-order agent quits tracking due to malfunction. In the case of the multiagent tracking systems that are subjected to agent failure followed by the agent replacement, a modified nonsingular terminal sliding mode control scheme and event-triggered coordination strategies were proposed in \cite{lijing}. In \cite{lululi}-\cite{Jianhua},  consensus recovery methods are proposed to compensate for the undesirable effects caused by the removal of agents while retaining the consensus property. The consensus problem for high-order MASs with switching topologies and time-varying delays is studied in \cite{cui}. Here, the consensus problem is converted into an $L_2$-$L_\infty$ control problem employing the tree-type transformation approach. Also, the consensus with the prescribed $L_2$-$L_\infty$ performance is ensured through sufficient conditions that are derived using linear matrix inequalities (LMIs).  In \cite{liu},  the consensus problem of MAS with switching topologies is transformed into an $H_\infty$ control problem. The sufficient condition is derived in terms of LMIs to ensure consensus of the MAS. Following this, a distributed dynamic output feedback protocol is developed where the system matrix of the protocol is designed by solving two LMIs. Moreover, a distributed algorithm is developed in \cite{deyuan} using an iterative learning rule for the consensus tracking control of MASs with switching topologies and disturbances. The LMI-based necessary and sufficient conditions for the convergence of the consensus tracking objectives are also presented. Further solutions to the consensus problem with switching topologies employing LMIs can be found in \cite{qin}-\cite{peng}. The consensus problems of MAS with fixed/switching topologies and time-delays is discussed in  \cite{olftai}. In this paper,   a  Lyapunov function-based disagreement function is used to study the convergence characteristics of consensus protocols.  In \cite{saboori}, the sufficient conditions to design a distributed protocol for the consensus of identical linear time-invariant (LTI) MASs subjected to bounded external disturbance, switching topologies, and directed communication network graph are proposed. These conditions are based on $L_2$ gain and RMS bounded disturbances. The Lyapunov stability theory is then used to investigate the stability characteristics of the proposed controllers. Maria Elena Valcher et al. \cite{maria} describes each agent of the MAS using a single-input stabilizable state-space model and then investigate the consensus problem under arbitrary switching for identical MAS with switching communication topology. Also, the consensusability of this  MAS is illustrated by constructing a common quadratic positive definite Lyapunov function  describing the evolution of the disagreement vector for the switched system. Guanghui Wen et al. \cite{Guanghui} discusses the distributed $H_\infty$ consensus problem of MASs with higher-order linear dynamics and switching directed topologies. It is demonstrated here that if the protocol's feedback gain matrix is properly designed and the coupling strength among neighboring agents is greater than a derived positive value, then distributed $H_\infty$ consensus the problem can be solved. The exponential state consensus problem for hierarchical multi-agent dynamical systems with switching topology and inter-layer communication delay is addressed in \cite{zhaoxia}. In this paper, the stability theory of switched systems and graph theory of hierarchical network topology are utilized to derive sufficient conditions for accomplishing the exponential hierarchical average consensus.
The robust consensus of linear MASs with agents subject to heterogeneous additive stable perturbations is addressed in \cite{xianwei}. To design dynamic output-feedback protocols, two methods based on an algebraic Riccati equation and some scalar/matrix inequalities are proposed. Moreover, in \cite{yangliu}, the sufficient condition in terms of LMIs is derived for the robust $H_\infty$ consensus control of MASs with model parameter uncertainties and external disturbances. Further,  the traditional $H_\infty$ controller design is utilized in \cite{plin} to design a consensus protocol for the MAS with second-order dynamics  that is subjected to parameter uncertainties and external disturbances. Also, the asymptotical convergence of agents along with desired $H_\infty$ performance is assured through suitable sufficient conditions. For a class of second-order multi-agent dynamic systems with disturbances and unmodeled agent dynamics, continuous distributed consensus protocols that enable global asymptotic consensus tracking are designed in \cite{Guoqiang}. These protocols are developed with the help of an identifier that estimates unknown disturbances and unmodeled agent dynamics.\par
The determination of a single controller that stabilizes a finite number of systems is referred to as the simultaneous stabilization (SS) problem \cite{Saek}-\cite{vidya}.  The SS problem of more than two systems has no closed-form solution due to its NP-hardness \cite{Gever}-\cite{onur}. Hence, iterative algorithms are utilized to solve SS problem. For example, an LMI-based iterative algorithm is developed in \cite{CAO1} and a bi-level optimization-based decomposition strategy is utilized in \cite{perez}  to solve the SS problem. The SS problem is solved in \cite{saif}-\cite{jinjgcd} by first determining the sufficiency condition for the existence of the simultaneously stabilizing controller.  Then, this condition is solved using a robust stabilization controller that is synthesized around the central plant (system).  In \cite{saif}, the central plant is obtained by solving a 2-block optimization problem. However, the central plant is identified in \cite{jinsmc}-\cite{jinjgcd} using the maximum $\nu$-gap metric of the systems that requires SS.  The definition of this  $\nu-$gap metric-based central plant and the maximum $\nu$-gap metric of the systems is given in Section \ref{PL}.  It is shown in \cite{huy}-\cite{yuebing} that the state consensus problem of  MAS with  $N$ identical LTI agents can be expressed as the SS problem of $N-1$ independent systems.  In these papers, the consensus problem of MAS has been studied using an LMI-based SS approach.  One needs to note that the methods discussed in  the preceding articles do not generate a single distributed consensus protocol that achieves consensus of MAS with attrition and inclusion of higher-order uncertain agents and switching topologies.\par
In this paper, the existing MAS is supposed to have  $N$ agents. From this MAS,  either \textit{attrition} of $P$ agents or \textit{inclusion} of $M$ agents at a time is considered. Subsequently, the problem of finding a single robust distributed dynamic state-feedback consensus protocol that achieves consensus of MAS with \textit{attrition} and \textit{inclusion} of LTI higher-order uncertain homogeneous agents and switching topologies is stated as the  Robust  Attrition-Inclusion (RAI) consensus problem. Also, the protocol that solves the RAI consensus problem is referred to as the Robust Attrition-Inclusion Distributed Dynamic (RAIDD) consensus protocol. In addition, the actual dynamics of every agent considered here is uncertain.  The nominal linear dynamics of each agent is identical.  Moreover, the uncertainty in the actual linear dynamics of each agent is also assumed to be homogeneous.  This uncertainty is represented by bounded perturbations in the system and input matrices of the state-space model of the nominal linear dynamics. In this article, the RAIDD consensus protocol is synthesized in two steps. In the first step,  the sufficient condition for the existence of this protocol is obtained using the $\nu$-gap metric-based SS method.  Next,  the RAIDD consensus protocol is attained using the Glover-McFarlane robust stabilization method presented in \cite{glover2} if the sufficient condition is satisfied. The main contributions of this article are the following.
\begin{enumerate}
	\item To the best of the author’s knowledge, this is the first paper to propose a method for  generating a distributed consensus control that accomplishes consensus of MAS with the varying number of higher-order agents, switching topologies, and model uncertainties.
	\item The sufficient condition for the existence of the RAIDD consensus protocol, which dependents on the Hankel norm of right coprime factors and the $\nu$-gap metric-based central plant's maximum $\nu$-gap metric, is developed using the $\nu$-gap metric-based SS method.
	\item A RAIDD  consensus protocol is developed for the RAI consensus of four unmanned underwater vehicles (UUVs), with attrition and inclusion of one UUV. The performance of this protocol is validated by numerical simulations.
\end{enumerate}	
The rest of this paper is organized as  follows. The preliminaries and notation are given in Section \ref{PL}. In Section \ref{PS}, the problem statement is presented. The synthesize of the RAID consensus protocol is described in Section \ref{PR}. The  simulation results are discussed in Section \ref{SR}. In Section \ref{CL}, conclusions are summarized.\par	
\section{Preliminaries and Notation}\label{PL}	
This section introduces various definitions and notations, as well as some of the fundamental concepts of graph theory and $\nu$-gap metric-based simultaneous stabilization.
\subsection{Notation}
 In this paper, $\mathbb{R}$, $\mathbb{R}_{\geq 0}$, $\mathbb{R}^{n}$,  and $\mathbb{R}^{n \times m}$ denote the set of real numbers, the set of non-negative real numbers, the set of $n$-column real vectors, and the set of all real matrices of dimension $n \times m$, respectively.  $\mathbb{N}$ is the set of natural numbers,  $\mathbb{N}^{+}$ is the set of positive real numbers without zero, and $\mathbb{N}_{a}^{b}$ is set of natural numbers from $a$ to $b$ ($\mathbb{N}_{a}^{b}=\{a,\dots,b\}, a,b \in \mathbb{N}, a < b$).  $n_{C_2}$ denotes $n(n-1)/2$. $A \bigotimes B$ indicates the Kronecker product of the  $A$ and $B$ matrices. $\mathcal{C}_-$ symbolizes the open left-hand side of complex plane. The superscript $'T'$ denotes matrix/vector transpose. $A=(a_{ij}) \in \mathbb{R}^{n \times m}$ denotes a real matrix with $n$ rows and  $m$ columns. Here, $a_{ij}$  is the element at the $i$th row and $j$th column of the matrix. The zero matrix with  $n$ rows and  $m$ columns is represented using $\mathbf{0}{n,m}$. The $\max\{{}\}$ and $\min\{{}\}$ symbolize the maximum  and the minimum element of the set, respectively. The  $\lambda_{max}[A]$ represents the largest eigenvalue of the matrix, $A$. $\mathcal{RH_\infty}$ is the set of all proper and stable rational transfer function matrices. Likewise, $\mathcal{R}$ is the set of all proper  rational transfer  function matrices. For a transfer function matrix, $\mathbf{P}(s)$,  its  $H_\infty$ norm, determinant, Hankel norm, and winding number are denoted by  $\parallel\mathbf{P}(s)\parallel_\infty$,  det ($\mathbf{P}(s)$), $\parallel\mathbf{P}(s)\parallel_H$, and wno ($\mathbf{P}(s)$) respectively. Besides these, $\mathbf{{P}}(s)=:(A,B,C,D)$ denotes the short form of  $\mathbf{{P}}(s):=C(s\mathbf{I}-A)^{-1}B+D)$. Also,   $\mathbf{P}(s)^*$ represents  $\mathbf{P}^T(-s)$. 
 \subsection{Definitions}
	\newtheorem{defn}{Definition}[section]
\begin{defn}
	\textit{Generalized stability margin :}~~{\normalfont Consider 	a closed-loop (CL) system,  $[\mathbf{P}(s), \mathbf{K}]$, with the    controller, $\mathbf{K}$. Then, the generalized stability margin, $b_{\mathbf{P},\mathbf{K}}$ $\in$ $[0, 1]$, is defined as \cite{steele}	
		\begin{equation}
	b_{\mathbf{P},\mathbf{K}}=
	\begin{cases}
	{||\Upsilon||^{-1}_\infty} &\text{if}~ [\mathbf{P}(s), \mathbf{K}]~ \text{is internally stable}\\
	0&\text{otherwise}
	\end{cases}
	\label{eq:RCF1_1_1}
	\end{equation}	
where $\Upsilon=\left[ \begin{array}{c} {\mathbf{P}(s)}  \\{\mathbf{I}}  \end{array} \right ]\left(\begin{array}{c} {\mathbf{I-KP}(s)}  \end{array} \right )^{-1}\left[ \begin{array}{cc} {\mathbf{-I}} & {\mathbf{K}}  \end{array} \right ]$.		
%
%
	}
\end{defn}
\begin{defn}
	\textit{$v-$gap metric:}~~{\normalfont Consider two systems, $\mathbf{P}_1(s)$ and $\mathbf{P}_2(s)$. Let $[\mathbf{N}_1(s) \in \mathcal{RH_\infty}, \mathbf{M}_1(s) \in \mathcal{RH_\infty}]$ and $[\widetilde{\mathbf{N}}_1(s) \in \mathcal{RH_\infty}, \widetilde{\mathbf{M}}_1(s) \in \mathcal{RH_\infty}]$ be the right and left coprime factors of $\mathbf{P}_1(s)$, respectively. Likewise, $[\mathbf{N}_2(s) \in \mathcal{RH_\infty}, \mathbf{M}_2(s) \in \mathcal{RH_\infty}]$ and $[\widetilde{\mathbf{N}}_2(s) \in \mathcal{RH_\infty}, \widetilde{\mathbf{M}}_2(s) \in \mathcal{RH_\infty}]$ be the right and left coprime factors of $\mathbf{P}_1(s)$, respectively. Then, $v-$gap metric, $\delta_v(\mathbf{P}_1(j\omega),\mathbf{P}_2(j\omega)) \in [0,1] $, of  $\mathbf{P}_1(s)$ and $\mathbf{P}_2(s)$ is defined as \cite{82}
		\begin{equation}
		\delta_v(\mathbf{P}_1(j\omega),\mathbf{P}_2(j\omega)) = \begin{cases} \parallel\Phi(\mathbf{P}_1(j\omega),\mathbf{P}_2(j\omega))\parallel_\infty &\mbox{if } det \Theta(j\omega)  \\
		&\neq 0~ \text{and}~ wno\\
		& det(\Theta(j\omega))\\
		& = 0 \\
		1 & \mbox{otherwise } \end{cases}
		\label{gapm}
		\end{equation}
\noindent where $\Phi(\mathbf{P}_1(j\omega),\mathbf{P}_2(j\omega))=-\widetilde{\mathbf{N}}_2(s)\mathbf{M}_1(s) + \widetilde{\mathbf{M}}_2(s)\mathbf{N}_1(s)$ and $\Theta(j\omega)=\mathbf{N}^*_2(s)\mathbf{N}_1(s) + \mathbf{M}_2^*(s)\mathbf{M}_1(s)  $. 
	} 
\end{defn}
The $\nu$-gap metric measures the distance between systems, and if this distance is closer to zero, then any controller that works well with one system will also work well with the other.
\begin{defn}
	\textit{Maximum $\nu$-gap metric of the plant:}~~{\normalfont Let $\mathcal{Q}=\{\mathbf{P}_1(s),\dots,\mathbf{P}_f(s),\dots, \mathbf{P}_\xi(s)\}$ be a finite set of systems. Then, the maximum $\nu$-gap metric of $\mathbf{P}_i(s) \in \mathcal{Q}$, $\epsilon_{i}$, is defined as \cite{jinjgcd}
		\begin{equation}
	\begin{split}
	\epsilon_{i}=\max\big\{&\delta_v\big(\mathbf{P}_{i}(j\omega),\mathbf{P}_f(j\omega)\big)~\big|~ \mathbf{P}_{i}(s), \mathbf{P}_f(s)  \in \mathcal{Q}~~ \forall~f  \in \mathbb{N}_1^\xi \big\} . 
	\end{split}
	\label{epix}
	\end{equation}
}
\end{defn}
\begin{defn}
	\textit{$\nu$-gap metric-based central plant of $\mathcal{Q}$:}~~{\normalfont The $\nu$-gap metric-based central plant of $\mathcal{Q}$  is defined as the system whose maximum $\nu$-gap metric is the smallest among the maximum $\nu$-gap metrics of all the plants of   $\mathcal{Q}$  \cite{jinjgcd}.}
\end{defn} 
This definition implies that the $\nu$-gap metric-based central plant is the system that is closest to all other systems belonging to $\mathcal{Q}$ in terms of $\nu$-gap metric. In this paper, the $\nu$-gap metric-based central plant of a set and any parameters associated with it are denoted by the subscript `$cp$'.
\begin{defn}
	\textit{Hankel norm:}~~{\normalfont  Consider a stable system, $\mathbf{P}(s)$. Then, the Hankel norm of  $\mathbf{P}(s)$ is given by
		\begin{equation}
		\parallel \mathbf{P}(s) \parallel_H=\sqrt{\lambda_{max}[W_cW_o]}
		\label{hnorm}
		\end{equation}
		where $W_c$, $W_o$, and $\lambda_{max}[W_cW_o]$ are respectively the controllability Gramian and  the observability Gramian. 
	} 
\end{defn}
\subsection{Graph Theory for Formulating RAI Consensus Problem}
The primary focus of this paper is the synthesize of a RAIDD consensus protocol that achieves consensus of MAS with $N \in \mathbb{N}$, $(N-P) \in \mathbb{N} $, and $(N+M) \in \mathbb{N}$ agents as well as switching topologies. The communication network topologies among these agents are represented using Undirected Simple (US) graphs. These graphs do not have any loops or parallel edges. The number of US graphs that can be formed with $N$, $N-P$, and $N+M$ nodes is $2^{N_{C_2}} < \infty$, $2^{(N-P)_{C_2}} < \infty$, and $2^{(N+M)_{C_2}} < \infty$, respectively.  The graphs of MAS  need to be connected for achieving consensus. Following this, let  the number of Connected Undirected Simple (CUS) graphs of the MAS with $N$, $N-P$, and $N+M$ agents,  would respectively be   $k \leq 2^{N_{C_2}}$,   $r\leq2^{{N-P}_{C_2}}$, and $p\leq2^{{N+M}_{C_2}}$.  The CUS graphs of the MAS with $N$, $N-P$, and $N+M$ agents  are  denoted by $\breve{\mathcal{G}}_h~\forall~h~\in~\mathbb{N}_1^k$, $\grave{\mathcal{G}}_h~\forall~h~\in~\mathbb{N}_1^r$,
and  $\hat{\mathcal{G}}_h~\forall~h~\in~\mathbb{N}_1^p$, respectively. The node set of $\breve{\mathcal{G}}_h$, $\grave{\mathcal{G}}_h$,
and  $\hat{\mathcal{G}}_h$ is $\breve{\mathcal{V}}=\{1,\dots,N\}$, $\grave{\mathcal{V}}=\{1,\dots,N-P\}$, and $\hat{\mathcal{V}}=\{1,\dots,N+M\}$, respectively. Likewise, $\breve{\varepsilon}_h=\{(i,j)~|~i,j~\in~\breve{\mathcal{V}}\}$, $\grave{\varepsilon}_h=\{(i,j)~|~i,j~\in~\grave{\mathcal{V}}\}$, and $\hat{\varepsilon}_h=\{(i,j)~|~i,j~\in~\hat{\varepsilon}\}$ are the edge set of $\breve{\mathcal{G}}_h$, $\grave{\mathcal{G}}_h$,
and  $\hat{\mathcal{G}}_h$, respectively. The adjacency matrix of $\breve{\mathcal{G}}_h$, $\grave{\mathcal{G}}_h$,
and  $\hat{\mathcal{G}}_h$ is denoted by $\breve{\mathcal{A}}_h \in \mathbb{R}^{(N \times N)}$, $\grave{\mathcal{A}}_h \in \mathbb{R}^{((N-P) \times (N-P))}$, and $\hat{\mathcal{A}}_h \in \mathbb{R}^{((N+M) \times (N+M))}$, respectively. These matrices are defined as $\breve{\mathcal{A}}_h =(\breve{a}_{h_{ij}})$, $\grave{\mathcal{A}}_h =(\grave{a}_{h_{ij}}) $, and $\hat{\mathcal{A}}_h =(\hat{a}_{h_{ij}})$ where $(\breve{a}_{h_{ij}})$, $(\grave{a}_{h_{ij}}) $, and $(\hat{a}_{h_{ij}})$ are given by
\begin{equation}
\breve{a}_{h_{ij}}=
\begin{cases}
1 & i \neq j~ \text{and}~ (i,j) \in \breve{\varepsilon}_h\\
0 & i = j ~\text{or}~ (i,j) \notin \breve{\varepsilon}_h
\end{cases}
\end{equation}

\begin{equation}
\grave{a}_{h_{ij}}=
\begin{cases}
1 & i \neq j ~\text{and}~ (i,j) \in \grave{\varepsilon}_h\\
0 & i = j ~\text{or} ~(i,j) \notin \grave{\varepsilon}_h
\end{cases}
\end{equation}
	
\begin{equation}
\hat{a}_{h_{ij}}=
\begin{cases}
1 & i \neq j ~\text{and}~ (i,j) \in \hat{\varepsilon}_h\\
0 & i = j ~\text{or} ~(i,j) \notin \hat{\varepsilon}_h
\end{cases}
\end{equation}
The neighboring set of the node $i$ of $\breve{\mathcal{G}}_h$, $\grave{\mathcal{G}}_h$,
and  $\hat{\mathcal{G}}_h$ is denoted by $\mathcal{N}_i(\breve{\mathcal{G}}_{h\in\mathbb{N}_1^k})$, $\mathcal{N}_i(\grave{\mathcal{G}}_{h\in\mathbb{N}_1^r})$, and $\mathcal{N}_i(\hat{\mathcal{G}}_{h\in\mathbb{N}_1^p})$, respectively. These sets are defined as $\mathcal{N}_i(\breve{\mathcal{G}}_{h\in\mathbb{N}_1^k})=\{j~|~(i,j)~\in~\breve{\varepsilon}_h\}$, $\mathcal{N}_i(\grave{\mathcal{G}}_{h\in\mathbb{N}_1^r})=\{j~|~(i,j)~\in~\grave{\varepsilon}_h\}$, and $\mathcal{N}_i(\hat{\mathcal{G}}_{h\in\mathbb{N}_1^p})=\{j~|~(i,j)~\in~\hat{\varepsilon}_h\}$. Now, the graph Laplacian matrix of $\breve{\mathcal{G}}_h$, $\grave{\mathcal{G}}_h$,
and  $\hat{\mathcal{G}}_h$ is defined as $\breve{L}_h \in \mathbb{R}^{(N \times N)}=(\breve{l}_{h_{ij}})$, $\grave{L}_h \in \mathbb{R}^{((N-P) \times (N-P))} =(\grave{l}_{h_{ij}})$, and $\hat{L}_h \in \mathbb{R}^{((N+M) \times (N+M))}=(\hat{l}_{h_{ij}})$, respectively. Here, $\breve{l}_{h_{ij}}$,  $\grave{l}_{h_{ij}}$, and $\hat{l}_{h_{ij}}$ are given by
	
\begin{equation}
\breve{l}_{h_{ij}}=
\begin{cases}
\Sigma_{j \in \mathcal{N}_i(\breve{\mathcal{G}}_{h\in\mathbb{N}_1^k})}\breve{a}_{h_{ij}} & i = j\\
-\breve{a}_{h_{ij}} & i \neq j 
\end{cases}
\end{equation}

\begin{equation}
\grave{l}_{h_{ij}}=
\begin{cases}
\Sigma_{j \in \mathcal{N}_i(\grave{\mathcal{G}}_{h\in\mathbb{N}_1^r})}\grave{a}_{h_{ij}} & i = j\\
-\grave{a}_{h_{ij}} & i \neq j 
\end{cases}
\end{equation}

\begin{equation}
\hat{l}_{h_{ij}}=
\begin{cases}
\Sigma_{j \in \mathcal{N}_i(\hat{\mathcal{G}}_{h\in\mathbb{N}_1^p})}\hat{a}_{h_{ij}} & i = j\\
-\grave{a}_{h_{ij}} & i \neq j 
\end{cases}
\end{equation}	
Note that all the row-sum of the Laplacian matrices	are zero. Hence, zero is an eigenvalue of these matrices with eigenvector, $\mathbf{1}=[1,1,\dots,1]^T$. Besides this, the rank of $\breve{L}_h$, $\grave{L}_h$, and $\hat{L}_h$ is $N-1$, $N-P-1$, $N+M-1$, respectively, as the graphs associated with these matrices are undirected and connected. Moreover, these matrices are positive semi-definite real symmetric. In these case,  the zero eigenvalue of  $\breve{L}_h$, $\grave{L}_h$, and $\hat{L}_h$ has multiplicity of one and their non-zero eigenvalues can be ordered increasingly as $0 < \breve{\lambda}_{h_2}, \leq \breve{\lambda}_{h_3}, \dots, \leq \breve{\lambda}_{h_N}$, $0 < \grave{\lambda}_{h_2}, \leq \grave{\lambda}_{h_3}, \dots, \leq \grave{\lambda}_{h_{N-P}}$, and $0 < \hat{\lambda}_{h_2}, \leq \hat{\lambda}_{h_3}, \dots, \leq \hat{\lambda}_{h_{N+M}}$, respectively.\par

\subsection{$\nu$-gap Metric-Based Simultaneous Stabilization for Solving RAI Problem}\label{SSCP}
The $\nu$-gap metric-based SS method proposed in \cite{jinsmc} and \cite{jinjgcd} serves as the foundation for synthesizing the RAIDD consensus protocol. To explain this method, let $\mathbf{P}_{cp}(s) \in \mathcal{Q}$ be  the $\nu$-gap metric-based central plant of $\mathcal{Q}$ and $\epsilon_{cp}$ be the the maximum $\nu$-gap metric of  $\mathbf{P}_{cp}(s)$. Now, the concept of the $\nu$-gap metric-based SS method is stated as the following.
\begin{itemize}
	\item A single controller can provide similar CL characteristics to all the plants in $\mathcal{Q}$ if that controller is synthesized around $\mathbf{P}_{cp}(s)$. This is because  $\mathbf{P}_{cp}(s)$ is closest to all other plants belonging to  $\mathcal{Q}$  as $\epsilon_{cp}$ is the smallest among the maximum $\nu$-gap metrics of all other plants of $\mathcal{Q}$.
	\item The possibility of providing similar CL characteristics to all the plants in $\mathcal{Q}$ with a stabilizing controller of $\mathbf{P}_{cp}(s)$ increases when $\epsilon_{cp}$ approaches zero.
\end{itemize}
In the $\nu$-gap metric-based SS method, the simultaneous stabilizing controller is generated  by solving the sufficient condition  that depends on $\mathbf{P}_cp(s)$ and $\epsilon_{cp}$.  This condition   is given as \cite{jinjgcd}
\begin{align}
b_{\mathbf{P}_{cp},\mathbf{K}}>\epsilon_{{cp}}
\label{sscondk1x}
\end{align}
where $b_{\mathbf{P}_{cp},\mathbf{K}}$ and $\epsilon_{{cp}}$ are given by (\ref{eq:RCF1_1_1}) and (\ref{epix}), respectively.   The identification of   $\mathbf{P}_{cp}(s)$ and $\epsilon_{{cp}}$ is required for solving \eqref{sscondk1x}. This is accomplished by following the steps given below. 
\begin{itemize}
	\itema \textbf{Step~1:} Find  the maximum $\nu$-gap metrics of all the plants of $\mathcal{Q}$ using (\ref{epix}). 
	\itemb \textbf{Step~2:} Identify the smallest value among the maximum $\nu$-gap metrics. This gives $\epsilon_{{cp}}$ and  the plant  associated with this value is $\mathbf{P}_{cp}(s)$. 
\end{itemize}
For more details about the $\nu-$gap metric-based SS method, one needs to refer to the supporting material of \cite{jinsmc}.
\section{Problem Statement}\label{PS}
The RAI consensus problem is formulated in this section. For this purpose, consider a MAS of identical but uncertain higher-order agents.  The uncertainty in the actual dynamics of these agents is also identical and parametric in nature. The nominal and actual dynamics of these agents are described by LTI systems. Following this, let $\mathbf{P}_i(s)$ be the transfer function of the nominal dynamics of $i$th agent of the MAS. Also, assume that all the states of $\mathbf{P}_i(s)$ are measured. The state-space form of $\mathbf{P}_i(s)$ is then given as
	\begin{equation}
	\mathbf{P}_i(s):
	\begin{cases}
	\mathbf{\dot{x}}_i=&A\mathbf{x}_i+B\mathbf{u}_i\\
	\mathbf{y}_i=&C\mathbf{x}_i
	\end{cases}
	\label{eq:1}
	\end{equation}
where $\mathbf{x}_f$ $\in$ $\mathbb{R}^{n}$, $\mathbf{u}_f$ $\in$ $\mathbb{R}^{m}$, $\mathbf{y}_f$ $\in$ $\mathbb{R}^{n}$, $A \in \mathbb{R}^{n \times n}$, $B \in  \mathbb{R}^{n \times m}$, and $C=\mathbf{I}$ $\in$ $\mathbb{R}^{n \times n}$
represent state vector, input vector, output vector, system matrix, input matrix, and output matrix of the $i$th agent, respectively.  Moreover, $n$ and $m$ are the number of states and inputs of the $i$th agent, respectively.  Furthermore, $A$ is not Hurwitz and ($A, B$) is stabilizable.    Let $\mathbf{\bar{P}}_i(s)$ be the transfer function of actual (perturbed system)  dynamics  of $i$th agent. The state-space form of $\mathbf{\bar{P}}_i(s)$ is given as
	\begin{equation}
\mathbf{\bar{P}}_i(s):
\begin{cases}
\mathbf{\dot{x}}_i=&\bar{A}\mathbf{x}_i+\bar{B}\mathbf{u}_i\\
\mathbf{y}_i=&C\mathbf{x}_i
\end{cases}
\label{eq:2}
\end{equation}
where $\bar{A}~\in~\mathbb{R}^{n \times n}=A+\Delta A$ and $\bar{B} ~\in~\mathbb{R}^{n \times m}=B+\Delta B$ are the system and input matrices, respectively. Here, the $\Delta A ~\in~\mathbb{R}^{n \times n}$ and $\Delta B  ~\in~\mathbb{R}^{n \times m}$ are the perturbations of $A$ and $B$ matrices of  the nominal dynamics, respectively, due to the parametric uncertainties in the system dynamics. When $\Delta A=\mathbf{0}_{n,n}$ and $\Delta B =\mathbf{0}_{n,m}$, then \eqref{eq:2} represents nominal dynamics of the $i$th agent.\par
 In this paper,  we consider three operational scenarios for the consensus of MAS with \textit{attrition} and \textit{inclusion} of agents.
These scenarios are the following.
\begin{enumerate}
	\item \textbf{scenario 1:} There is no \textit{attrition} and \textit{inclusion} of agents in this scenario. The MAS operates with a fixed number of agents, say $N  \in \mathbb{N}_2^O$ agents, in the whole operating time.
	\item \textbf{scenario 2:} The MAS begins its operation with $N$ agents and  $P \in \mathbb{N}^{+}$  agents are later removed at a given time instant. Consequently,   the number of agents is not constant and varies between $N$ and $N-P$ over the course of the operation.  Also, note that the maximum value of  $P$ is $N-2$.
	
	\item \textbf{scenario 3:} The MAS commences its operations with $N$ agents. Afterward, at a given time instant, $M \in \mathbb{N}^+<\infty$ agents were included to the MAS. In this case, the number of agents  vaires between $N$ and $N+M$. Note that the dynamics of $M$ agents is identical to those of $N$ existing agents.
	
\end{enumerate}
Moreover, the communication network topologies between agents of the MAS are allowed to switch from time to time so that the network graph remains connected. Also, we assume that the graphs of the communication network topologies formed after the \textit{attrition} or \textit{inclusion} of agents from the MAS remain connected. Consequently, when the MAS with 
\begin{enumerate}
	\item 
	$N$ agents operates over a  time interval, then the CUS graphs and corresponding Laplacian matrices belong to the sets, $\{\breve{\mathcal{G}}_1, \dots,\breve{\mathcal{G}}_h,\dots,\breve{\mathcal{G}}_k\}$ and  $\{\breve{\mathcal{L}}_1, \dots,\breve{\mathcal{L}}_h,\dots,\breve{\mathcal{L}}_k\}$, respectively,  switch at desired time instants.
	
	\item $N-P$ agents operates over a  time interval, then the CUS graphs and corresponding Laplacian matrices belong to the sets, $\{\grave{\mathcal{G}}_1, \dots,\grave{\mathcal{G}}_h,\dots,\grave{\mathcal{G}}_r\}$ and  $\{\grave{\mathcal{L}}_1, \dots,\grave{\mathcal{L}}_h,\dots,\grave{\mathcal{L}}_r\}$, respectively,  switch at desired time instants.
	
	\item $N+M$ agents operates over a  time interval, then the CUS graphs and corresponding Laplacian matrices belong to the sets, $\{\hat{\mathcal{G}}_1, \dots,\hat{\mathcal{G}}_h,\dots,\grave{\mathcal{G}}_r\}$ and  $\{\hat{\mathcal{L}}_1, \dots,\hat{\mathcal{L}}_h,\dots,\hat{\mathcal{L}}_r\}$, respectively,  switch at desired time instants.
\end{enumerate}

In view of the aforementioned three scenarios and switching topologies,  the RAIDD consensus protocol needs to achieve consensus of the MAS with $N$, $N-P$, and  $N+M$ uncertain agents that are connected using the communication network topologies whose graphs are $\breve{\mathcal{G}}_h~\forall~h~\in~\mathbb{N}_1^k$, $\grave{\mathcal{G}}_h~\forall~h~\in~\mathbb{N}_1^r$, and $\hat{\mathcal{G}}_h~\forall~h~\in~\mathbb{N}_1^p$, respectively. Now, consider the   distributed dynamic protocol, $\mathbf{K}_i(s)$, whose state-space form is given	by
\begin{equation}
\mathbf{K}_i(s):
\begin{cases}
\mathbf{\dot{v}}_i=K_A\mathbf{v}_i+K_B \mathbf{\delta}_i \\
\mathbf{u}_i=K_C\mathbf{v}_i+K_D \mathbf{\delta}_i 
\end{cases}
\label{eq:3}
\end{equation}
where $\mathbf{v}_i, \mathbf{\delta}_i, K_A, K_B, K_C $, and $K_D$ are the state vector, input vector, system matrix, input matrix, output matrix, and feed-forward matrix of appropriate dimension, respectively. Here, $\mathbf{\delta}_i$ is defined as
\begin{align}
\mathbf{\delta}_i=
\begin{cases}
\Sigma_{j \in \mathcal{N}_i(\breve{\mathcal{G}}_{h\in\mathbb{N}_1^k})}\breve{a}_{h_{ij}}(\mathbf{x}_i-\mathbf{x}_j) &if ~\psi=N\\
\Sigma_{j \in \mathcal{N}_i(\grave{\mathcal{G}}_{h\in\mathbb{N}_1^r})}\grave{a}_{h_{ij}}(\mathbf{x}_i-\mathbf{x}_j) &if ~\psi=N-P\\
\Sigma_{j \in \mathcal{N}_i(\hat{\mathcal{G}}_{h\in\mathbb{N}_1^p})}\hat{a}_{h_{ij}}(\mathbf{x}_i-\mathbf{x}_j) &if ~\psi=N+M\\
\end{cases}
\end{align}
where  $\psi   \in \mathbb{N}_2^O $ ($2$ $<$ $O \in$ $\mathbb{N}$ $<$ $\infty$) is the number of agents of the MAS during its operating time. Let $\mathbf{x}_a$, $\mathbf{x}_b$, and $\mathbf{x}_c$  are defined as $\mathbf{x}_a$=$[\mathbf{x}_1^T,\dots,\mathbf{x}_N^T]^T$, $\mathbf{x}_b$=$[\mathbf{x}_1^T,\dots,\mathbf{x}_{N-P}^T]^T$, $\mathbf{x}_c$ = $[\mathbf{x}_1^T,\dots,\mathbf{x}_{N+M}^T]^T$, $\mathbf{v}_a$ = $[\mathbf{v}_1^T,\dots,\mathbf{v}_N^T]^T$, $\mathbf{v}_b$ = $[\mathbf{v}_1^T,\dots,\mathbf{v}_{N-P}^T]^T$, and  $\mathbf{v}_c$ = $[\mathbf{v}_1^T,\dots,\mathbf{v}_{N+M}^T]^T$, respectively. Then, the CL systems of the MAS with switching topologies are obtained by interconnecting the agents whose dynamics given in \eqref{eq:2} using  \eqref{eq:3}. These CL systems associated with $N$, $N-P$, and  $N+M$ uncertain agents  are given by
	\begin{equation}
	\begin{aligned}
	\begin{bmatrix}
	\mathbf{\dot{x}}_a \\
	\mathbf{\dot{v}}_a
	\end{bmatrix}
	=\begin{bmatrix}
	\mathbf{I}_N \bigotimes \bar{A} + \breve{\mathcal{L}}_h \bar{B}K_D& \mathbf{I}_N \bigotimes \bar{B} K_C \\
	\breve{\mathcal{L}}_h \bigotimes K_B& \mathbf{I}_N \bigotimes K_A
	\end{bmatrix}\begin{bmatrix}
	\mathbf{x}_a \\
	\mathbf{v}_a
	\end{bmatrix}\\~\forall~h~\in~\mathbb{N}_1^k,
	\end{aligned}
	\label{eq:4}
	\end{equation}
	\begin{equation}
	\begin{aligned}
	\begin{bmatrix}
	\mathbf{\dot{x}}_b \\
	\mathbf{\dot{v}}_b
	\end{bmatrix}
	=\begin{bmatrix}
	\mathbf{I}_{N-P} \bigotimes \bar{A} + \grave{\mathcal{L}}_h \bar{B}K_D& \mathbf{I}_{N-P} \bigotimes \bar{B} K_C \\
	\grave{\mathcal{L}}_h \bigotimes K_B& \mathbf{I}_{N-P} \bigotimes K_A
	\end{bmatrix}\begin{bmatrix}
	\mathbf{x}_b \\
	\mathbf{v}_b
	\end{bmatrix}\\~\forall~h~\in~\mathbb{N}_1^r,
	\end{aligned}
	\label{eq:6}
	\end{equation}
and 
	\begin{equation}
	\begin{aligned}
	\begin{bmatrix}
	\mathbf{\dot{x}}_c \\
	\mathbf{\dot{v}}_c
	\end{bmatrix}
	=\begin{bmatrix}
	\mathbf{I}_{N+M} \bigotimes \bar{A} + \hat{\mathcal{L}}_h \bar{B}K_D& \mathbf{I}_{N+M} \bigotimes \bar{B} K_C \\
	\hat{\mathcal{L}}_h \bigotimes K_B& \mathbf{I}_{N+M} \bigotimes K_A
	\end{bmatrix}\begin{bmatrix}
	\mathbf{x}_c \\
	\mathbf{v}_c
	\end{bmatrix}\\~\forall~h~\in~\mathbb{N}_1^p,
	\end{aligned}
	\label{eq:8}
	\end{equation}
respectively.
Subsequently, the protocol given in \eqref{eq:3} achieves:
\begin{enumerate}
	\item consensus of the MAS of $N$ uncertain agents  with  $\breve{\mathcal{G}}_h~\forall~h~\in~\mathbb{N}_1^k$ when the CL systems given in \eqref{eq:4}
 satisfy 
	\begin{align}
	\lim_{t \to \infty} (\mathbf{x}_i-\mathbf{x}_j)=0~\forall~i, j~\in~\mathbb{N}_1^N
		\label{eq:5}
	\end{align}
	\item consensus of the MAS of $N-P$ uncertain agents  with  $\grave{\mathcal{G}}_h~\forall~h~\in~\mathbb{N}_1^r$ when the CL systems given in \eqref{eq:6}
 satisfy 
	\begin{align}
	\lim_{t \to \infty} (\mathbf{x}_i-\mathbf{x}_j)=0~\forall~i, j~\in~\mathbb{N}^{N-P}_1
		\label{eq:7}
	\end{align}
	\item consensus of the MAS of $N+M$ uncertain agents  with  $\hat{\mathcal{G}}_h~\forall~h~\in~\mathbb{N}_1^p$ when the CL systems given in  \eqref{eq:8}
	 satisfy 
	\begin{align}
	\lim_{t \to \infty} (\mathbf{x}_i-\mathbf{x}_j)=0~\forall~i, j~\in~\mathbb{N}^{N+M}_1
		\label{eq:9}
	\end{align}
\end{enumerate}
Considering the CL systems given by  \eqref{eq:4}, \eqref{eq:6}, and \eqref{eq:8}, the protocol given by \eqref{eq:3} becomes the RAIDD consensus protocol when \eqref{eq:3} achieves \eqref{eq:5}, \eqref{eq:7}, and \eqref{eq:9}, respectively. Therefore, the RAI consensus problem  is stated as the following. Determine the existence conditions of the $K_A$,  $K_B$, $K_C$, and $K_D$  such that the CL systems given in \eqref{eq:4}, \eqref{eq:6}, and \eqref{eq:8} satisfy \eqref{eq:5}, \eqref{eq:7}, and \eqref{eq:9}, respectively.
 \section{Synthesize of Robust Attrition-Inclusion Distributed  Consensus Protocol}\label{PR}
The sufficient condition for the existence of $K_A$,  $K_B$, $K_C$, and $K_D$,  as well as the design procedure for the RAIDD consensus protocol,  are proposed in this section. To develop this sufficient condition, we define the following. \par
\begin{enumerate}
	\item Let  $\mathcal{P}$ be a finite set that contains  all the eigenvalues of 	$\breve{\mathcal{L}}_h~\forall~ h\in\mathbb{N}_1^k$, $\grave{\mathcal{L}}_h~\forall~ h\in\mathbb{N}_1^r$, and $\hat{\mathcal{L}}_h~\forall~ h\in\mathbb{N}_1^p$ except their first eigenvalue, zero. The $i$th element of $\mathcal{P}$ is denoted by $\lambda_{i}$. Also, the cardinality of $\mathcal{P}$ is $\textbf{\textit{n}}(\mathcal{P})=\xi$  where $\xi=k(N-1)+r(N-P-1)+p(N+M-1)$. Now,   $\mathbf{\hat{P}}_{i}(s) ~\forall~i \in \mathbb{N}_1^{\xi}$  are defined as
	\begin{equation}
\mathbf{\hat{P}}_{i}(s):
	\begin{cases}
	\mathbf{\dot{\hat{x}}}_i=A\mathbf{\hat{x}}_{i}+\lambda_{i} B\mathbf{\hat{u}}_{i}\\
	\mathbf{\hat{y}}_{i}=C\mathbf{\hat{x}}_{i};~\forall~i \in \mathbb{N}_1^{\xi}
	\end{cases}
	\label{eq:10}
	\end{equation}
	where $\mathbf{\hat{x}}_{i} \in \mathbb{R}^n$, $\mathbf{\hat{u}}_{i} \in \mathbb{R}^m$, and $\mathbf{\hat{y}}_{i}  \in \mathbb{R}^n$ are the state, input, and output vectors, respectively. Also, ($A, \lambda_{i}B$) is stabilizable.
		\item 
		Define  $\mathcal{Q}$ as $\mathcal{Q}=\{\mathbf{\hat{P}}_i(s)~|~\mathbf{\hat{P}}_i(s)=:(A,\lambda_{i}B,C,D=\mathbf{0}_{n,m})~\forall~ i \in \mathbb{N}_1^\xi\}$. The maximum $\nu$-gap metric of the $i$th system belongs to $\mathcal{Q}$     is given by 
	\begin{equation}
	\begin{split}
\epsilon_i=&\max\{\delta_{\nu}(\hat{\mathbf{P}}_i(j\omega), \hat{\mathbf{P}}_f(j\omega))~|~\hat{\mathbf{P}}_i(j\omega), \hat{\mathbf{P}}_f(j\omega)~\in \\&\qquad \mathcal{Q}, \forall~f ~\in~\mathbb{N}_1^\xi	\}\in [0,1].
\end{split}
\label{eq:13}
\end{equation}
Also, the   central plant of $\mathcal{Q}$  and its  maximum $\nu$-gap metric be $\hat{\mathbf{P}}_{cp}(s) \in \mathcal{Q}$ and  $\epsilon_{{cp}}$, respectively. Furthermore, the normalized right coprime factors of $\hat{\mathbf{P}}_{cp}(s)$ are $\hat{\mathbf{N}}_{cp}(s) \in \mathcal{RH_\infty}$ and $\hat{\mathbf{M}}_{cp}(s) \in \mathcal{RH_\infty}$ with det($\hat{\mathbf{M}}_{cp}(s) \neq 0$).
 \item Let  $\mathbf{\acute{P}}_i(s)~$ $\forall$ $i \in \mathbb{N}_1^\xi$ be the perturbed systems  of $\mathbf{\hat{P}}_i(s) \in {\mathcal{Q}}~\forall~i~\in~\mathbb{N}_1^\xi$, respectively. These  systems  arise due to the perturbation of  $A$ and $B$ matrices  mentioned in \eqref{eq:10} to form $\bar{A}=A+\Delta A$ and $\bar{B}=B+\Delta B$. Subsequently,  the function, $\Psi:\mathbb{R}^{n \times n} \times \mathbb{R}^{n \times m} \rightarrow [0,1] $ is defined for  $dom(\Psi)=\{(\Delta A, \Delta B)~|~\Delta A \in \mathbb{R}^{n \times n}, \Delta B \in \mathbb{R}^{n \times m}\}$ as
 \begin{equation}
 \begin{split}
 \Psi=&\max\{\delta_{\nu}(\mathbf{\hat{P}}_{cp}(j\omega),\mathbf{\acute{P}}_i(j\omega))~|~\mathbf{\hat{P}}_{cp}(s)~\in~\mathcal{Q}, \\&\qquad ~\mathbf{\acute{P}}_i(s) =:(\bar{A}, \lambda_{i}\bar{B}, C,D )~\forall~ i  \in~\mathbb{N}_1^\xi\}.
 \end{split}
 \label{eq:15}
 \end{equation}
Also, note that \eqref{eq:13}, \eqref{eq:15},  and $dom(\Psi)$ indicate that    $\Psi=\epsilon_{{cp}}$ 	when $\Delta A=\mathbf{0}_{n,n}$ and $\Delta B=\mathbf{0}_{n,m}$.
\end{enumerate}\par
In the following theorem, the sufficient condition for the existence of $K A$, $K B$, $K C$, and $K D$ is proposed.
	\newtheorem{theorem}{Theorem}[section]
	\begin{theorem}
	The $K_A$,  $K_B$, $K_C$, and $K_D$  exist such that the closed-loop systems given in \eqref{eq:4}, \eqref{eq:6}, and \eqref{eq:8} satisfy \eqref{eq:5}, \eqref{eq:7}, and \eqref{eq:9}, respectively, if the condition,
	\begin{equation}
\sqrt{(1-\parallel [\mathbf{\hat{N}}_{{cp}}(s) ~\mathbf{\hat{M}}_{{cp}}(s)]^T\parallel_H^2)}>\Psi,
\label{h1}
\end{equation}
		is true.
		\label{thm1}
	\end{theorem}
	\begin{proof}
	\textbf{Case~a}:  Let   $\Delta A$ and $ \Delta B$ be $\mathbf{0}_{n,n}$ and $\mathbf{0}_{n,m}$, respectively. When $\Delta A = \mathbf{0}_{n,n}$ and $ \Delta B=\mathbf{0}_{n,m}$, then \textit{Proposition 2} mentioned in  \cite{huy} suggests that the CL systems   given in \eqref{eq:4}, \eqref{eq:6}, and \eqref{eq:8}  satisfy \eqref{eq:5}, \eqref{eq:7}, and \eqref{eq:9}, respectively, if all the systems belong to $\mathcal{Q}$ can be simultaneously stabilized using a   full state feedback controller, $\mathbf{\hat{K}}(s)$, which is given as
	\begin{equation}
	\mathbf{\hat{K}}(s):
	\begin{cases}
	\mathbf{\dot{\hat{v}}}&=K_A\mathbf{\hat{v}}+K_B \mathbf{\hat{x}}_{i} \\
	\mathbf{\hat{u}}_{i}&=K_C\mathbf{\hat{v}}+K_D \mathbf{\hat{x}}_{i}
	\end{cases}
	\label{eq:17}
	\end{equation}
where $\mathbf{\hat{v}}$ is the state vector with appropriate dimension. Consequently, the consensus problem of MAS with $N$, $N-P$, and $N+M$  agents with their nominal dynamics (given in \eqref{eq:1})  is equivalent to the following SS problem. Determine a full state feedback controller of the form given in \eqref{eq:17} that simultaneously stabilizes all the systems belonging to $\mathcal{Q}$. Let us assume $\mathbf{\hat{P}}_{cp}$ and $\epsilon_{{cp}}$ are identified
for developing the existence condition for the RAIDD consensus protocol based on  the concept of   $\nu$-gap metric-based SS method proposed in \cite{jinsmc}-\cite{jinjgcd}. Then, the  sufficient condition for the SS all the systems belong to $\mathcal{Q}$  is given as 
	\begin{equation}
	b_{\mathbf{\hat{P}}_{cp},\mathbf{\hat{K}}}>\epsilon_{{cp}}
	\label{sscondk1}
	\end{equation}
Equation \eqref{eq:RCF1_1_1} implies that a controller needs to be obtained in the first place to solve \eqref{sscondk1}. Hence,  \eqref{sscondk1} does not indicate the existence of a simultaneous stabilizing controller without synthesizing a controller.		
To derive the controller independent sufficient condition for the SS of all the systems belonging to $\mathcal{Q}$, let 	$b_{\mathbf{\hat{P}}_{cp},\mathbf{\hat{K}}}^{max}$ be the maximum generalized stability margin  of $\mathbf{\hat{P}}_{cp}(s)$. This margin  indicates the largest infinity norm of  left/right coprime factor perturbations for which $[\mathbf{\hat{P}}_{cp}(s), \mathbf{\hat{K}}(s)] $ remains stable. From \cite{glover1}, $b_{\mathbf{\hat{P}}_{cp},\mathbf{\hat{K}}}^{max}$  is given by 
	\begin{equation}
	b_{\mathbf{\hat{P}}_{cp},\mathbf{\hat{K}}}^{max}=\sqrt{(1-\parallel [\mathbf{\hat{N}}_{{cp}}(s) ~\mathbf{\hat{M}}_{{cp}}(s)]^T\parallel_H^2)}
	\label{sscondk2}
	\end{equation}
Further,  assume $\mathcal{Q}$ as an uncertainty  set with   $\mathbf{\hat{P}}_{cp}(s) \in \mathcal{Q}$ as its nominal system and the systems  belonging to $\mathcal{Q}\setminus\{\mathbf{\hat{P}}_{cp}(s)\}$ as the   perturbed systems of $\mathbf{\hat{P}}_{cp}(s)$. Let   $\mathbf{\hat{P}}_{f}(s)$ $\in$ $\mathcal{Q} \setminus \{\mathbf{\hat{P}}_{cp}(s)\}$ be a perturbed system of $\mathbf{\hat{P}}_{cp}(s)$. Also, $\epsilon_{{}_{{cp}f}}$ be the least upper bound on the normalized right coprime factor perturbations of $\mathbf{\hat{P}}_{cp}(s)$ to form  $\mathbf{\hat{P}}_f(s)$. Additionally,  $\mathbf{\Delta}_{\hat{N}_{{cp}f}}(s) \in \mathcal{RH_\infty}^{}$ and  $\mathbf{\Delta}_{\hat{M}_{{cp}f}}(s) \in \mathcal{RH_\infty}^{}$ be the normalized right coprime factor perturbations of $\mathbf{\hat{N}}_{cp}(s)$ and $\mathbf{\hat{M}}_{cp}(s)$, respectively.  These perturbations satisfy the condition given by
\begin{equation}
\big|\big|[\begin{array}{cc}\mathbf{\Delta}_{\hat{N}_{{cp}f}}(s)&  \mathbf{\Delta}_{\hat{M}_{{cp}f}}(s)\end{array}]^{T}\big|\big|_\infty\leq\epsilon_{{}_{{cp}f}}.
\label{eq:per1}
\end{equation}
Subsequently, $\mathbf{\hat{P}}_{cp}(s)$ and $\mathbf{\hat{P}}_f(s)$ are defined as
	\begin{align}
	\mathbf{\hat{P}}_{cp}(s)&=\mathbf{\hat{N}}_{cp}(s)\mathbf{\hat{M}}_{cp}^{-1}(s)\label{copr}\\
	\mathbf{\hat{P}}_f(s)&=\big(\mathbf{\hat{N}}_{cp}(s)+\mathbf{\Delta}_{\hat{N}_{{cp}f}}(s)\big)\big(\mathbf{\hat{M}}_{cp}(s)+ \mathbf{\Delta}_{\hat{M}_{{cp}f}}(s)\big)^{-1}
	\label{copr1}
	\end{align}
Now, there always exists a full state feedback controller of the form given in \eqref{eq:17} that stabilizes both $\mathbf{\hat{P}}_{cp}(s)$ and $\mathbf{\hat{P}}_{f}(s)$ when the condition given by 
\begin{equation}
b_{\mathbf{\hat{P}}_{cp},\mathbf{\hat{K}}}^{max}=\sqrt{(1-\parallel [\mathbf{\hat{N}}_{{cp}}(s) ~\mathbf{\hat{M}}_{{cp}}(s)]^T\parallel_H^2)}>\epsilon_{{}_{cpf}}
\label{eq:cond}
\end{equation}
 holds \cite{glover2}. 
Even though $\mathcal{Q}$ is considered as a uncertainty set,  the systems belong to it  are known. Therefore, the condition given in  \eqref{eq:per1} becomes 
\begin{equation}
  \epsilon_{{}_{{cp}f}}=\big|\big|[\begin{array}{cc}\mathbf{\Delta}_{\hat{N}_{{cp}f}}(s)&  \mathbf{\Delta}_{\hat{M}_{{cp}f}}(s)\end{array}]^{T}\big|\big|_\infty. 
\end{equation}
Also, the relation between $\delta_v(\mathbf{\hat{P}}_{cp}(j\omega),\mathbf{\hat{P}}_{f}(j\omega))$  and their normalized right coprime factor perturbations is given as \cite{feyel}
\begin{equation}
\delta_v(\mathbf{\hat{P}}_{cp}(j\omega),\mathbf{\hat{P}}_{f}(j\omega))=\big|\big|[\begin{array}{cc}\mathbf{\Delta}_{\hat{N}_{{cp}f}}(s)&  \mathbf{\Delta}_{\hat{M}_{{cp}f}}(s)\end{array}]^{T}\big|\big|_\infty. 
\label{eq:per2}
\end{equation}
Let    the largest infinity norm of  the right coprime factor perturbations between  $\mathbf{\hat{P}}_{cp}(s)$ and the systems belong to $\mathcal{Q}\setminus\{\mathbf{\hat{P}}_{cp}(s)\}$ be $\epsilon_{{}_{cpf}}^{max}$. Following \eqref{eq:per1} and \eqref{eq:per2}, $\epsilon_{{}_{cpf}}^{max}$  can be written as  \cite{jinjgcd}
\begin{equation}
\begin{split}
\epsilon_{{}_{cpf}}^{max}=&\max\{\delta_v(\mathbf{\hat{P}}_{cp}(j\omega),\mathbf{\hat{P}}_{f}(j\omega))~|~\mathbf{\hat{P}}_{cp}(s), \mathbf{\hat{P}}_{f}(s) \in \mathcal{Q}, \\ &~\qquad \forall~f \in \mathbb{N}_1^\xi\}. 
\end{split}
\label{eq:per3}
\end{equation}
Using the definition of maximum $\nu$-gap metric, we can write \eqref{eq:per3} as
\begin{equation}
\epsilon_{{}_{cpf}}^{max}=\epsilon_{{cp}}.
\label{eq:per4}
\end{equation}
Substituting \eqref{eq:per4} in \eqref{eq:cond} results in
\begin{equation}
	b_{\mathbf{\hat{P}}_{cp},\mathbf{\hat{K}}}^{max}=\sqrt{(1-\parallel [\mathbf{\hat{N}}_{{cp}}(s) ~\mathbf{\hat{M}}_{{cp}}(s)]^T\parallel_H^2)}> \epsilon_{{cp}}
	\label{sscondk3}
	\end{equation} 
When \eqref{sscondk3}	holds, then  there always exists a full state feedback controller  of the form given in \eqref{eq:17} that  simultaneously stabilizes all the plants of $\mathcal{Q}$ as $\epsilon_{{cp}} \geq \epsilon_{{}_{cpf}}~\forall~f~\in~\mathbb{N}_1^\xi$. Hence, there exists $K_A$, $K_B$, $K_C$, and $K_D$ such that that the CL systems   given in \eqref{eq:4}, \eqref{eq:6}, and \eqref{eq:8} with $\Delta A=\mathbf{0}_{n,n}$ and $ \Delta B=\mathbf{0}_{n,m}$ satisfy \eqref{eq:5}, \eqref{eq:7}, and \eqref{eq:9}, respectively. Moreover, it is important to note that validating \eqref{sscondk3} does not require any controller. \par 
\noindent \textbf{Case b:}  Here, let $\Delta A\neq\mathbf{0}_{n,n}$ and $ \Delta B\neq\mathbf{0}_{n,m}$. Furthermore,  consider two ball of systems, $\mathcal{B}(\mathbf{\hat{P}}_{cp}(s), b_{\mathbf{\hat{P}}_{cp},\mathbf{\hat{K}}}^{max})=\{\mathbf{G}(s)~|~\delta_{\nu}(\mathbf{\hat{P}}_{cp}(s),\mathbf{{G}}(s))<b_{\mathbf{\hat{P}}_{cp},\mathbf{\hat{K}}}^{max}\}$ and $\mathcal{B}(\mathbf{\hat{P}}_{cp}(s), \epsilon_{{cp}})=\{\mathbf{\hat{G}}(s)~|~\delta_{\nu}(\mathbf{\hat{P}}_{cp}(s),\mathbf{\hat{G}}(s))\leq\epsilon_{{cp}}\}$. The $\nu$-gap metrics between  $\mathbf{\hat{P}}_{cp}(s)$ and all the systems belong to $ \mathcal{B}(\mathbf{\hat{P}}_{cp}(s), \epsilon_{{cp}})$ is less than or equal to $\epsilon_{{cp}}$ which is less than $\sqrt{(1-\parallel [\mathbf{\hat{N}}_{{cp}}(s) ~\mathbf{\hat{M}}_{{cp}}(s)]^T\parallel_H^2)}$. Hence, if \eqref{sscondk3} holds, then there exists a full state feedback controller that simultaneously stabilizes all the systems belong to $\mathcal{Q}$, $\mathcal{B}(\mathbf{\hat{P}}_{cp}(s), \epsilon_{{cp}})$, and $\mathcal{B}(\mathbf{\hat{P}}_{cp}(s), b_{\mathbf{\hat{P}}_{cp},\mathbf{\hat{K}}}^{max})$. Subsequently, for the existence of a full state feedback controller that simultaneously stabilizes all the perturbed systems, $\mathbf{\acute{P}}_i(s)~\forall~i \in \mathbb{N}_1^\xi$, necessitates 
	\begin{align}
	\mathbf{\acute{P}}_i(s) \in  \mathcal{B}(\mathbf{\hat{P}}_{cp}(s), b_{\mathbf{\hat{P}}_{cp},\mathbf{\hat{K}}}^{max})~\forall~i \in \mathbb{N}_1^\xi.
	\label{eq:18}
	\end{align}
For \eqref{eq:18} to hold, the $\nu$-gap metrics between $\mathbf{\hat{P}}_{cp}(s)$ and 	$\mathbf{\acute{P}}_i(s)$  $\forall$ $ i \in \mathbb{N}_1^\xi$ need to be less than $\sqrt{(1-\parallel [\mathbf{\hat{N}}_{{cp}}(s) ~\mathbf{\hat{M}}_{{cp}}(s)]^T\parallel_H^2)}$, i.e., $\sqrt{(1-\parallel [\mathbf{\hat{N}}_{{cp}}(s) ~\mathbf{\hat{M}}_{{cp}}(s)]^T\parallel_H^2)} > \delta_\nu(\mathbf{\hat{P}}_{cp}(s), \mathbf{\acute{P}}_i(s)) ~\forall~i \in \mathbb{N}_1^\xi$. The  condition given in \eqref{h1} needs to be true for this to happen. Note that the condition given in \eqref{h1} and the condition given in \eqref{sscondk3} will be the same when $\Delta A=\mathbf{0}_{n,n}, \Delta B =\mathbf{0}_{n,m}$ as  ($\Delta A=\mathbf{0}_{n,n}, \Delta B =\mathbf{0}_{n,m}) \in dom(\Psi)$. Hence, there exists a full state feedback controller that simultaneously stabilizes all the systems, $\mathbf{{P}}_i(s)~\forall~i \in \mathbb{N}_1^\xi$, and its perturbed systems, $\mathbf{\acute{P}}_i(s)~\forall~i \in \mathbb{N}_1^\xi$, if the condition given in \eqref{h1} holds true.  Therefore, $K_A$,  $K_B$, $K_C$, and $K_D$  exist such that the closed-loop systems given in \eqref{eq:4}, \eqref{eq:6}, and \eqref{eq:8} satisfy \eqref{eq:5}, \eqref{eq:7}, and \eqref{eq:9}, respectively, if  the condition given in \eqref{h1} holds. This establishes the proof.
		\end{proof}
  To realize sufficient condition given in \eqref{h1} requires $\mathbf{\hat{P}}_{cp}(s)$, $\epsilon_{cp}$, $\Delta A$, and $\Delta B$.  Here, $\mathbf{\hat{P}}_{cp}(s)$ and $\epsilon_{cp}$  is identified by following the steps given in Section \ref{SSCP}.  Once $\mathbf{P}_{cp}(s)$ is identified, then $\mathbf{\Delta}_{\hat{N}_{{cp}f}}(s) $ and  $\mathbf{\Delta}_{\hat{M}_{{cp}f}}(s)$ are obtained and thereafter $\sqrt{(1-\parallel [\mathbf{\hat{N}}_{{cp}}(s) ~\mathbf{\hat{M}}_{{cp}}(s)]^T\parallel_H^2)}$ is computed.  In this paper, we consider real parameter perturbations and therefore the   $\Delta A$  and $\Delta B$ are specified through the individual upper and lower bound on the elements of $\Delta A$  and $\Delta B$.  These bounds define a set of  $(\Delta A, \Delta B)$ as given by
 \begin{align}
 \begin{split}
\Xi=&\{( \Delta A,\Delta B)~|~\Delta A \in \mathbb{R}^{n \times n},\Delta B \in \mathbb{R}^{n \times m}, b_{ij}\geq \Delta A_{ij}  \\\qquad & \geq -a_{ij}, d_{iw}\geq \Delta B_{ij} \geq -c_{iw}, \forall ~ i,j \in \mathbb{N}_1^n, w \in \mathbb{N}_1^m \}
\end{split}
 \end{align}
where $a_{ij} \in \mathbb{R}_{\geq 0}$, $b_{ij} \in \mathbb{R}_{\geq 0} $, $c_{iw} \in \mathbb{R}_{\geq 0}$, and $d_{iw}\in \mathbb{R}_{\geq 0}$. It is possible to compute the value of $\Psi$ associated with each element of $\Xi$ and see if these values satisfy the condition given in \eqref{h1}. Those  values of $\Psi$ that satisfy the condition given in \eqref{h1}  yield upper and lower bounds on the elements of $\Delta A$ and $\Delta B$ for which consensus of $N$, $N-P$, and $N+M$ uncertain agents can be achieved. Consequently, the sufficient condition stated by \eqref{h1} is tractable as $\sqrt{(1-\parallel [\mathbf{\hat{N}}_{{cp}}(s) ~\mathbf{\hat{M}}_{{cp}}(s)]^T\parallel_H^2)}$ and $\Psi$  are determinable.
Now,  $\mathbf{\hat{K}}(s)$ is obtained utilizing the Glover-McFarlane method proposed in \cite{glover2} using  Matlab function, $\textit{ncfsys}$,  when the condition given in \eqref{h1} holds for the desired set of ($\Delta A, \Delta B)$. $\textit{ncfsys}$ finds a $\mathbf{\hat{K}}(s)$ that achieves the maximum generalized stability margin that is equal to $\sqrt{(1-\parallel [\mathbf{\hat{N}}_{{cp}}(s) ~\mathbf{\hat{M}}_{{cp}}(s)] ^T\parallel_H^2)}$. Hence, $\mathbf{\hat{K}}(s)$ stabilizes all the systems belonging to $\mathcal{B}(\mathbf{\hat{P}}_{cp}(s), b_{\mathbf{\hat{P}}_{cp},\mathbf{\hat{K}}}^{max})$. Once $\mathbf{\hat{K}}(s)$ is attained, $\mathbf{K}(s)$ is  established using the state-space matrices of $\mathbf{\hat{K}}(s)$. 
Eventually, the  RAIDD consensus protocol given in \eqref{eq:3}   can be synthesized by following the steps given below.
 \begin{enumerate}
\itema \textbf{Step~1:} Find the values of the maximum $\nu$-gap metrics of all the plants using (\ref{eq:10}). 
\itemb \textbf{Step~2:} Identify the smallest value of the maximum $\nu$-gap metrics. This gives $\epsilon_{{cp}}$ and the system associated with this value is the central plant. 
\itemc \textbf{Step~3:} \textbf{If}  $\sqrt{(1-\parallel [\mathbf{\hat{N}}_{{cp}}(s) ~\mathbf{\hat{M}}_{{cp}}(s)]^T\parallel_H^2)}>\epsilon_{{cp}} $  true, \textbf{then} \begin{enumerate}
 		\item Obtain the set of 
 		 $\Delta A$  and $\Delta B$ by varying  their elements  within  lower bound and upper bound. For this set,  compute corresponding set of  $\Psi$. 
 		 	\item \textbf{if} $\sqrt{(1-\parallel [\mathbf{\hat{N}}_{{cp}}(s) ~\mathbf{\hat{M}}_{{cp}}(s)]^T\parallel_H^2)}>\Psi~\forall~\Delta A, \Delta B$ \textbf{then} Synthesize $\mathbf{\hat{K}}(s)$ using \textit{ncfsys} and thereafter establish $\mathbf{K}(s)$  using the state-space matrices of $\mathbf{\hat{K}}(s)$.
 		 	\textbf{Else} \textbf{stop}.
 		 \end{enumerate}
\itemd \textbf{Step~4:}	 \textbf{Else} \textbf{stop}.
 \end{enumerate}
	\section{Simulation Results}\label{SR}
\begin{figure}[h!]
	\centering
	\subfigure[ \label{fig:g1}]{\includegraphics[width=0.8in, height=0.8in]{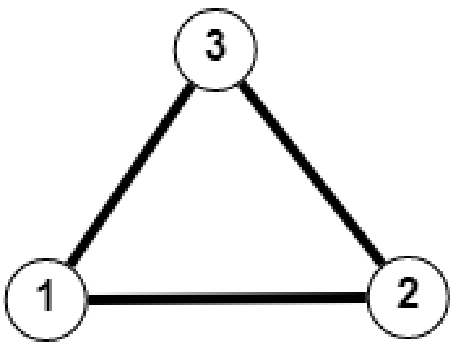}}
	\subfigure[ \label{fig:g2}]{\includegraphics[width=0.8in, height=0.8in]{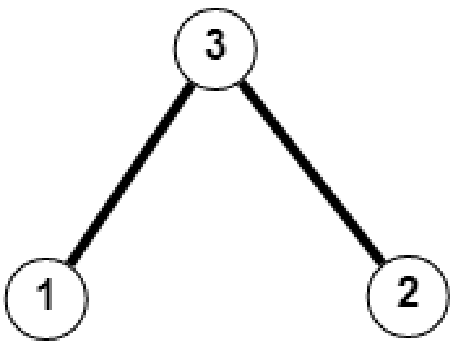}}
	\subfigure[ \label{fig:g3}]{\includegraphics[width=0.8in, height=0.8in]{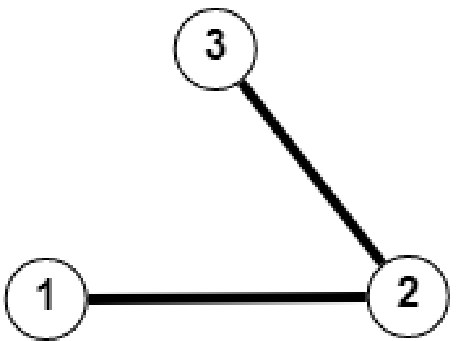}}
	\subfigure[ \label{fig:g4}]{\includegraphics[width=0.8in, height=0.8in]{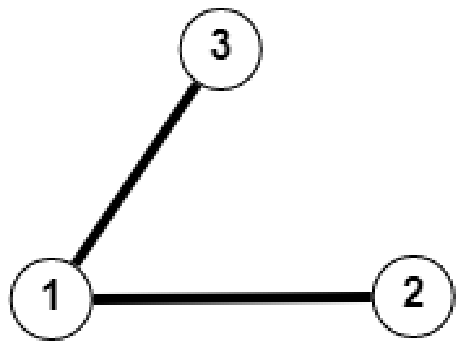}}
	\caption{CUS graphs associated with  3 UUVs}
\end{figure}

\begin{figure}[h!]
 	\centering
 	\subfigure[ \label{fig:g8}]{\includegraphics[width=0.9in, height=0.85in]{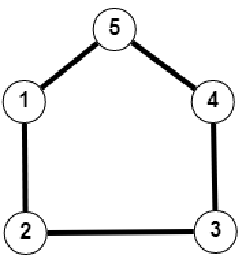}}
 	\subfigure[ \label{fig:g9}]{\includegraphics[width=0.95in, height=0.85in]{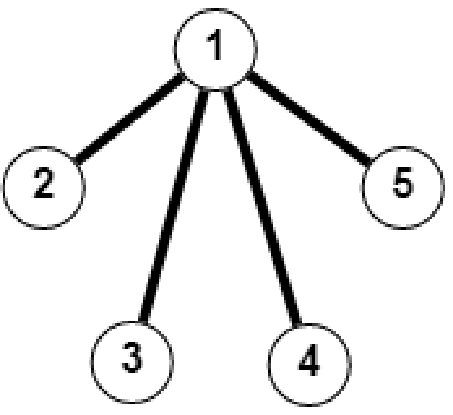}}
 	\subfigure[ \label{fig:g10}]{\includegraphics[width=0.9in, height=0.85in]{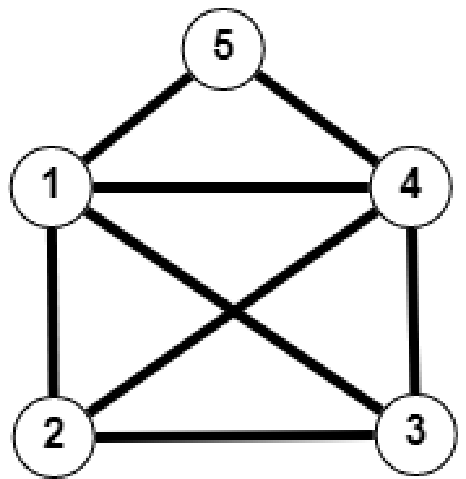}}
 	\caption{CUS graphs associated with 5 UUVs}
 \end{figure}

\begin{figure}[h!]
	\centering
	{\includegraphics[width=2.5in, height=1.65in]{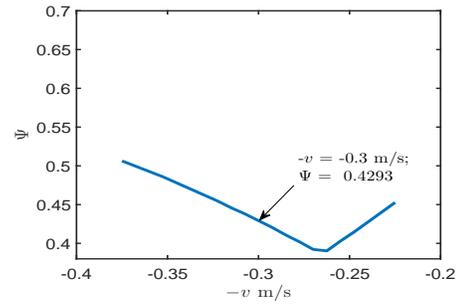}}
	\caption{Trajectory of $\Psi$ when  -$v$ is varied from  0.3750 to 0.2250}. 
	\label{fig:2}
\end{figure}

\begin{figure}[h!]
	\centering
	{\includegraphics[width=2.5in, height=1.65in]{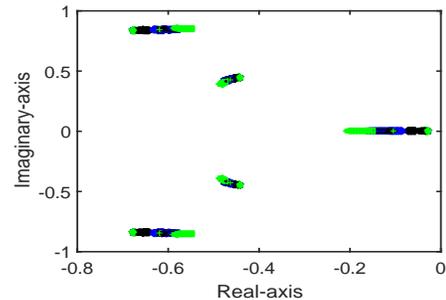}}
	\caption{Trajectories of all the eigenvalues of $[\mathbf{\acute{P}}_i(s), \mathbf{\hat{K}}(s)]~ \forall~i \in \mathbb{N}_1^{29}$ when -$v$ is varied from  0.3750 to 0.2250}. 
	\label{fig:3}
\end{figure}

\begin{figure}[H]
	\centering
	\subfigure[\textbf{case 1}: pitch angular velocity response \label{fig:4}]{\includegraphics[width=3.5in, height=1.65in]{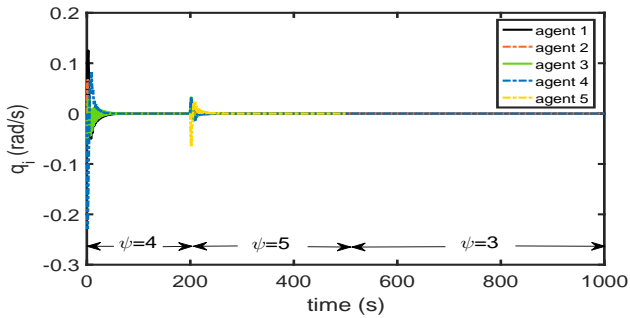}}
	\subfigure[\textbf{case 1}: pitch angle  response  \label{fig:5}]{\includegraphics[width=3.5in, height=1.65in]{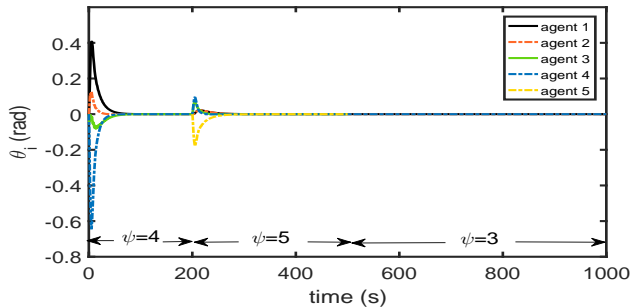}}
	\subfigure[\textbf{case 1}: depth response  \label{fig:6}]{\includegraphics[width=3.5in, height=1.65in]{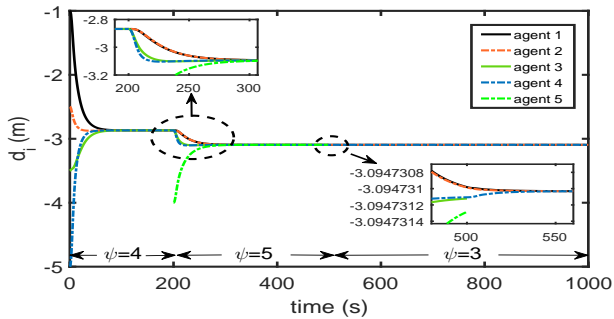}}
	\caption{Response of the CL MAS that begins its operation with four UUVs. Later, the fifth UUV is added to the CL MAS at the 200th~s and two UUVs are removed from the CL MAS at the 500th~s. Also, $v_0$=0.3~m/s and the communication network topologies of the CL MAS are switched at 1~s.}
	\label{fig:a}
\end{figure}

\begin{figure}[H]
		\centering
	\subfigure[\textbf{case 2}: pitch angular velocity response \label{fig:7}]{\includegraphics[width=3.5in, height=1.65in]{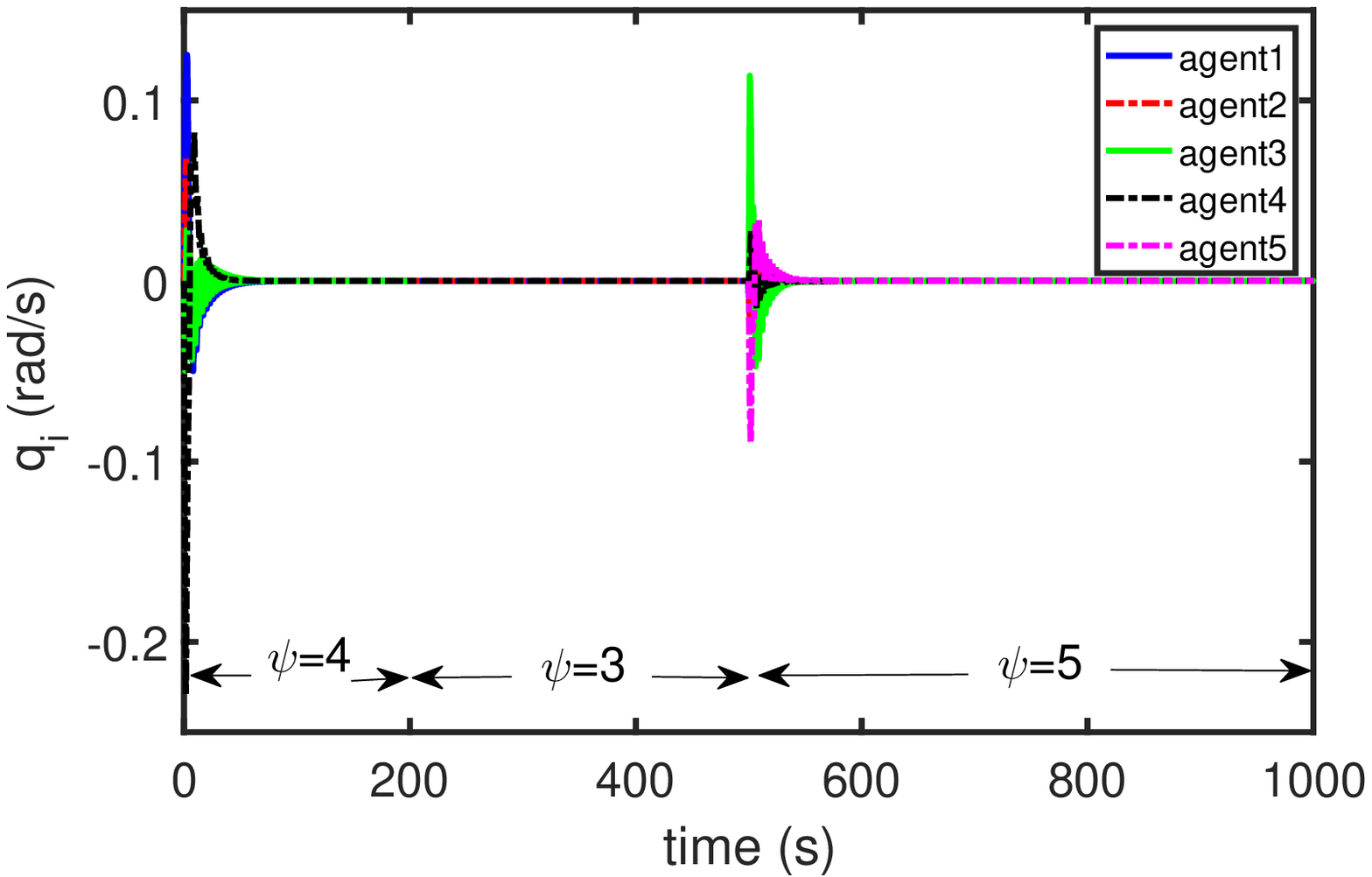}}
	\centering
	\subfigure[\textbf{case 2}: pitch angle  response  \label{fig:8}]{\includegraphics[width=3.5in, height=1.65in]{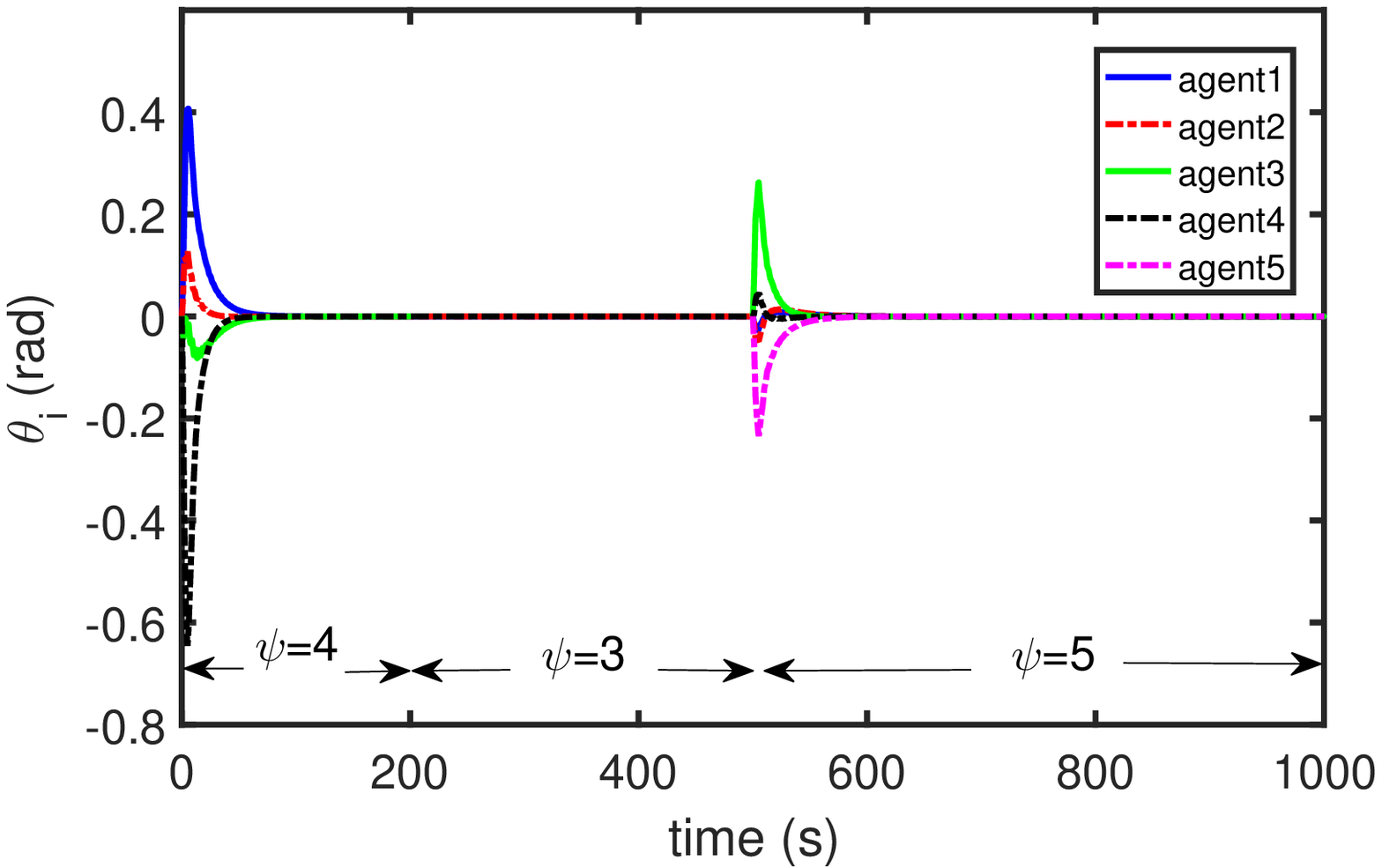}}
	\subfigure[\textbf{case 2}: depth  response  \label{fig:9}]{\includegraphics[width=3.5in, height=1.65in]{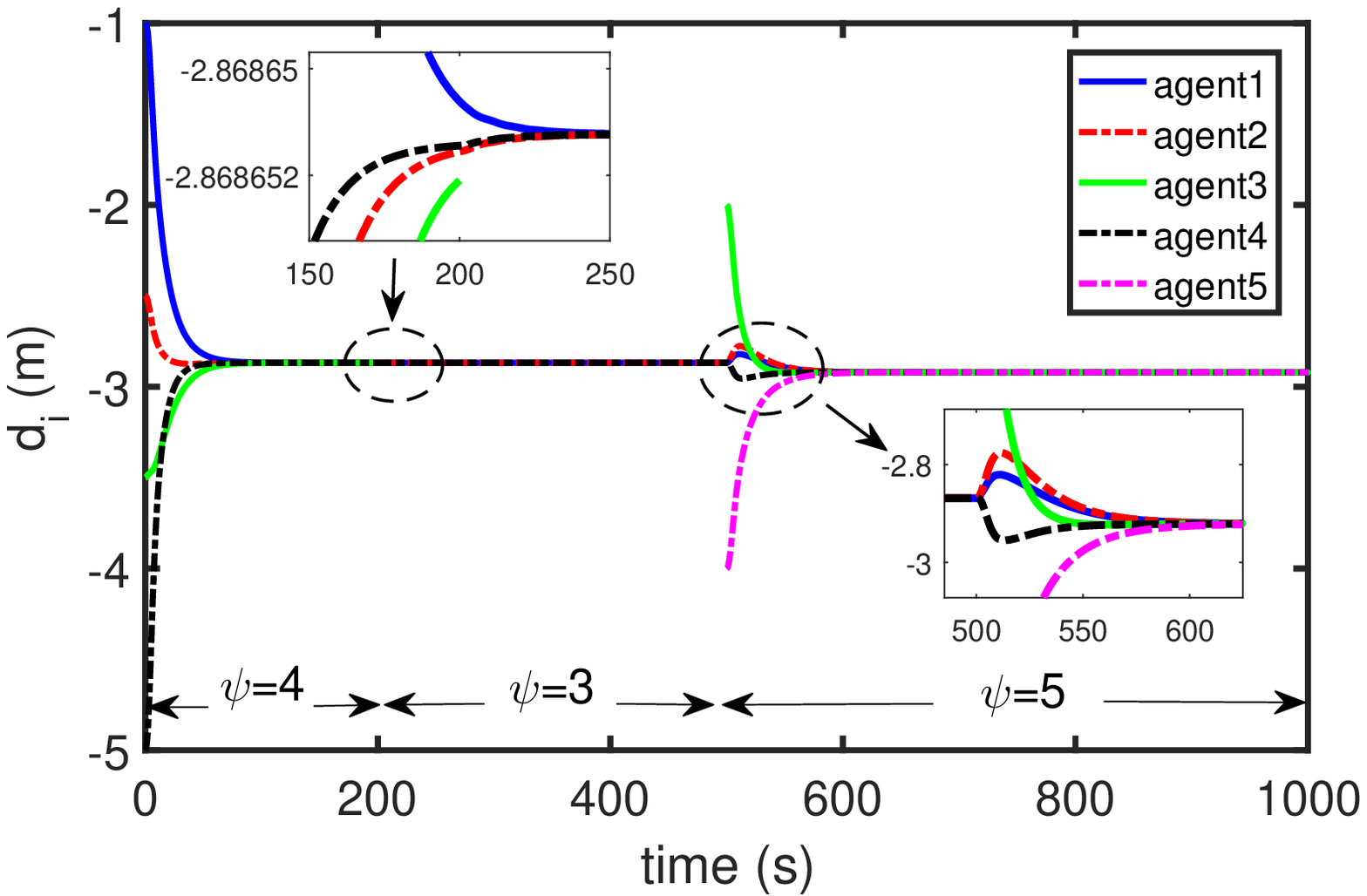}}
	\caption{Response of the CL MAS that begins its operation with four UUVs. Later, one UUV is removed from the CL MAS at the 200th~s and two UUVs are added to the CL MAS at the 500th~s. Also, $v_0$=0.3~m/s and the communication network topologies of the CL MAS are switched at 1~s.}
	\label{fig:b}
\end{figure}

\begin{figure}[H]
	\centering
	\subfigure[\textbf{case 3}: pitch angular velocity response \label{fig:10}]{\includegraphics[width=3.5in, height=1.65in]{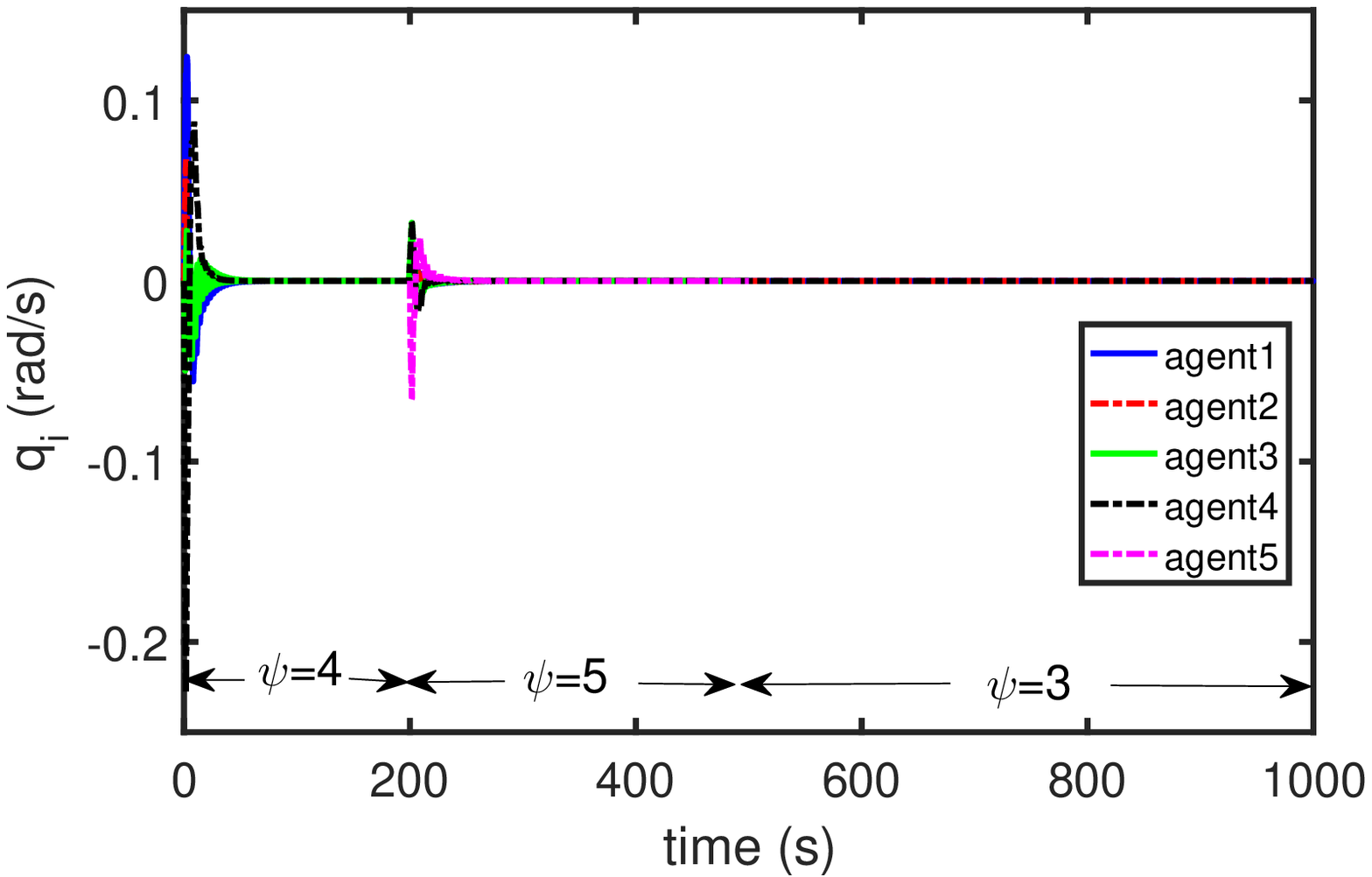}}
		\end{figure}

In this section, a RAIDD consensus protocol is synthesized for the consensus of MAS with $N=4$, $N-P=3$, and $N+M=5$ unmanned underwater vehicles (UUVs) whose    nominal dynamics  \cite{saboori} in state-space form   is described by \eqref{eq:1} where
	\begin{equation}
	A=\begin{bmatrix}
	-0.7&-0.3&0\\1&0&0\\0&-v_0&0
	\end{bmatrix},  	B=\begin{bmatrix}
	0.035\\0\\0
	\end{bmatrix}, 	C=\begin{bmatrix}
	1&0&0\\0&1&0\\0&0&1
	\end{bmatrix}.
	\label{SM}
	\end{equation}	
 The pitch angular velocity, pitch angle,  the depth, and the deflection of the control surface from the stern plane of the UUV are symbolized by $q_i$, $\theta_i$,  $d_i$, and $u_i$ , respectively. Then, the state vector, $\mathbf{x}_i$, and the input vector, $\mathbf{u}_i$, of  \eqref{eq:1}  are defined as $\mathbf{x}_i=[q_i, \theta_i,d_i]^T$ and 	$\mathbf{u}_i=[u_i]$, respectively. The uncertain parameter of the UUV's state-space model is the surge velocity ($v$).  The nominal value of $v$ is represented by $v_0$ and its value  is 0.3~m/s. The lower and upper bound of  $v$ are  0.225~m/s and 0.375~m/s, respectively. Hence, the $\Delta A$ in its interval matrix form is given by
  \begin{equation}
  \Delta A=\begin{bmatrix}
 0&0&0\\0&0&0\\0&\Delta v&0
  \end{bmatrix}; \Delta v \in [-0.075, +0.075].
  \label{eq:20}
  \end{equation}	
The input matrix has no uncertainty and hence $\Delta B=\mathbf{0}_{n,m}$. Also, the number of   CUS graphs considered for the synthesize of  RAIDD protocol is $k=3$, $p=4$, and $r=3$. These graphs are  shown in Figs. \ref{fig:g1}-\ref{fig:10}. 
 \begin{figure}[h!]
 	\centering
 	\subfigure[ \label{fig:g5}]{\includegraphics[width=0.9in, height=0.8in]{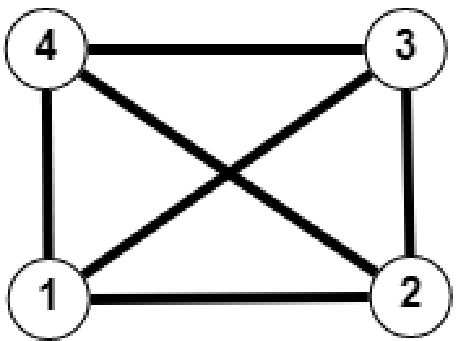}}
 	\subfigure[ \label{fig:g6}]{\includegraphics[width=0.9in, height=0.8in]{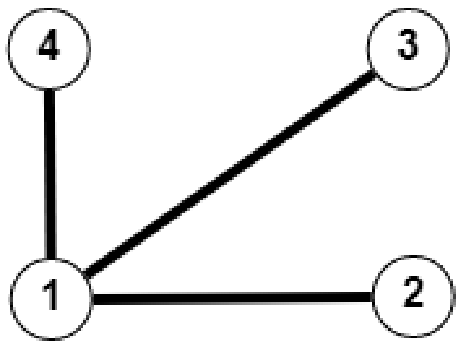}}
 	\subfigure[ \label{fig:g7}]{\includegraphics[width=0.9in, height=0.8in]{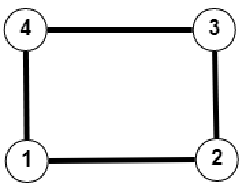}}
 	\caption{CUS graphs associated with 4 UUVs}
 \end{figure}
 The number of agents and the values of $k$, $p$, and $r$ suggest that 	$\xi$ is 29.  $\mathcal{P}$ is then formed using the  eigenvalues of 	$\breve{\mathcal{L}}_h~\forall~ h\in\mathbb{N}_1^3$, $\grave{\mathcal{L}}_h~\forall~ h\in\mathbb{N}_1^4$, and $\hat{\mathcal{L}}_h~\forall~ h\in\mathbb{N}_1^3$. Subsequently, the state-space forms of  $\mathbf{\hat{P}}_i(s)~\forall~i \in \mathbb{N}_1^{29}$  are determined using the matrices given in \eqref{SM} and the eigenvalues  belonging to $\mathcal{P}$. These systems  are thereafter used to constitute  $\mathcal{Q}$. Now following \textbf{Step~1} and \textbf{Step~2} described in Section \ref{PR},  $\epsilon_{{cp}}$ is   identified as 0.4293 and $\mathbf{\hat{P}}_{cp}(s) \in \mathcal{Q}$  as
 \begin{equation}
 \mathbf{\hat{P}}_{cp}(s):
 \begin{cases}
 \mathbf{\dot{x}}_i=A\mathbf{x}_{i}+2 B\mathbf{u}_{i}\\
 \mathbf{y}_{i}=C\mathbf{x}_{i}.
 \end{cases}
  \label{eq:19}
 \end{equation}
Using this $\mathbf{\hat{P}}_{cp}(s)$,  $\hat{\mathbf{N}}_{cp}(s)$ and $\hat{\mathbf{M}}_{cp}(s)$ are obtained, and   $\sqrt{(1-\parallel [\mathbf{\hat{N}}_{{cp}}(s) ~\mathbf{\hat{M}}_{{cp}}(s)]  ^T\parallel_H^2)} $ is computed to be  0.6539, which is   greater than 0.4293. Hence, the condition given by \eqref{sscondk3}  holds. Subsequently, $\Xi$ is formed using $\Delta A$ given in \eqref{eq:20}  and $\Delta B=\mathbf{0}_{n,m}$. For this $\Xi$, $\Psi$ is determined and  its values are less than 0.6539 as indicated by Fig. \ref{fig:2}.
Then, the state-space matrices of $\mathbf{\hat{K}}(s)$ are  computed using Glover-McFarlane method proposed in \cite{glover1} and are given as
\begin{align}
K_A=&\begin{bmatrix}
-0.3227&  -0.3283\\0.658  &-0.5469
\end{bmatrix},\\
K_B=&\begin{bmatrix}
0.01976 & -0.05098 &   0.4598\\-0.01496 &   0.1107  & -0.4072
\end{bmatrix},\\
K_C=&\begin{bmatrix}
-0.2959&  0.09703
\end{bmatrix},
K_D=\begin{bmatrix}
-0.003565 &   -0.2504  &     1.13
\end{bmatrix}
\label{eq:21}
\end{align}
All the closed-loop systems, $[\mathbf{\acute{P}}_i(s), \mathbf{\hat{K}}(s)]~ \forall~i \in \mathbb{N}_1^{29}$,  are stable because the trajectories of all their eigenvalues belong to $\mathcal{C}_-$ as shown in Fig. \ref{fig:3} even when $\Delta v$ is varied from $-0.075$ to $0.075$.  Hence, $\mathbf{\hat{K}}(s)$ simultaneously stabilizes $\mathbf{{P}}_i(s)~\forall~i \in \mathbb{N}_1^\xi$ and its perturbed systems.
Consequently,  $\mathbf{K}(s)$ is formed using the state-space matrices of    $\mathbf{\hat{K}}(s)$ given in \eqref{eq:21}. The effectiveness of this $\mathbf{K}(s)$  is evaluated using the four cases of simulations listed below.\\
\noindent \textbf{Case~1:} The nominal dynamics of the UUVs is  used in this case. The MAS begins its operations with four UUVS.  The fifth UUV is then added at the 200th~s. Following that, two UUVs are removed at the 500th~s.\\
\textbf{Case~2:} Here also nominal dynamics of the UUVs is used. Initially, the MAS begins its operation with 4 UUVs. Later,  one UUV is removed at 200th~s. Thereafter, at 500th~s, 4th and 5th UUVs are added.\\ 
\textbf{Case~3:} and \textbf{Case~4:} In these cases, \textbf{Case~1} and \textbf{Case~2} are repeated with uncertain dynamics  in which $v$  varied from 0.3750~m/s to 0.2550~m/s.\par
\begin{figure}[H]
	\subfigure[\textbf{case 3}: pitch angle  response  \label{fig:11}]{\includegraphics[width=3.5in, height=1.65in]{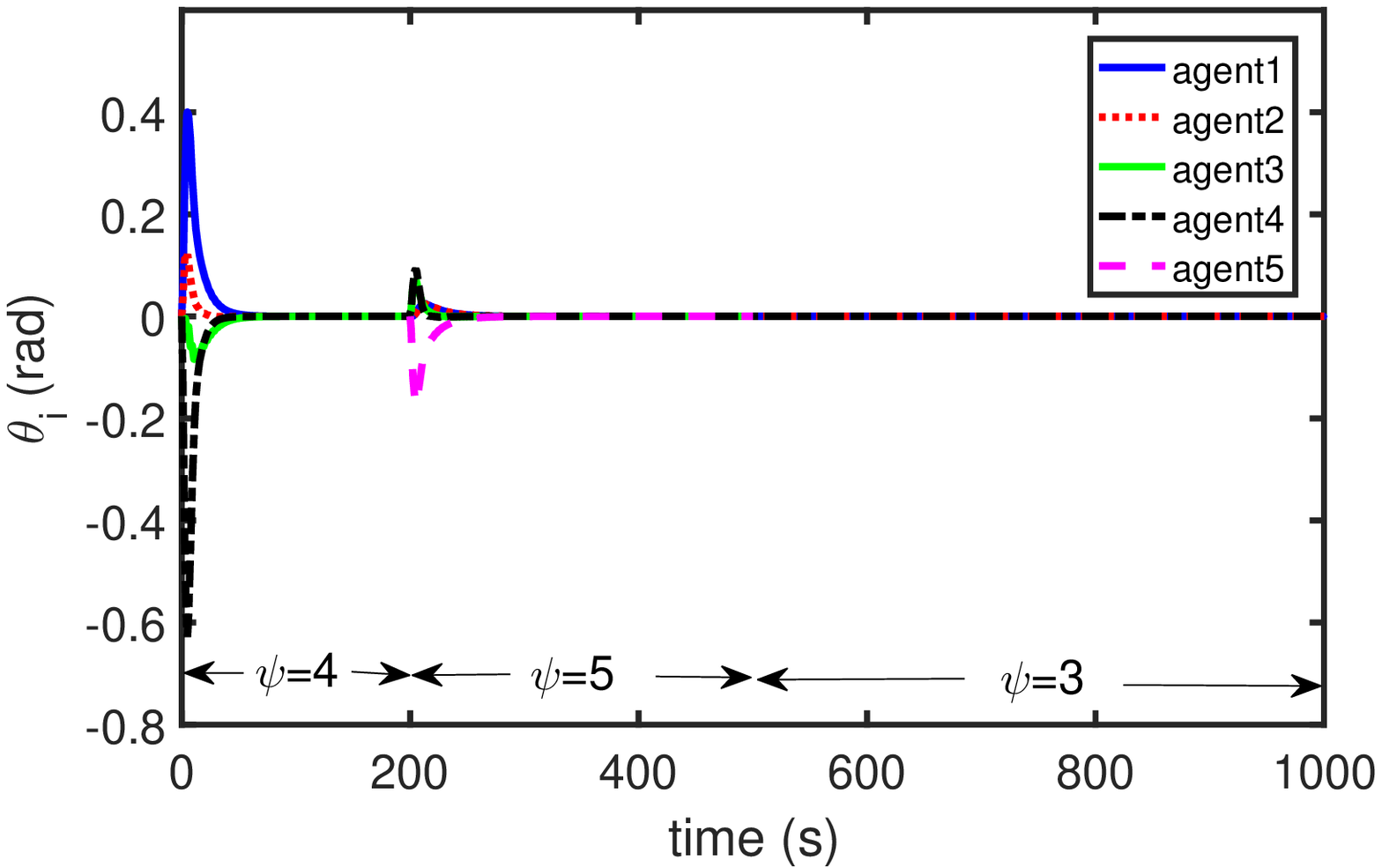}}
\centering
	\subfigure[\textbf{case 3}: depth response  \label{fig:12}]{\includegraphics[width=3.5in, height=1.65in]{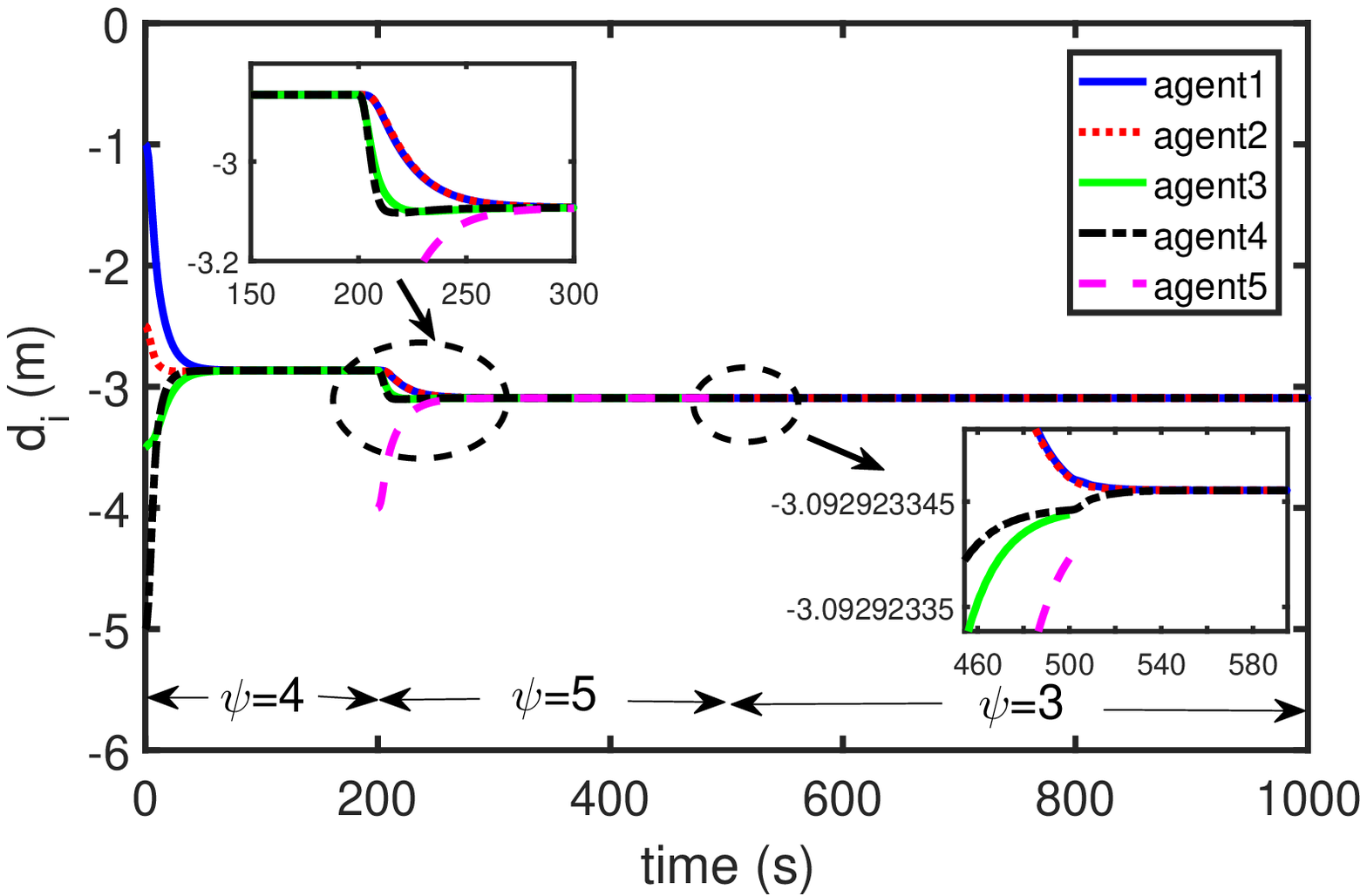}}
	\caption{Response of the CL MAS that begins its operation with four UUVs. Later, the fifth UUV is added to the CL MAS at the 200th~s and two UUVs are removed from the CL MAS at the 500th~s. Also, $v_0$=0.3750~m/s and the communication network topologies of the CL MAS are switched at 1~s.}
	\label{fig:c}
\end{figure}
\begin{figure}[H]
	\centering
	\subfigure[\textbf{case 3}: pitch angular velocity response \label{fig:16}]{\includegraphics[width=3.5in, height=1.65in]{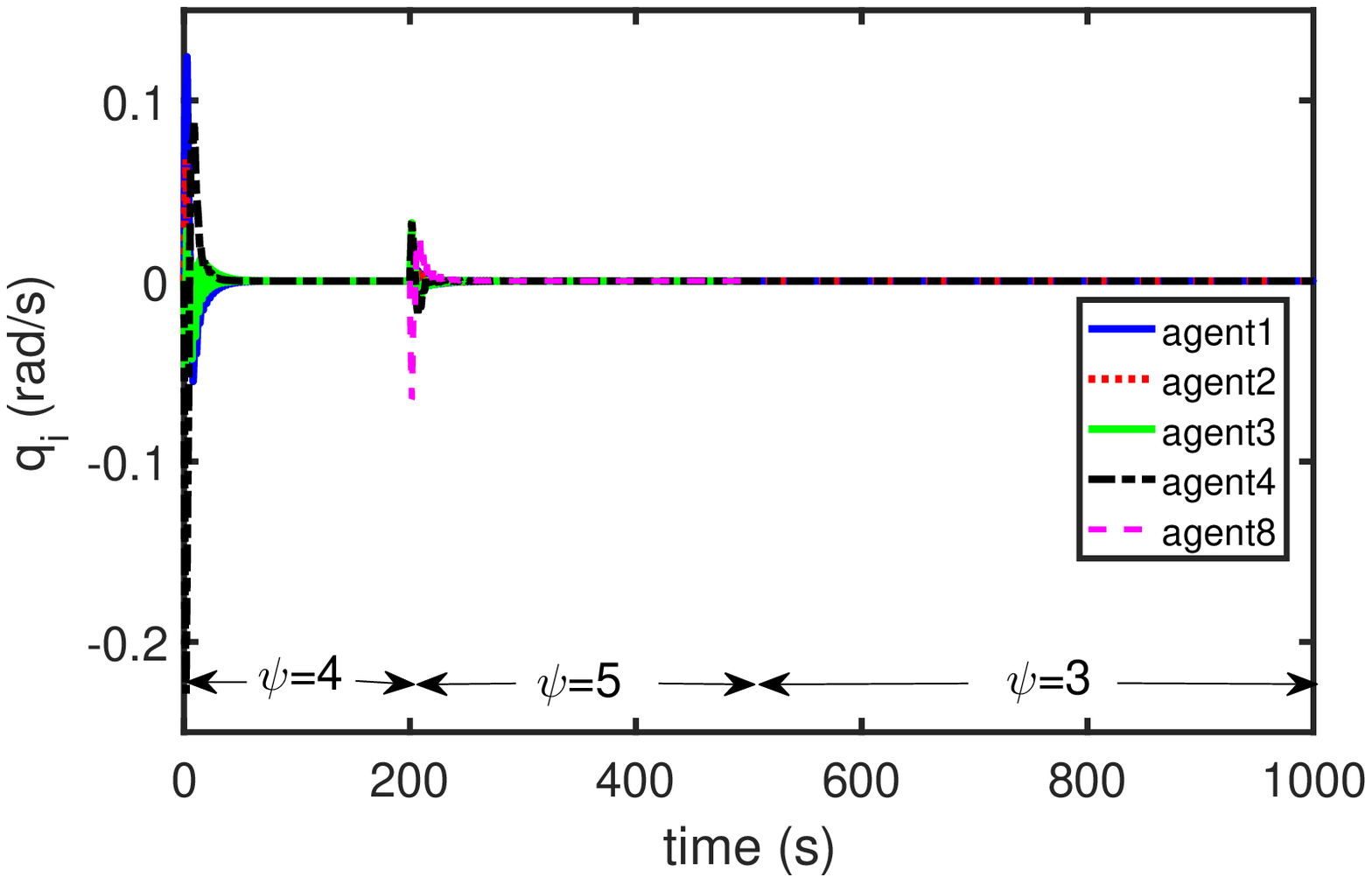}}
	\subfigure[\textbf{case 3}: pitch angle  response  \label{fig:17}]{\includegraphics[width=3.5in, height=1.65in]{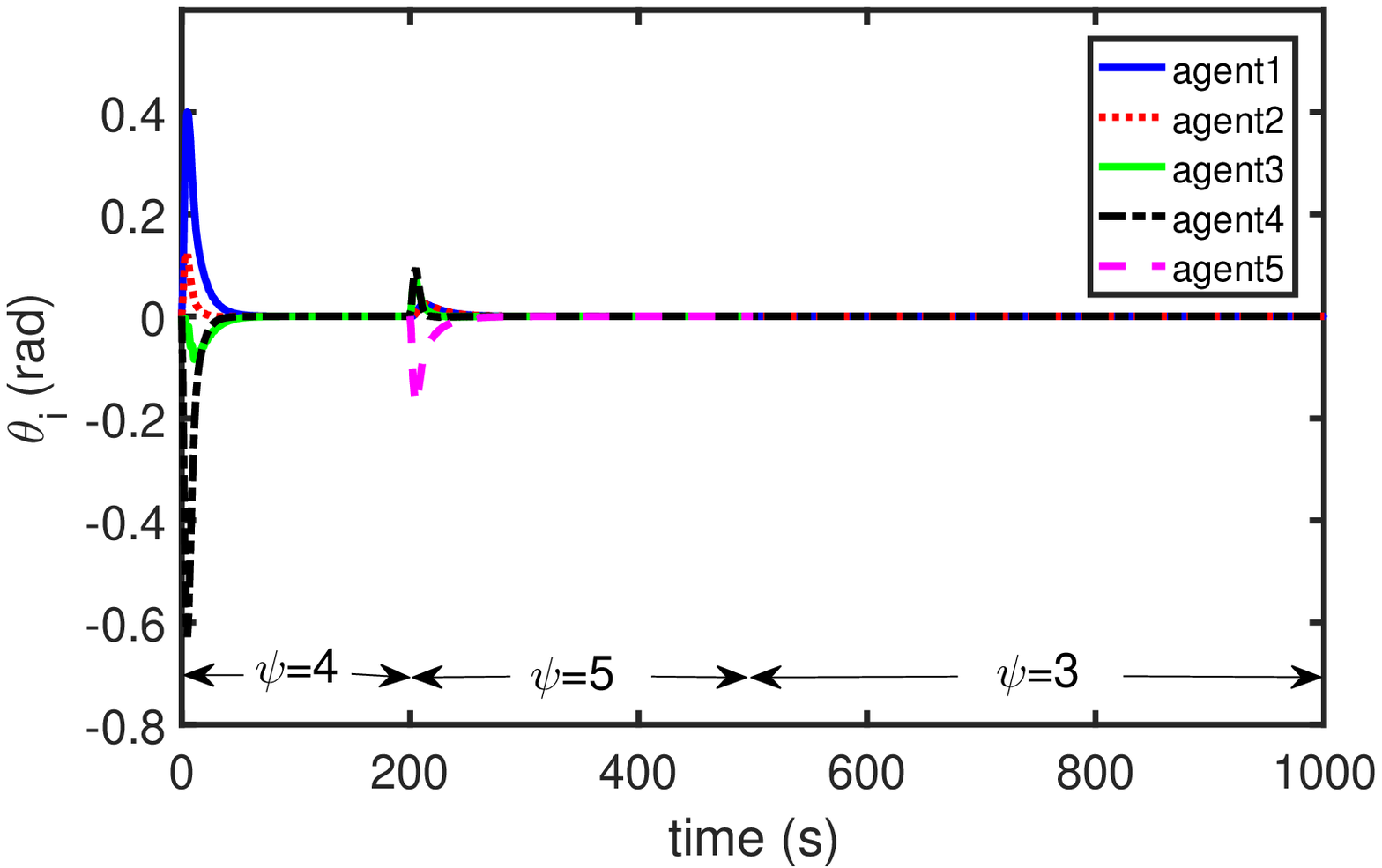}}
		\end{figure}
\begin{figure}[H]
	\subfigure[\textbf{case 3}: depth  response  \label{fig:18}]{\includegraphics[width=3.5in, height=1.65in]{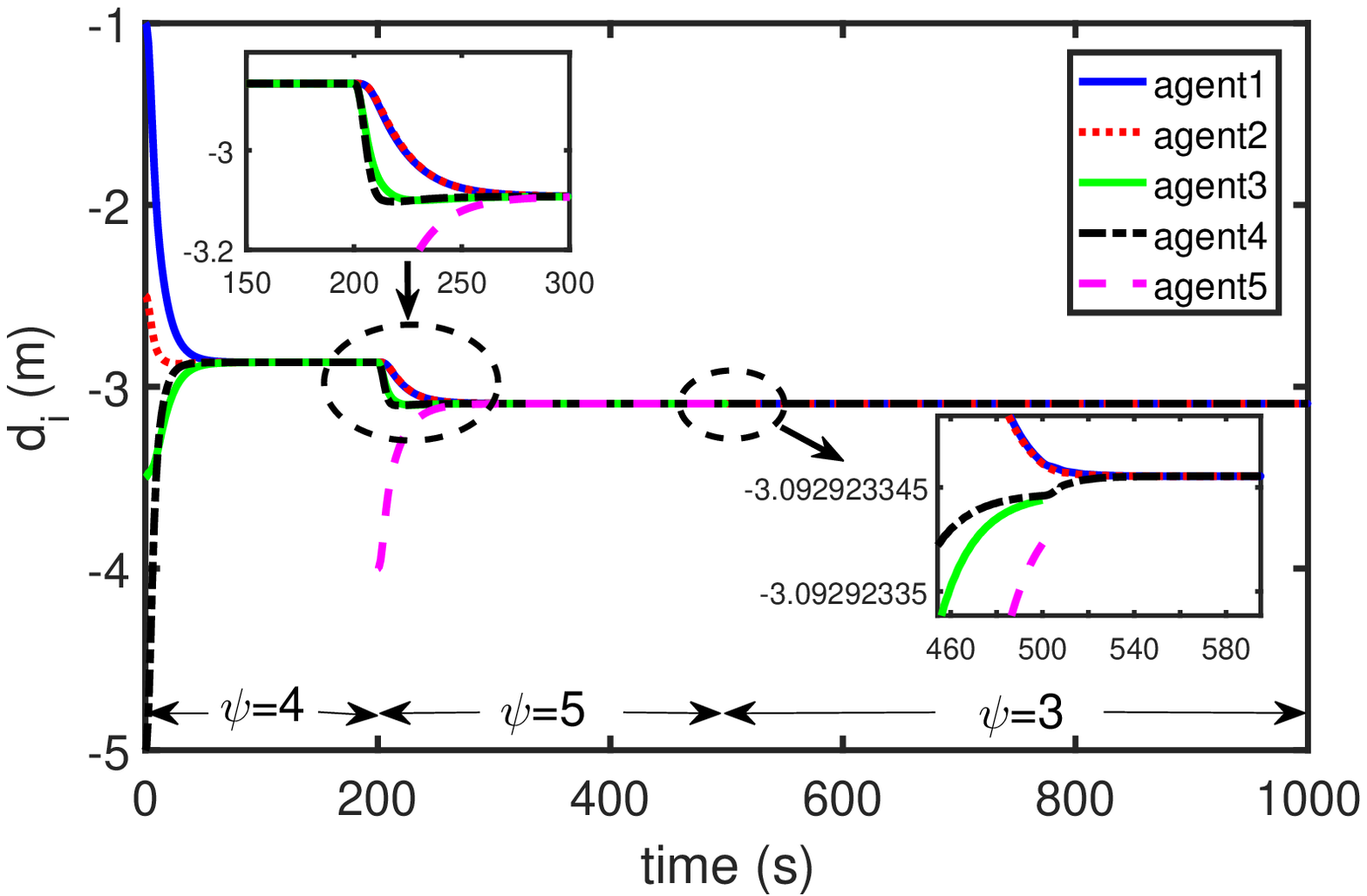}}
	\caption{Response of the CL MAS that begins its operation with four UUVs. Later, the fifth UUV is added to the CL MAS at the 200th~s and two UUVs are removed from the CL MAS at the 500th~s. Also, $v_0$=0.3450~m/s and the communication network topologies of the CL MAS are switched at 1~s.}
	\label{fig:d}
\end{figure}
\begin{figure}[H]
	\centering
	\subfigure[\textbf{case 3}: pitch angular velocity response \label{fig:19}]{\includegraphics[width=3.5in, height=1.65in]{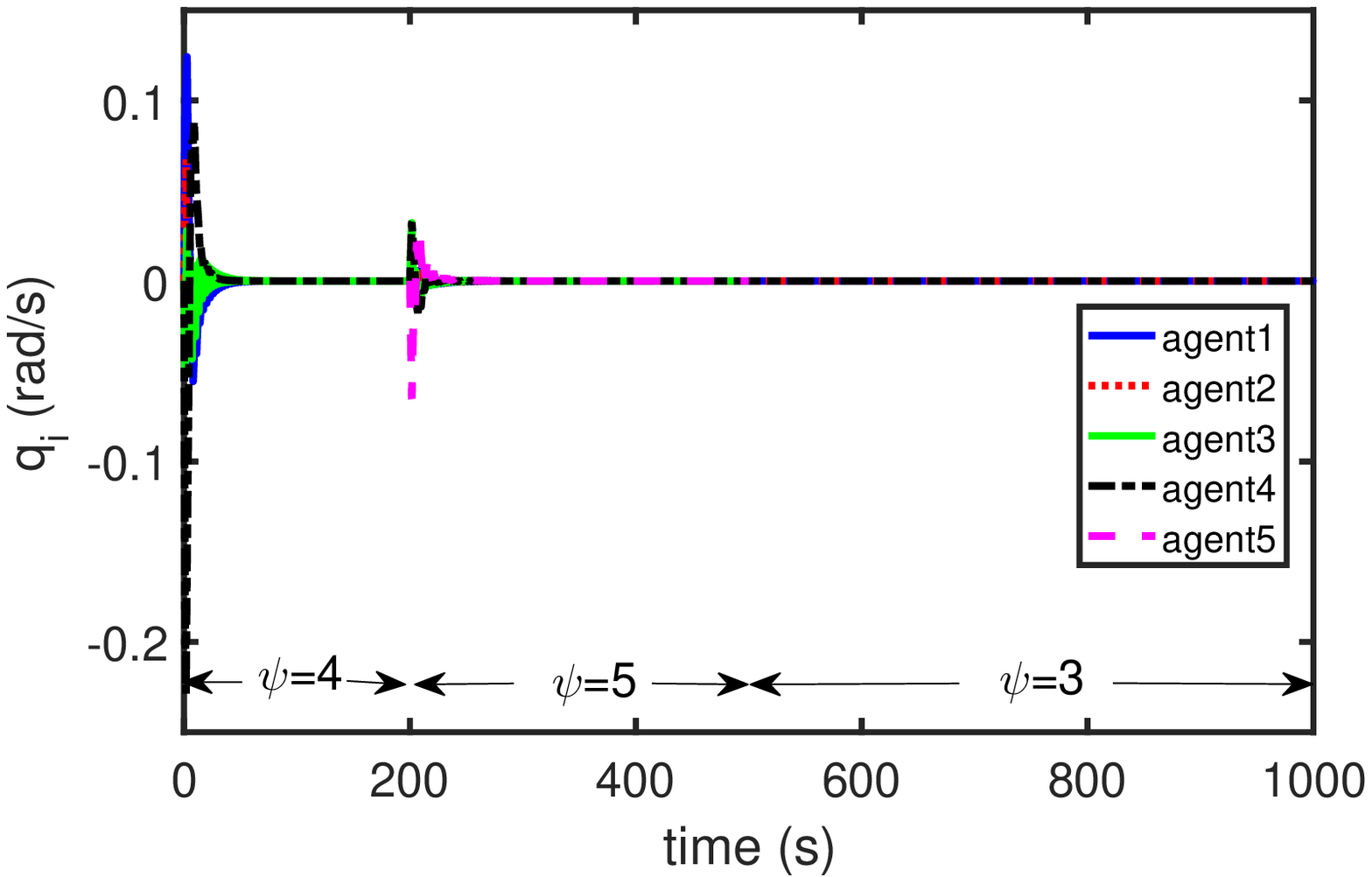}}
	\subfigure[\textbf{case 3}: pitch angle  response  \label{fig:20}]{\includegraphics[width=3.5in, height=1.65in]{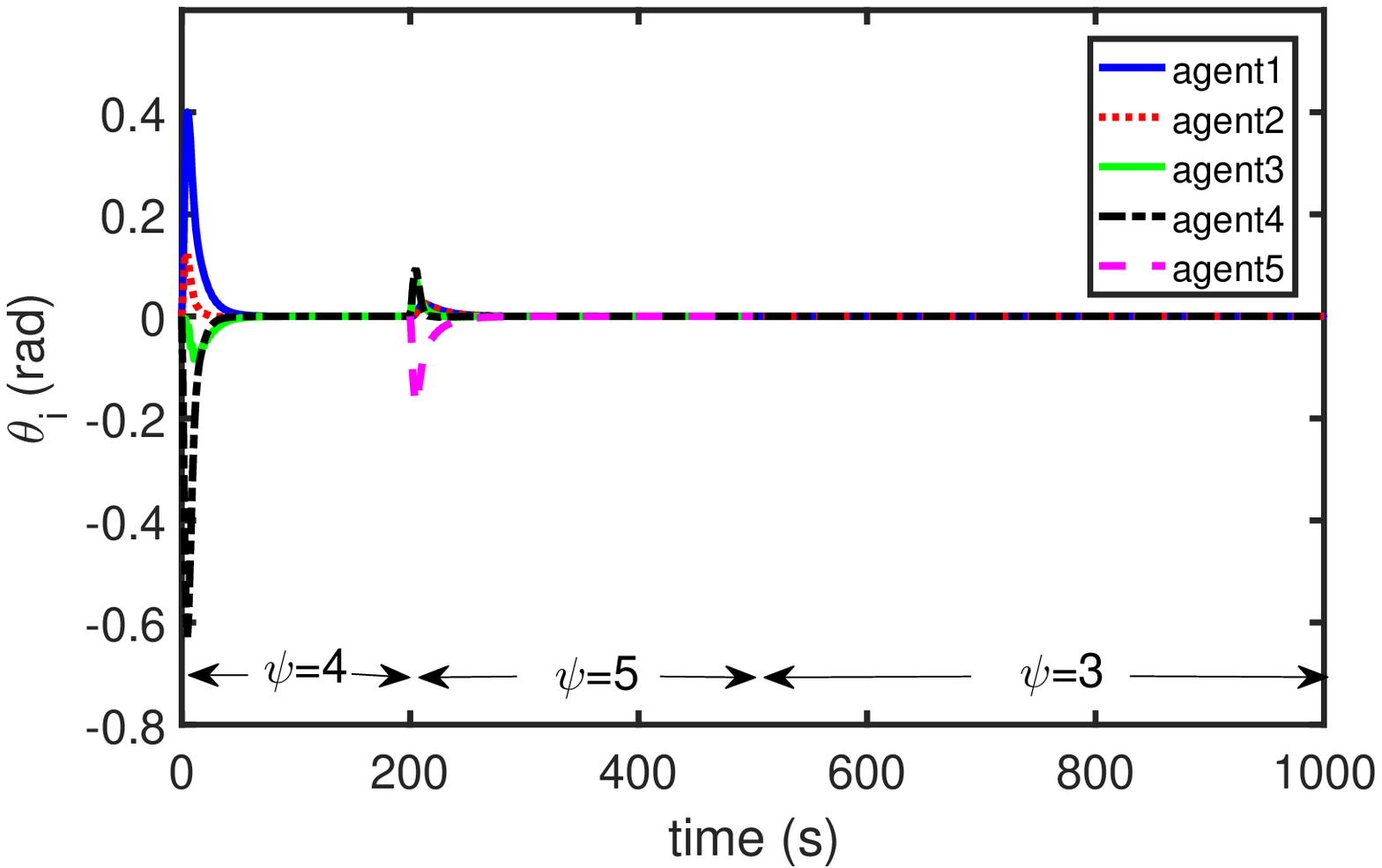}}
	\subfigure[\textbf{case 3}: depth response  \label{fig:21}]{\includegraphics[width=3.5in, height=1.65in]{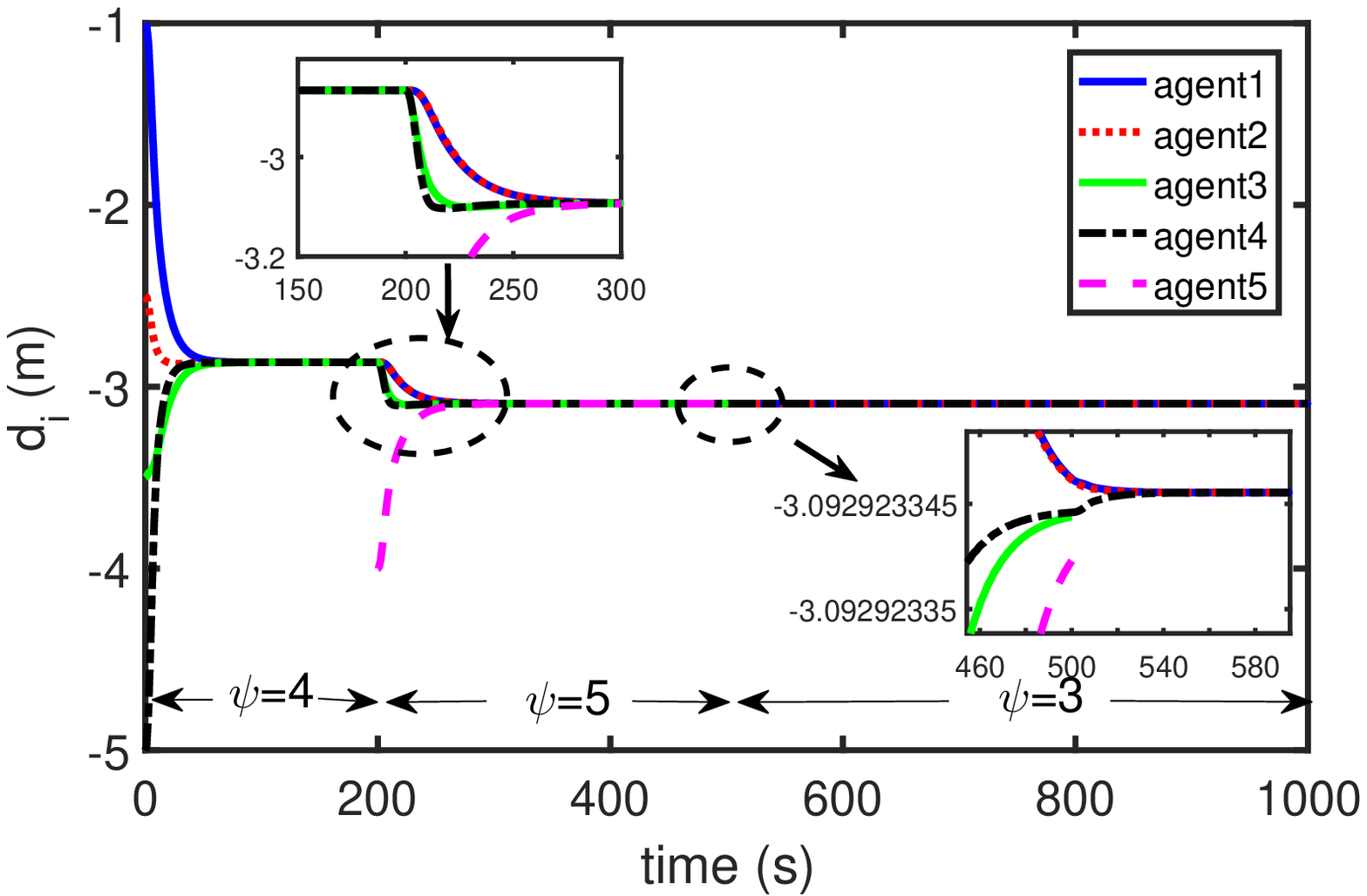}}
	\caption{Response of the CL MAS that begins its operation with four UUVs. Later, the fifth UUV is added to the CL MAS at the 200th~s and two UUVs are removed from the CL MAS at the 500th~s. Also, $v_0$=0.3150~m/s and the communication network topologies of the CL MAS are switched at 1~s.}
	\label{fig:e}
\end{figure}
\begin{figure}[H]
	\centering
	\subfigure[\textbf{case 3}: pitch angular velocity response \label{fig:22}]{\includegraphics[width=3.5in, height=1.65in]{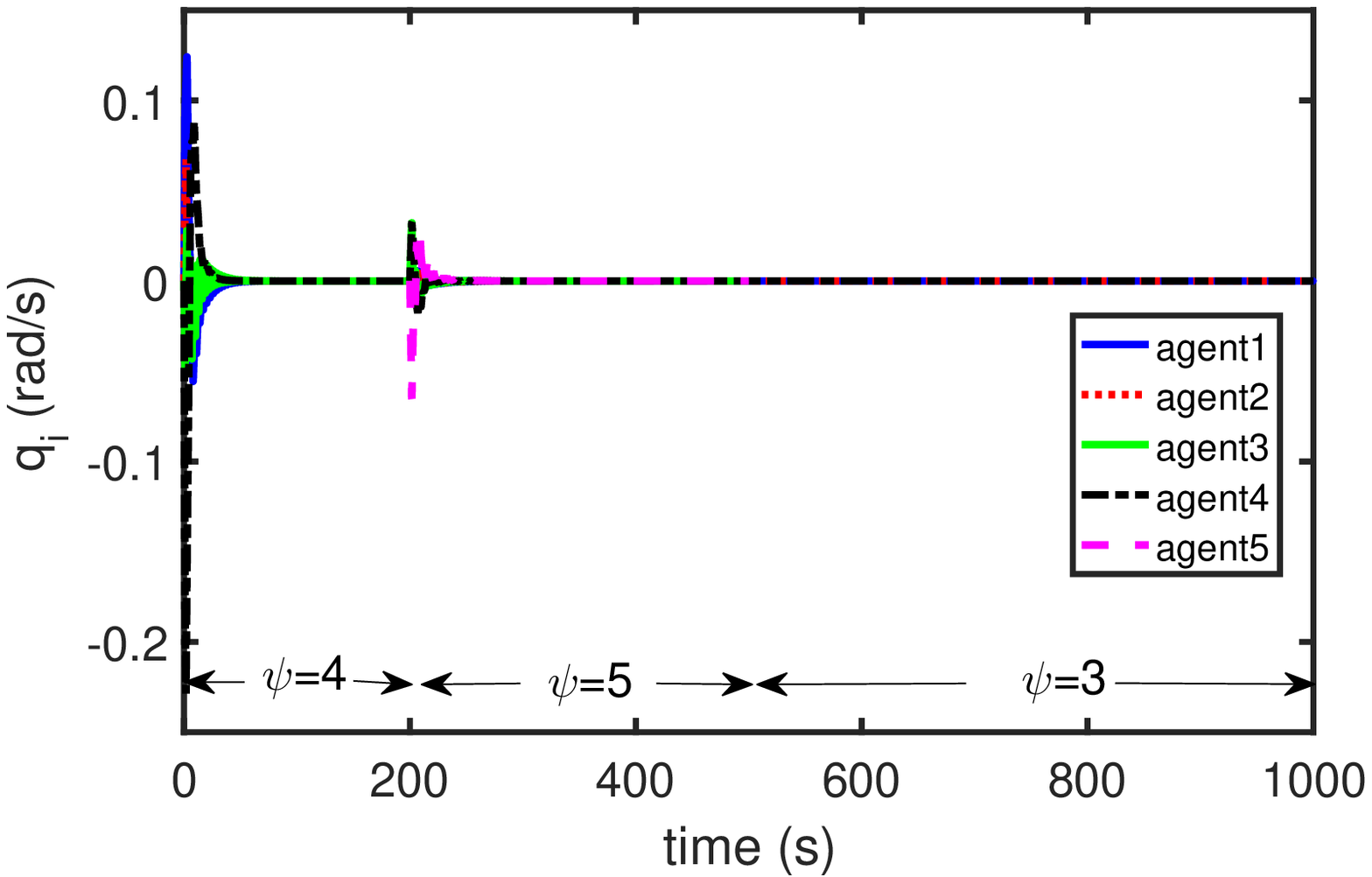}}
		\end{figure}
\begin{figure}[H]
\centering
	\subfigure[\textbf{case 3}: pitch angle  response  \label{fig:23}]{\includegraphics[width=3.5in, height=1.65in]{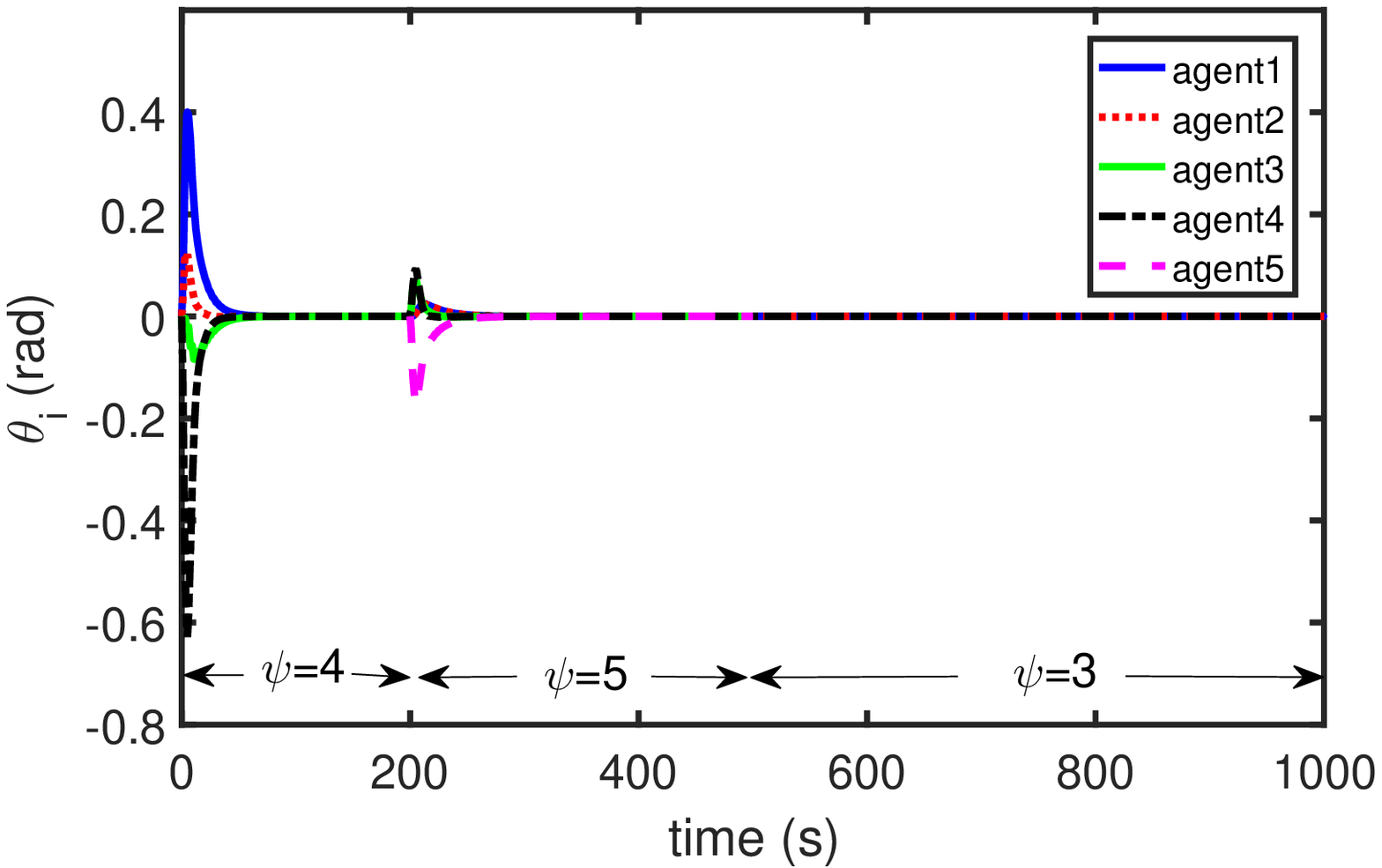}}
	\subfigure[\textbf{case 3}: depth  response  \label{fig:24}]{\includegraphics[width=3.5in, height=1.65in]{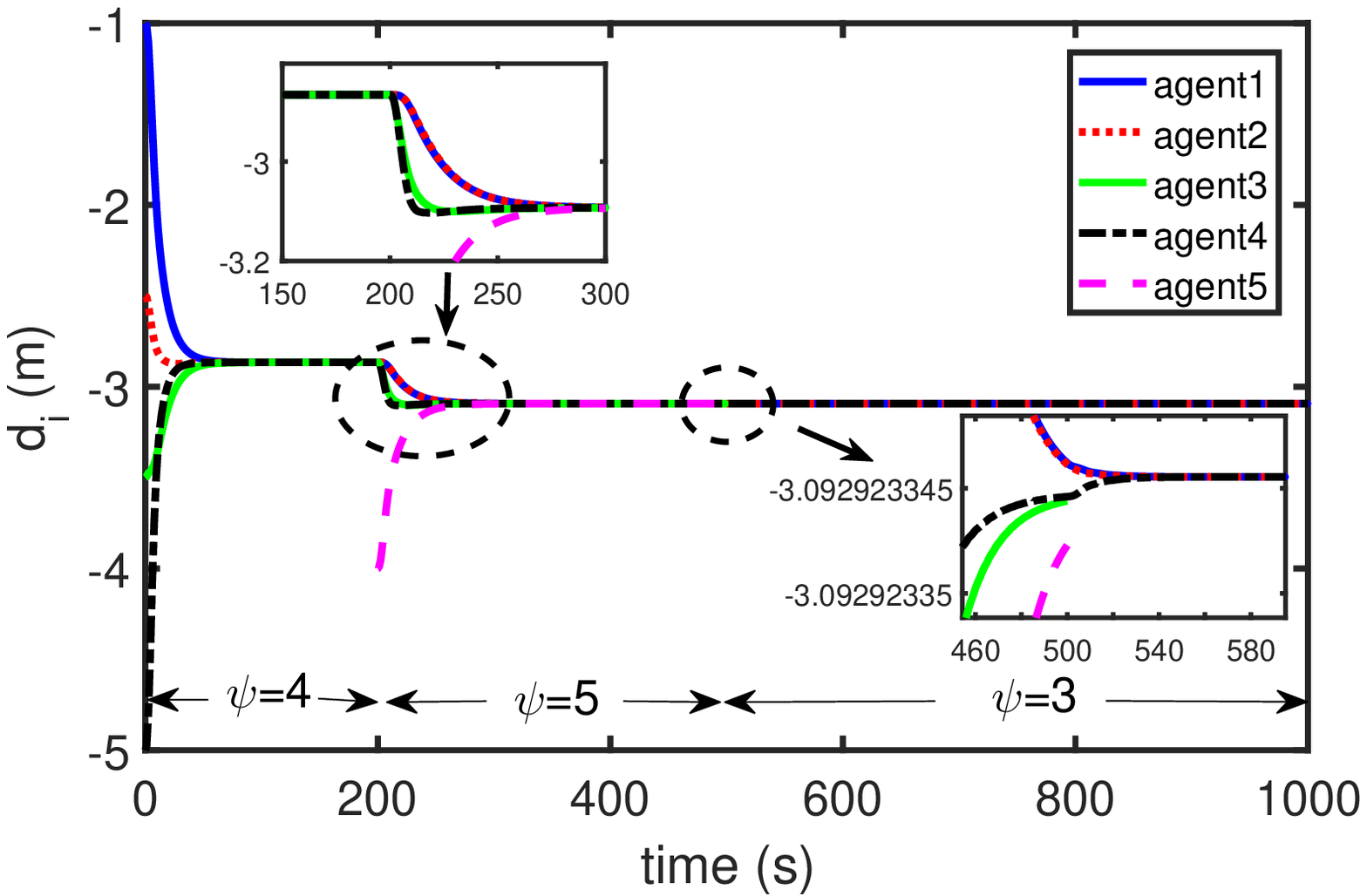}}
	\caption{Response of the CL MAS that begins its operation with four UUVs. Later, the fifth UUV is added to the CL MAS at the 200th~s and two UUVs are removed from the CL MAS at the 500th~s. Also, $v_0$=0.2850~m/s and the communication network topologies of the CL MAS are switched at 1~s.}
	\label{fig:f}
\end{figure}
\begin{figure}[H]
	\centering
	\subfigure[\textbf{case 3}: pitch angular velocity response \label{fig:25}]{\includegraphics[width=3.5in, height=1.65in]{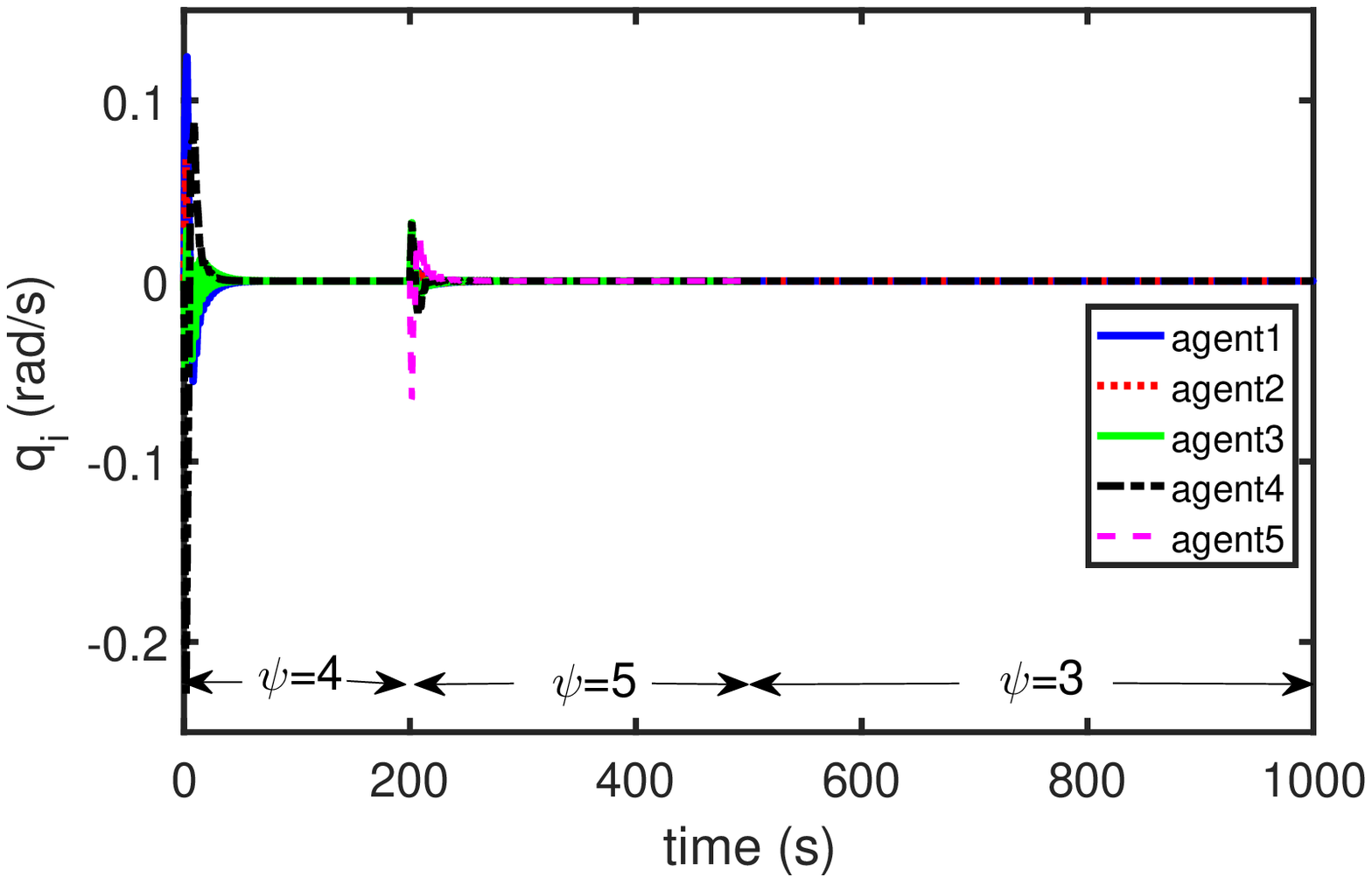}}
\end{figure}
\begin{figure}[H]
	\subfigure[\textbf{case 3}: pitch angle  response  \label{fig:26}]{\includegraphics[width=3.5in, height=1.65in]{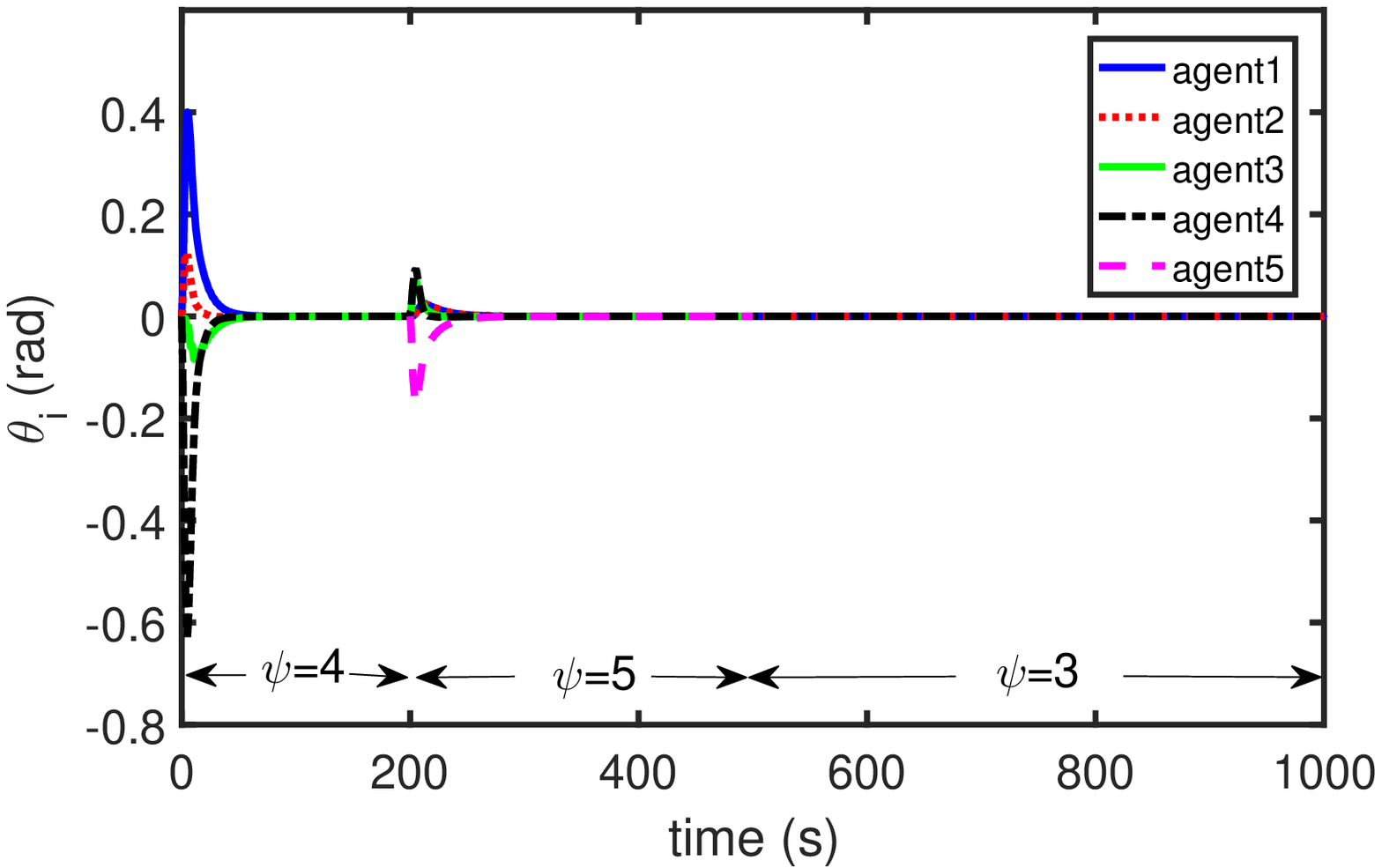}}
\centering
	\subfigure[\textbf{case 3}: depth response  \label{fig:27}]{\includegraphics[width=3.5in, height=1.65in]{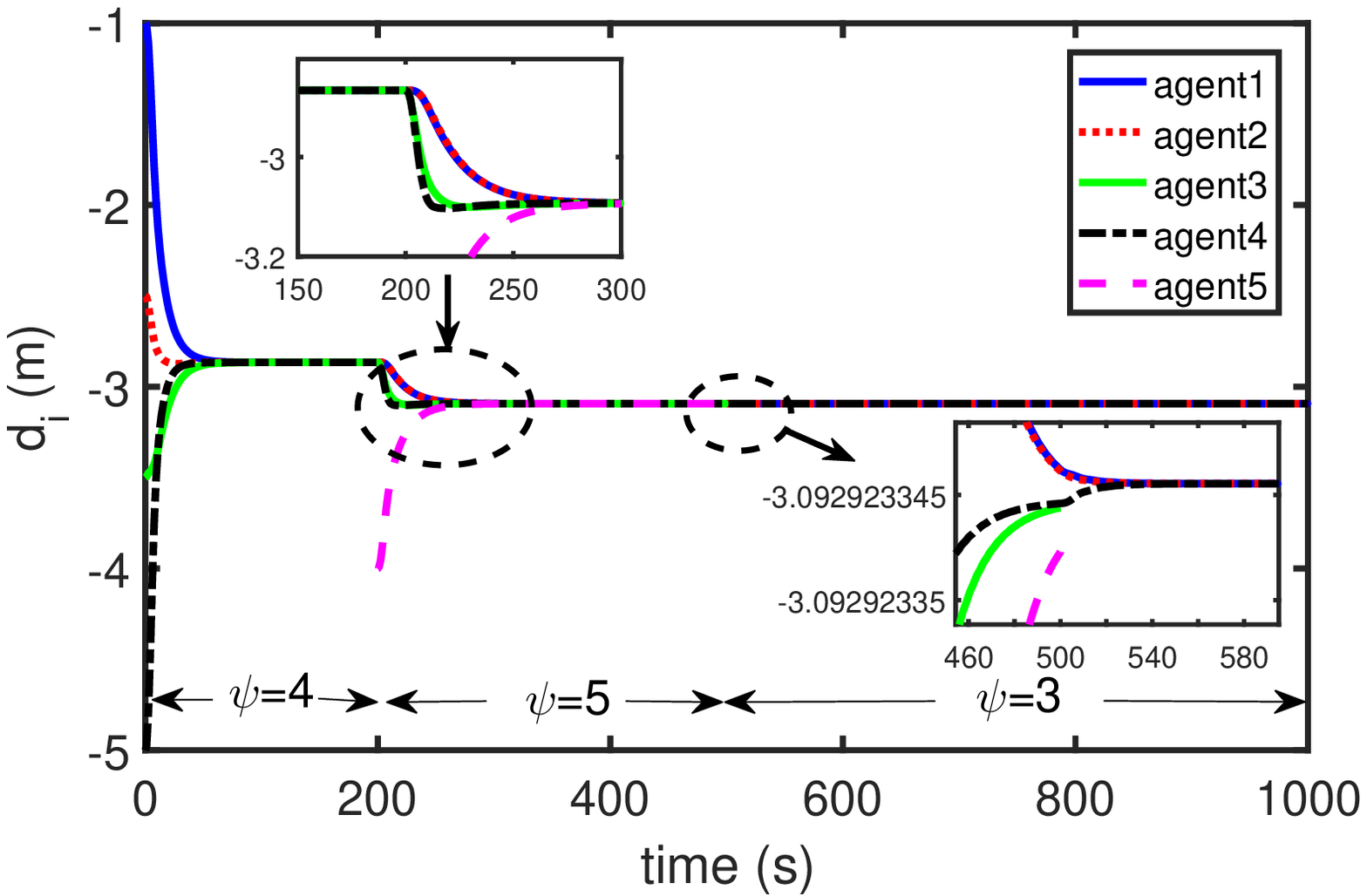}}
	\caption{Response of the CL MAS that begins its operation with four UUVs. Later, the fifth UUV is added to the CL MAS at the 200th~s and two UUVs are removed from the CL MAS at the 500th~s. Also, $v_0$=0.2550~m/s and the communication network topologies of the CL MAS are switched at 1~s.}
	\label{fig:g}
\end{figure}
\begin{figure}[H]
	\centering
	\subfigure[\textbf{case 3}: pitch angular velocity response \label{fig:28}]{\includegraphics[width=3.5in, height=1.65in]{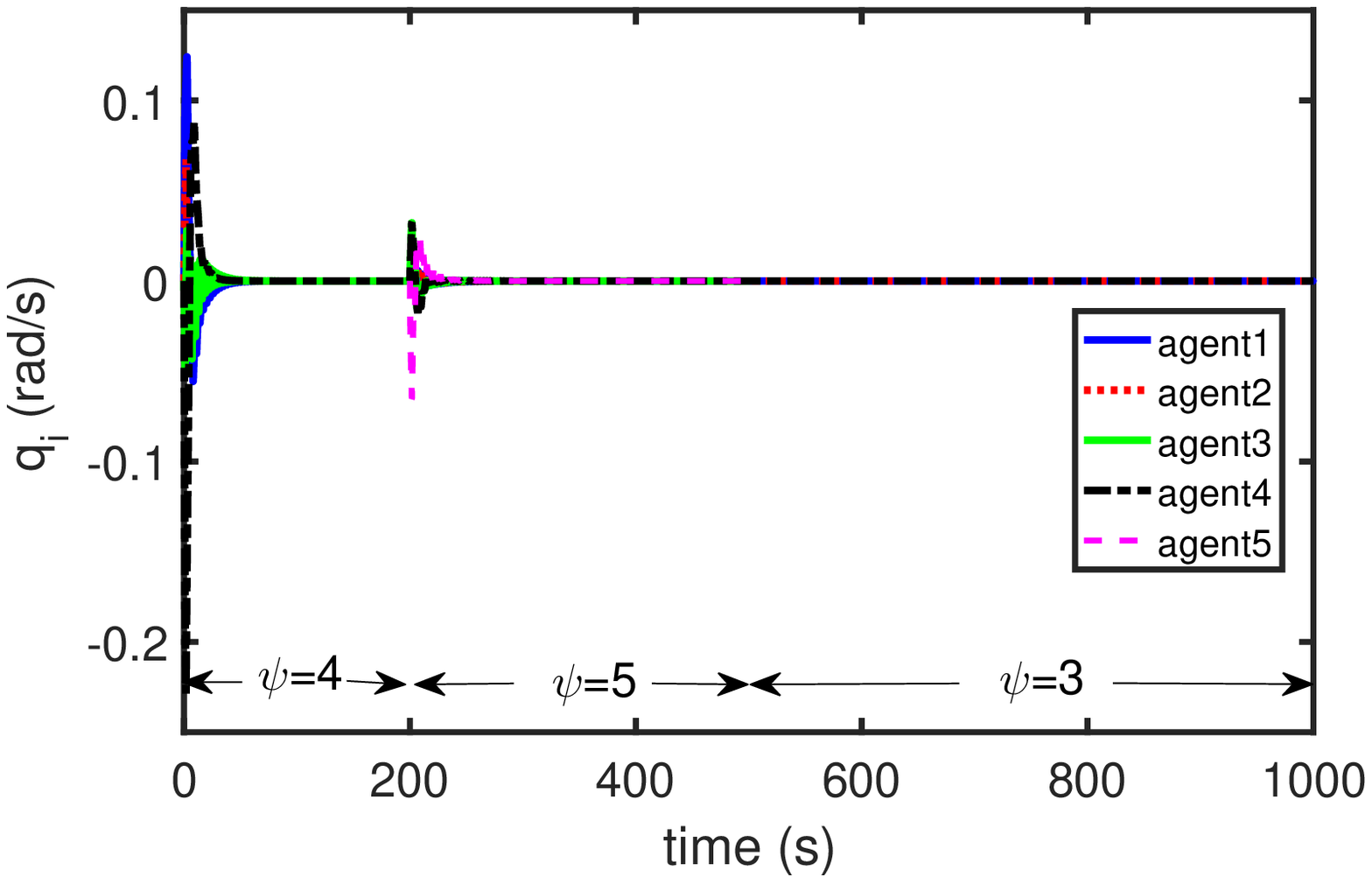}}
	\subfigure[\textbf{case 3}: pitch angle  response  \label{fig:29}]{\includegraphics[width=3.5in, height=1.65in]{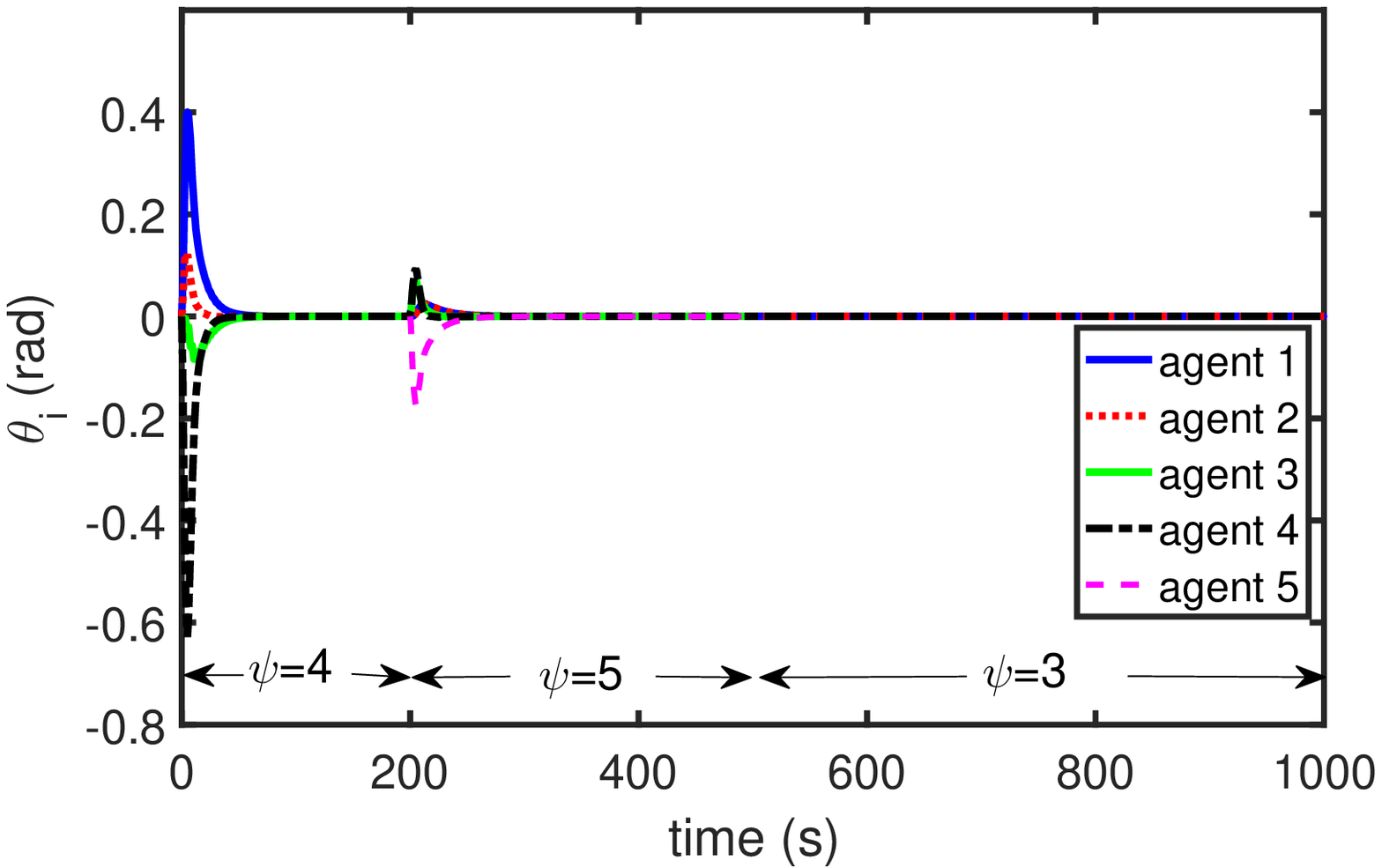}}
\end{figure}
\begin{figure}[H]
	\subfigure[\textbf{case 3}: depth  response  \label{fig:30}]{\includegraphics[width=3.5in, height=1.65in]{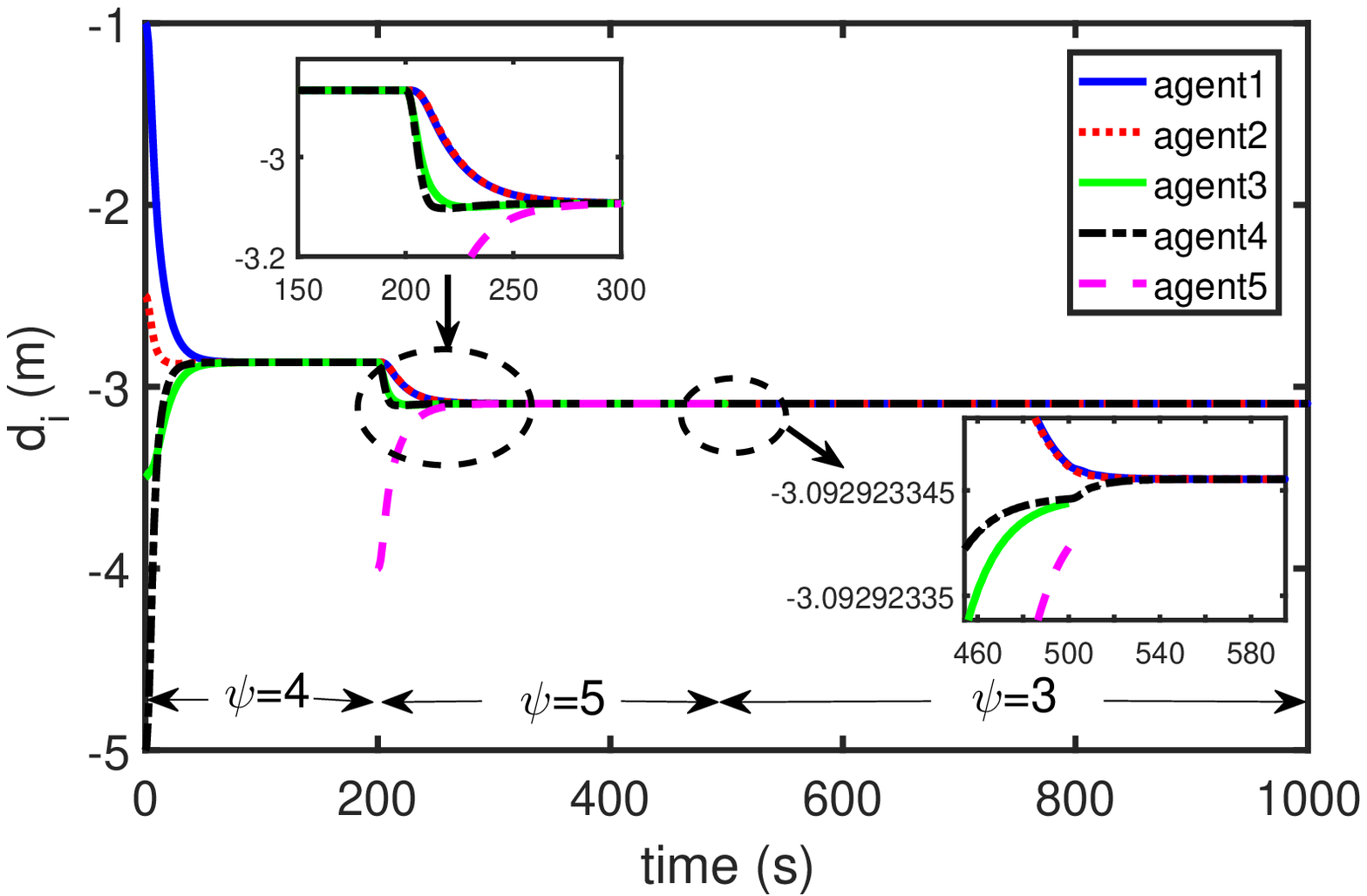}}
	\caption{Response of the CL MAS that begins its operation with four UUVs. Later, the fifth UUV is added to the CL MAS at the 200th~s and two UUVs are removed from the CL MAS at the 500th~s. Also, $v_0$=0.2250~m/s and the communication network topologies of the CL MAS are switched at 1~s.}
	\label{fig:h}
\end{figure}
\begin{figure}[H]
	\centering
	\subfigure[\textbf{case 4}: pitch angular velocity response \label{fig:31}]{\includegraphics[width=3.5in, height=1.65in]{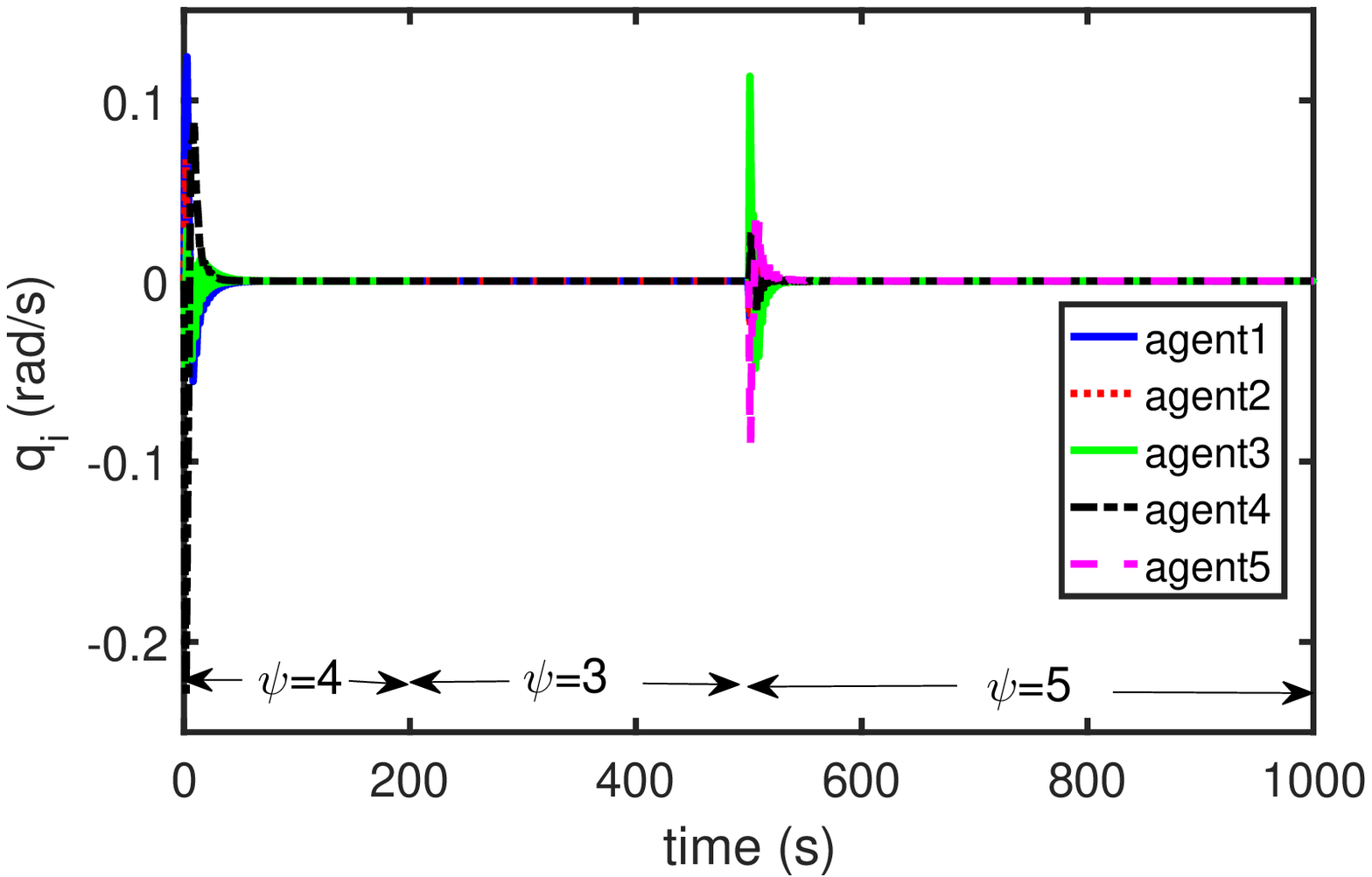}}
	\subfigure[\textbf{case 4}: pitch angle  response  \label{fig:32}]{\includegraphics[width=3.5in, height=1.65in]{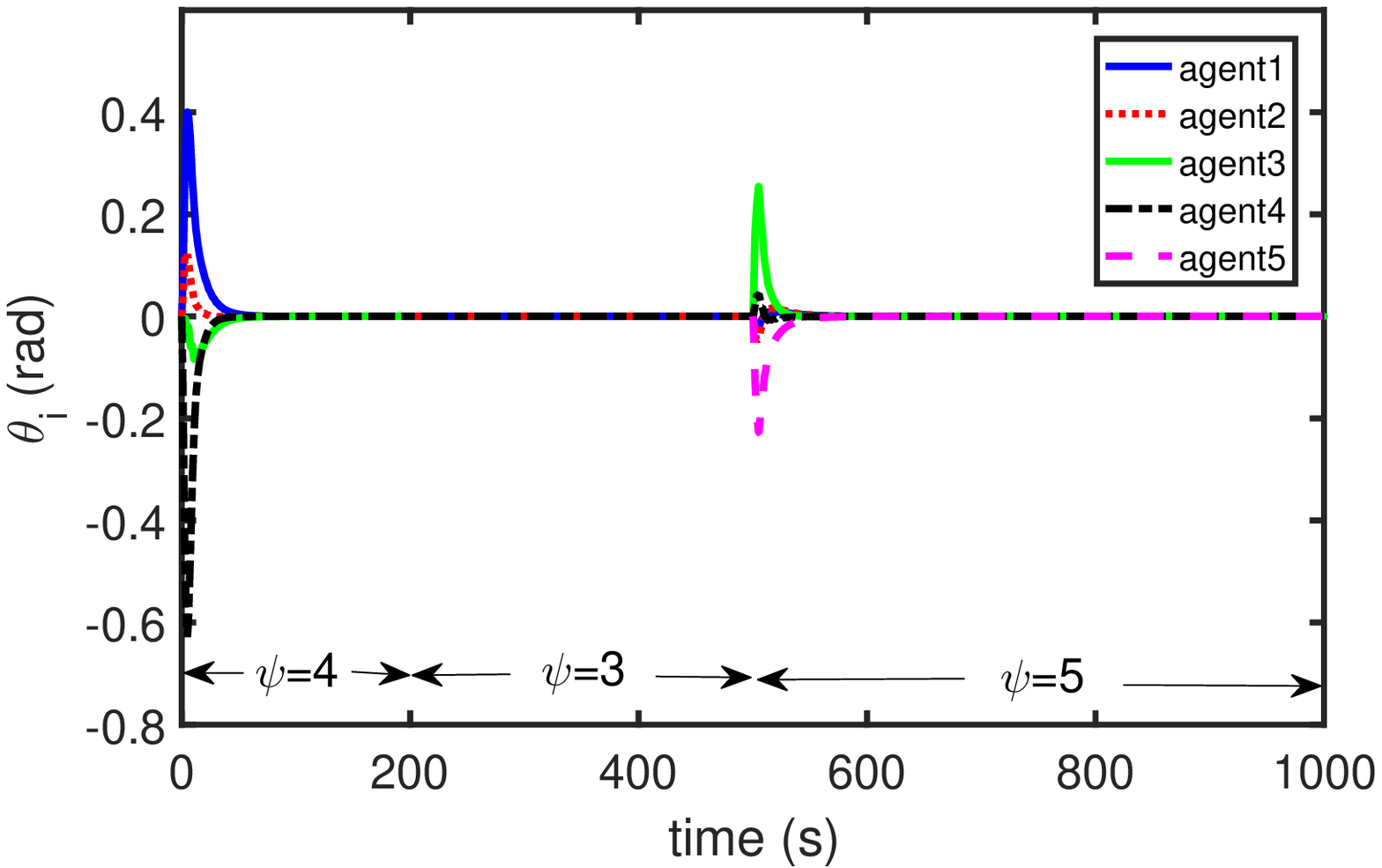}}
	\subfigure[\textbf{case 4}: depth response  \label{fig:33}]{\includegraphics[width=3.5in, height=1.65in]{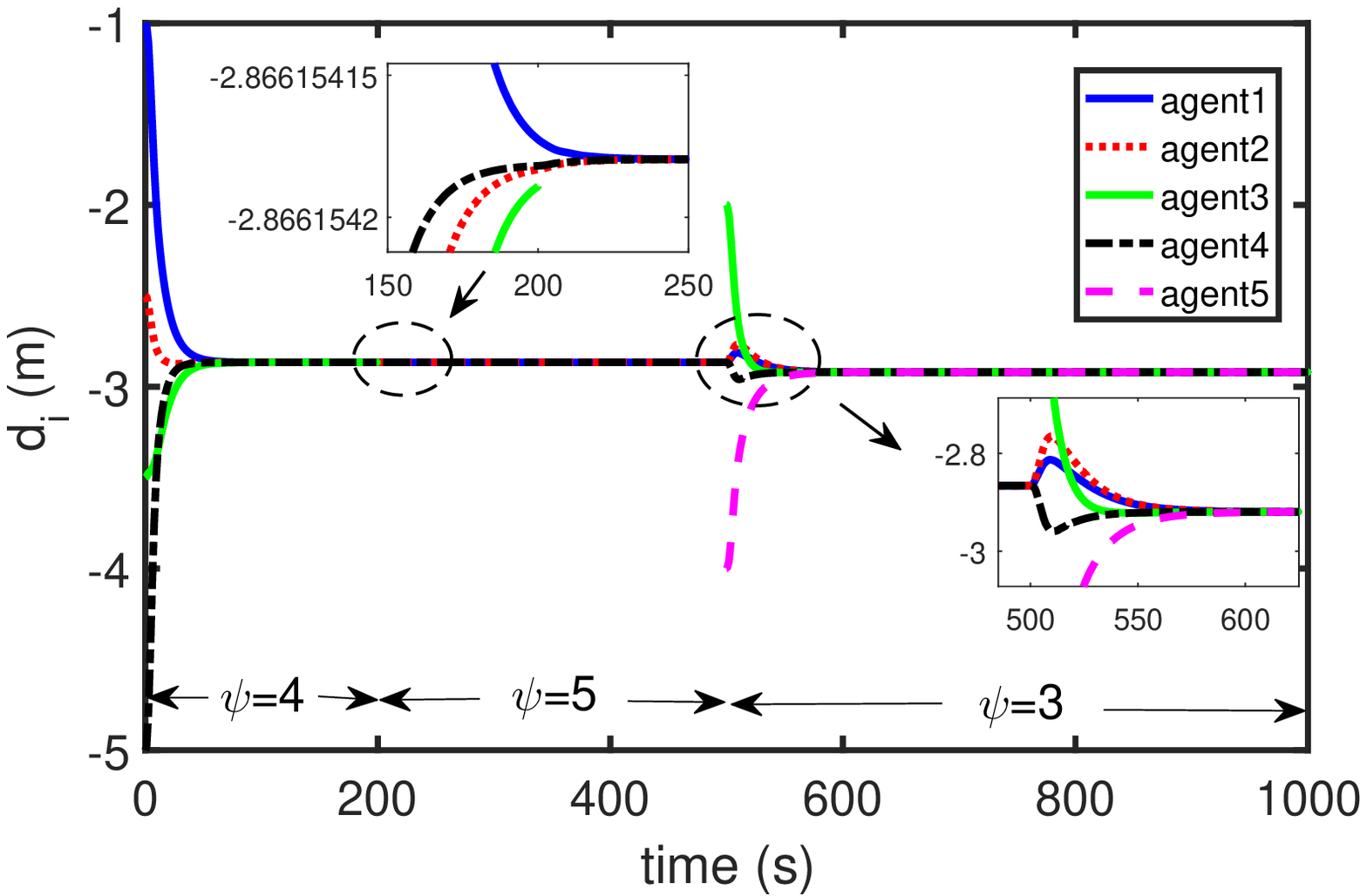}}
	\caption{Response of the CL MAS that begins its operation with four UUVs. Later, one UUV is removed from the CL MAS at the 200th~s and two UUVs are added to the CL MAS at the 500th~s. Also, $v_0$=0.3750~m/s and the communication network topologies of the CL MAS are switched at 1~s.}
	\label{fig:i}
\end{figure}
\begin{figure}[H]
	\centering
	\subfigure[\textbf{case 4}: pitch angular velocity response \label{fig:34}]{\includegraphics[width=3.5in, height=1.65in]{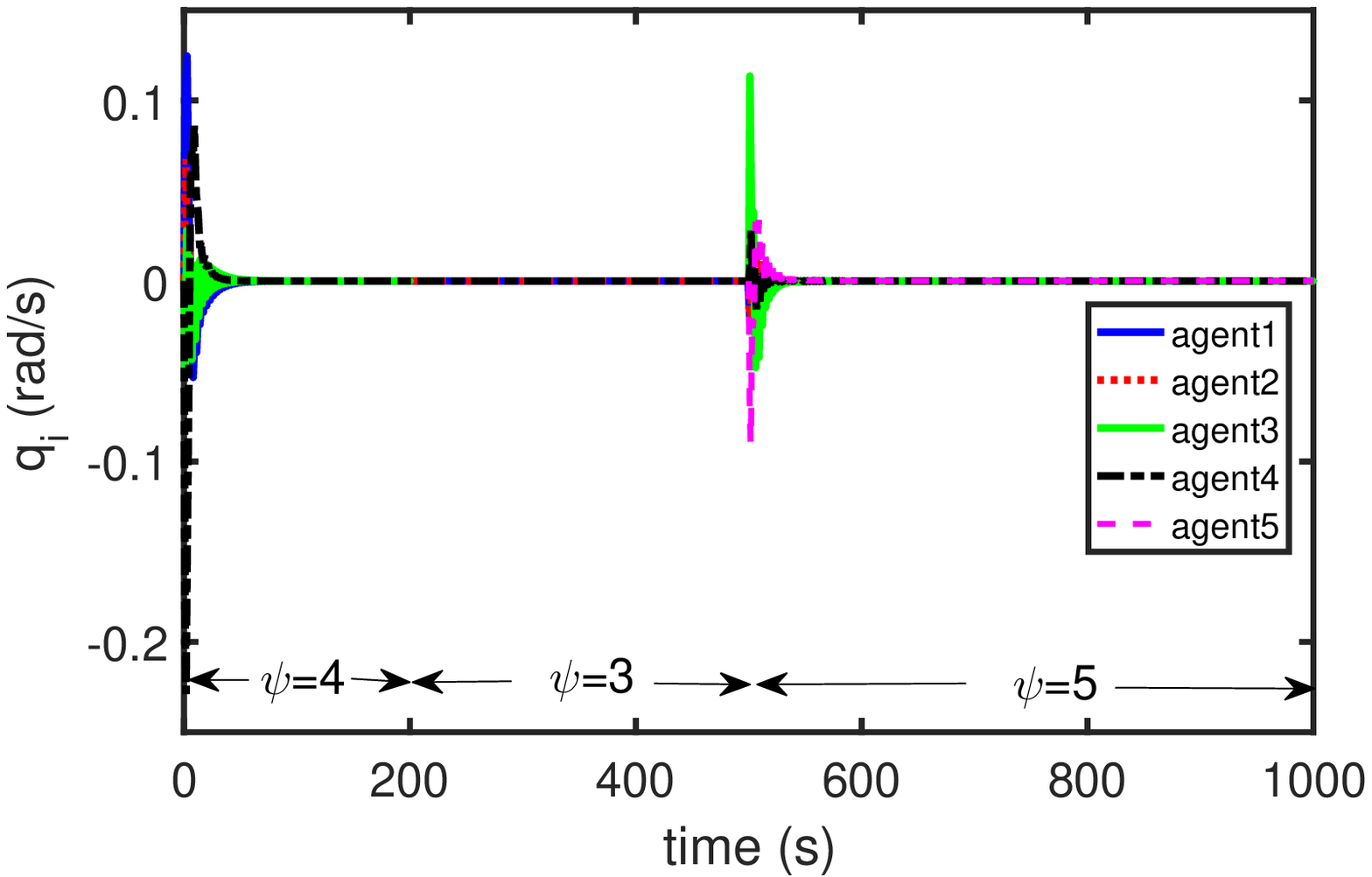}}
		\end{figure}
\begin{figure}[H]
\centering
	\subfigure[\textbf{case 4}: pitch angle  response  \label{fig:35}]{\includegraphics[width=3.5in, height=1.65in]{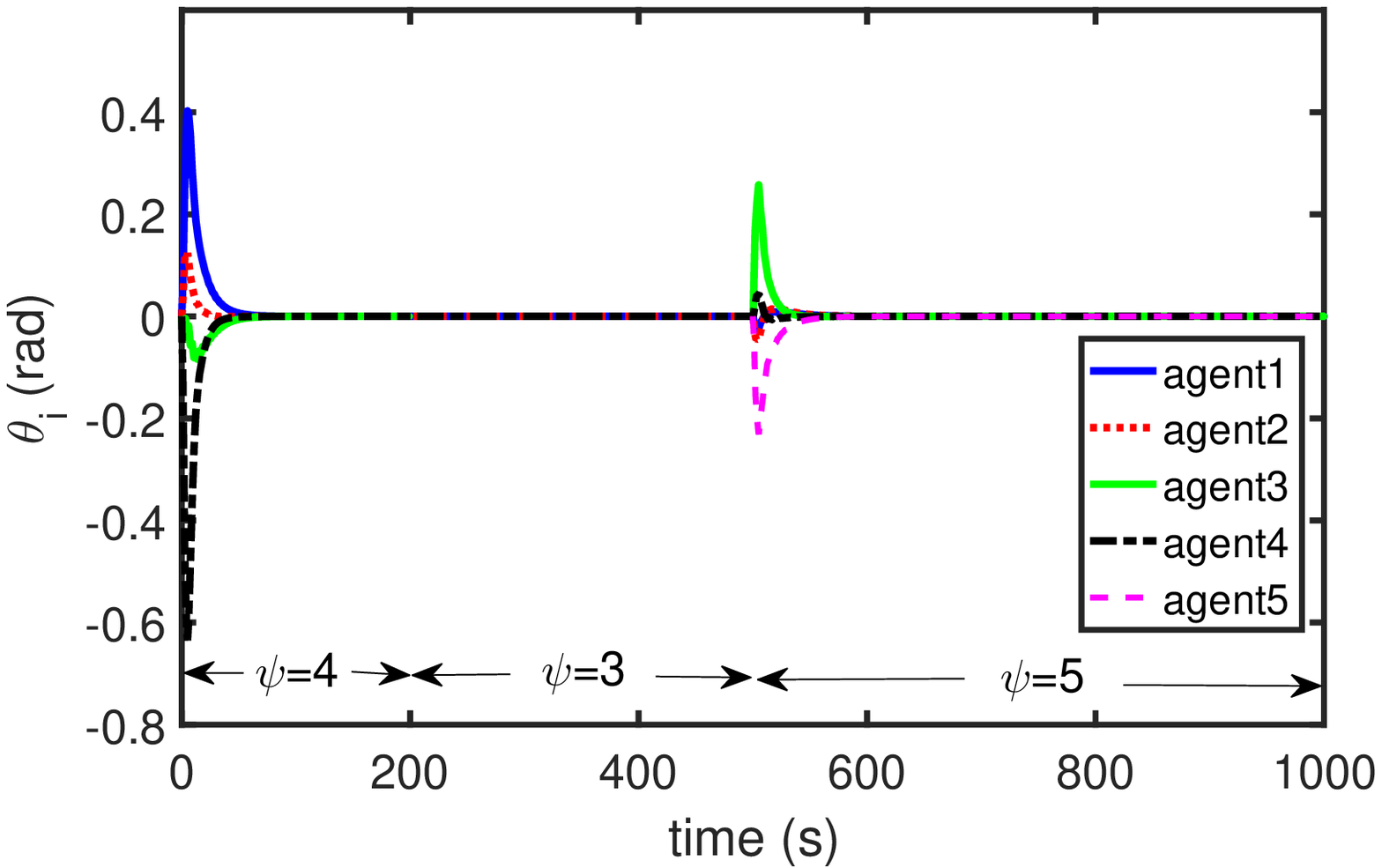}}
	\subfigure[\textbf{case 4}: depth  response  \label{fig:36}]{\includegraphics[width=3.5in, height=1.65in]{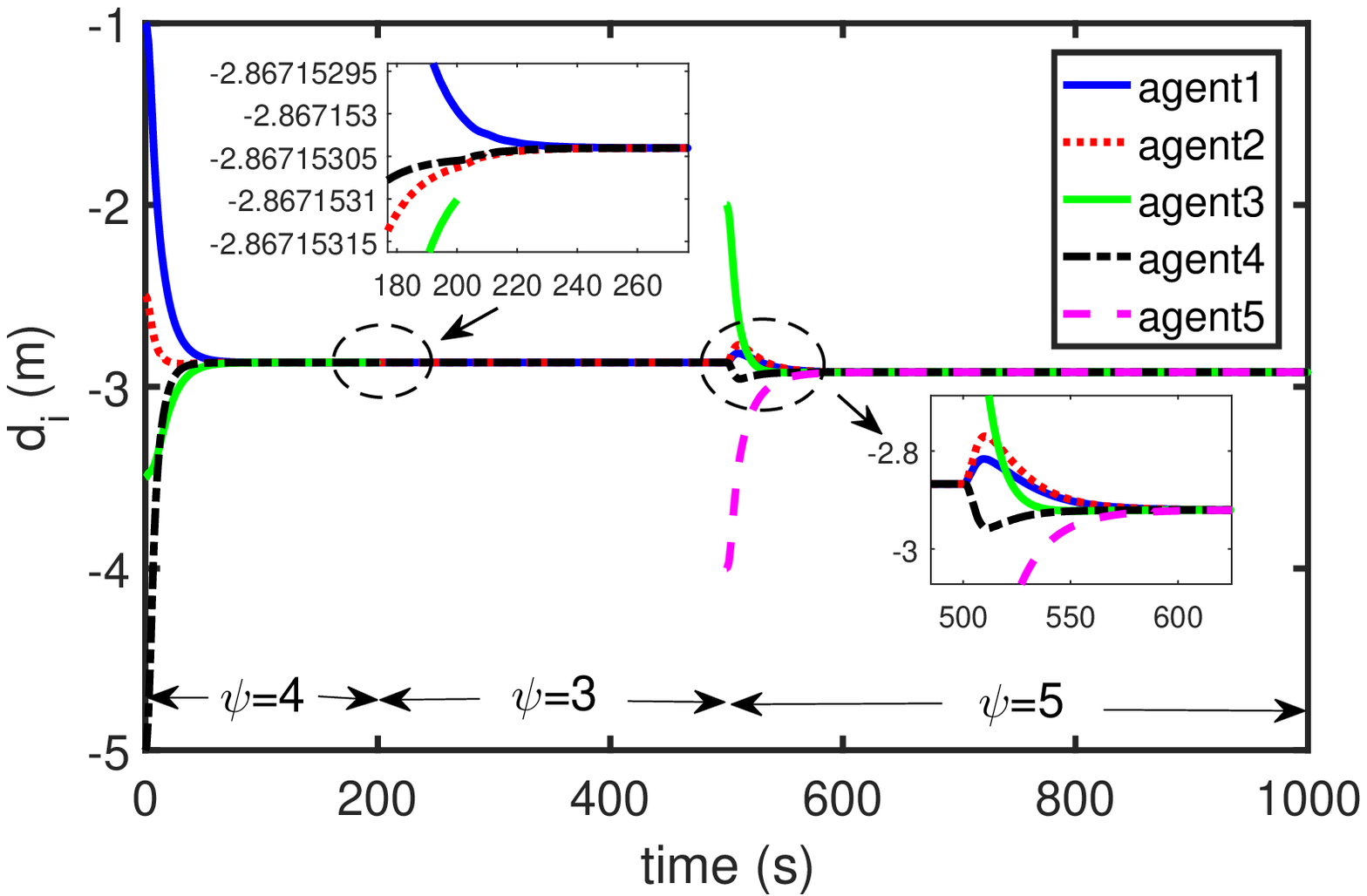}}
	\caption{Response of the CL MAS that begins its operation with four UUVs. Later, one UUV is removed from the CL MAS at the 200th~s and two UUVs are added to the CL MAS at the 500th~s. Also, $v_0$=0.3450~m/s and the communication network topologies of the CL MAS are switched at 1~s.}
	\label{fig:j}
\end{figure}
\begin{figure}[H]
	\centering
	\subfigure[\textbf{case 4}: pitch angular velocity response \label{fig:37}]{\includegraphics[width=3.5in, height=1.65in]{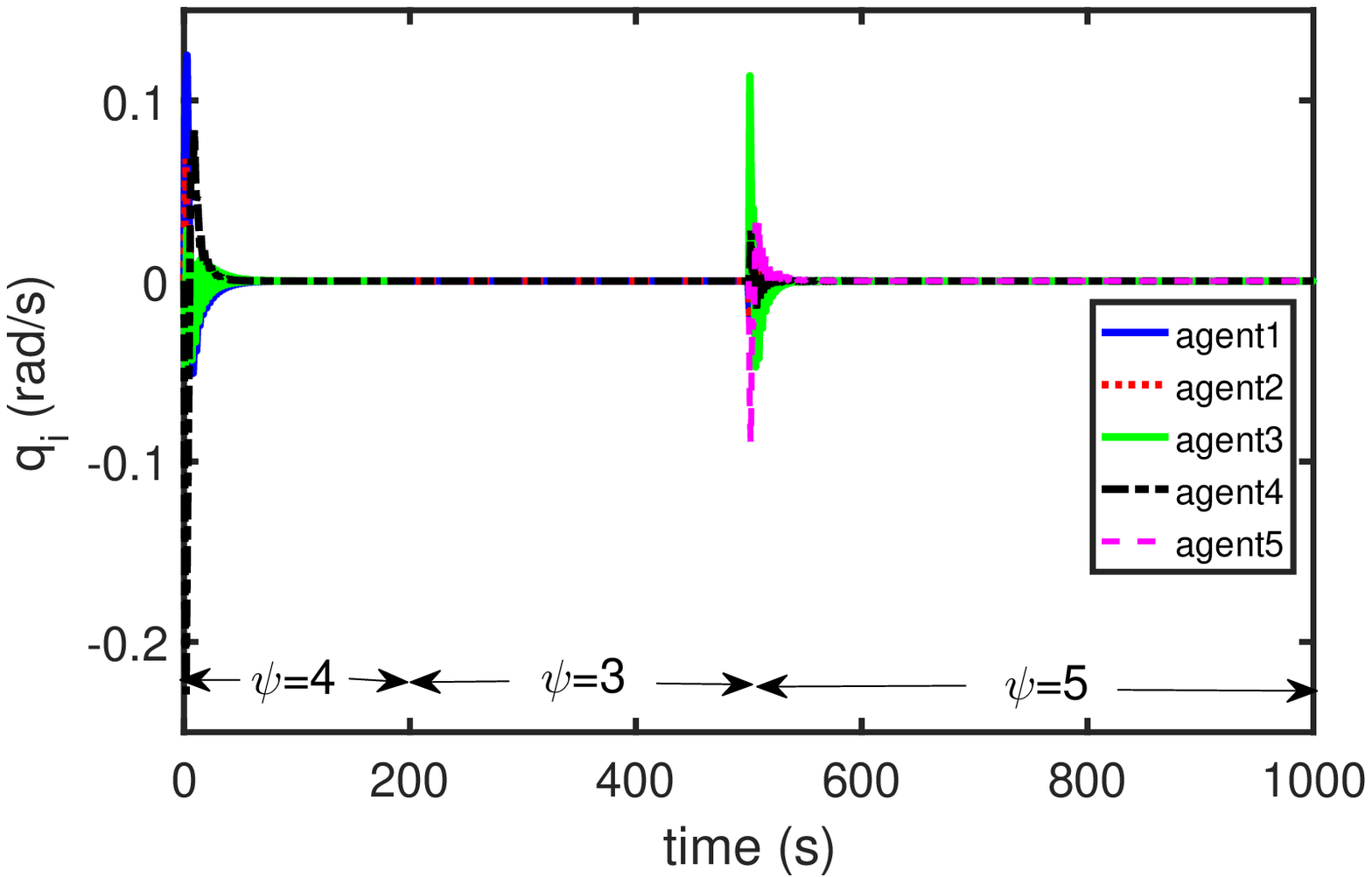}}
			\end{figure}
\begin{figure}[H]
	\subfigure[\textbf{case 4}: pitch angle  response  \label{fig:38}]{\includegraphics[width=3.5in, height=1.65in]{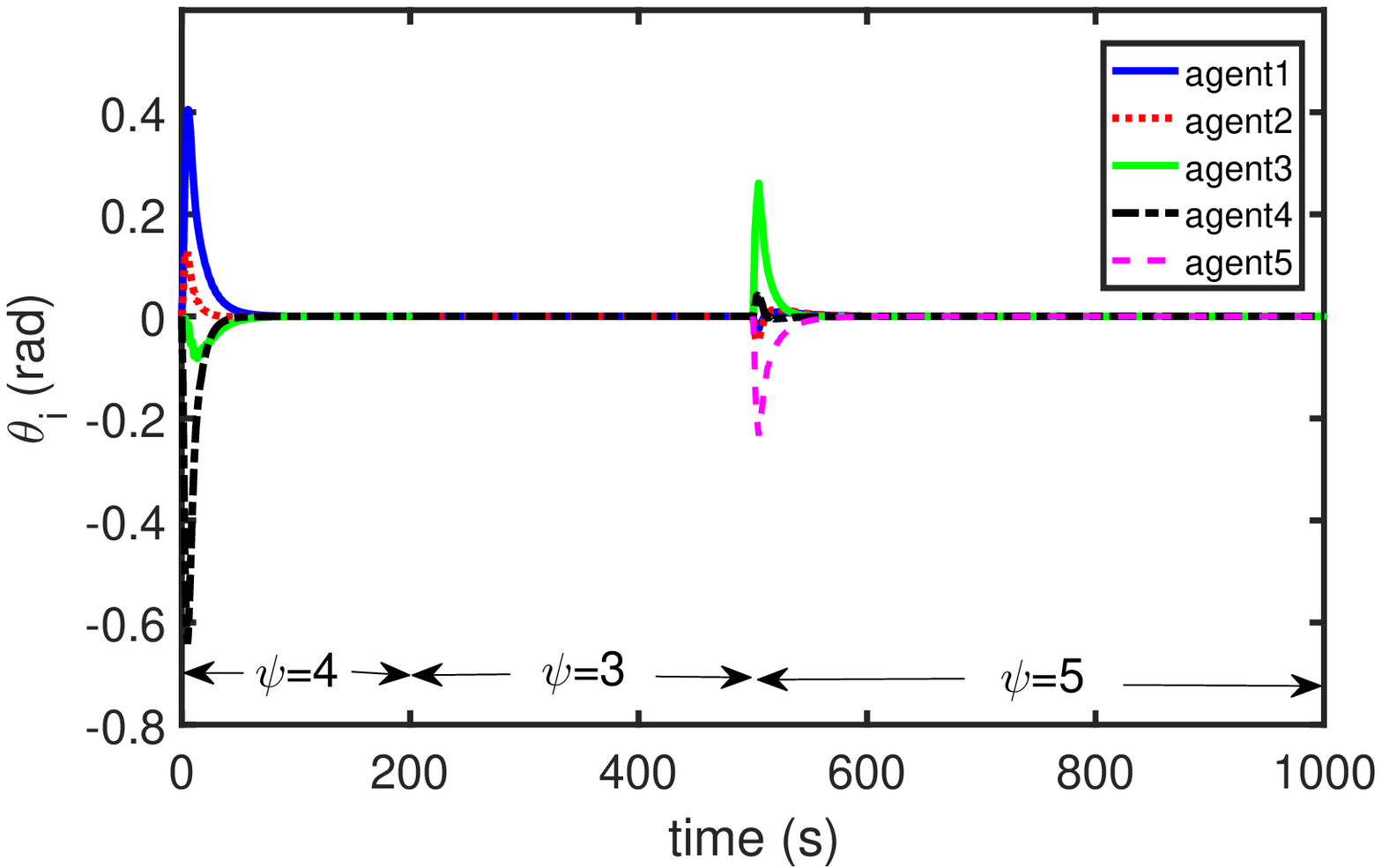}}
\centering
	\subfigure[\textbf{case 4}: depth response  \label{fig:39}]{\includegraphics[width=3.5in, height=1.65in]{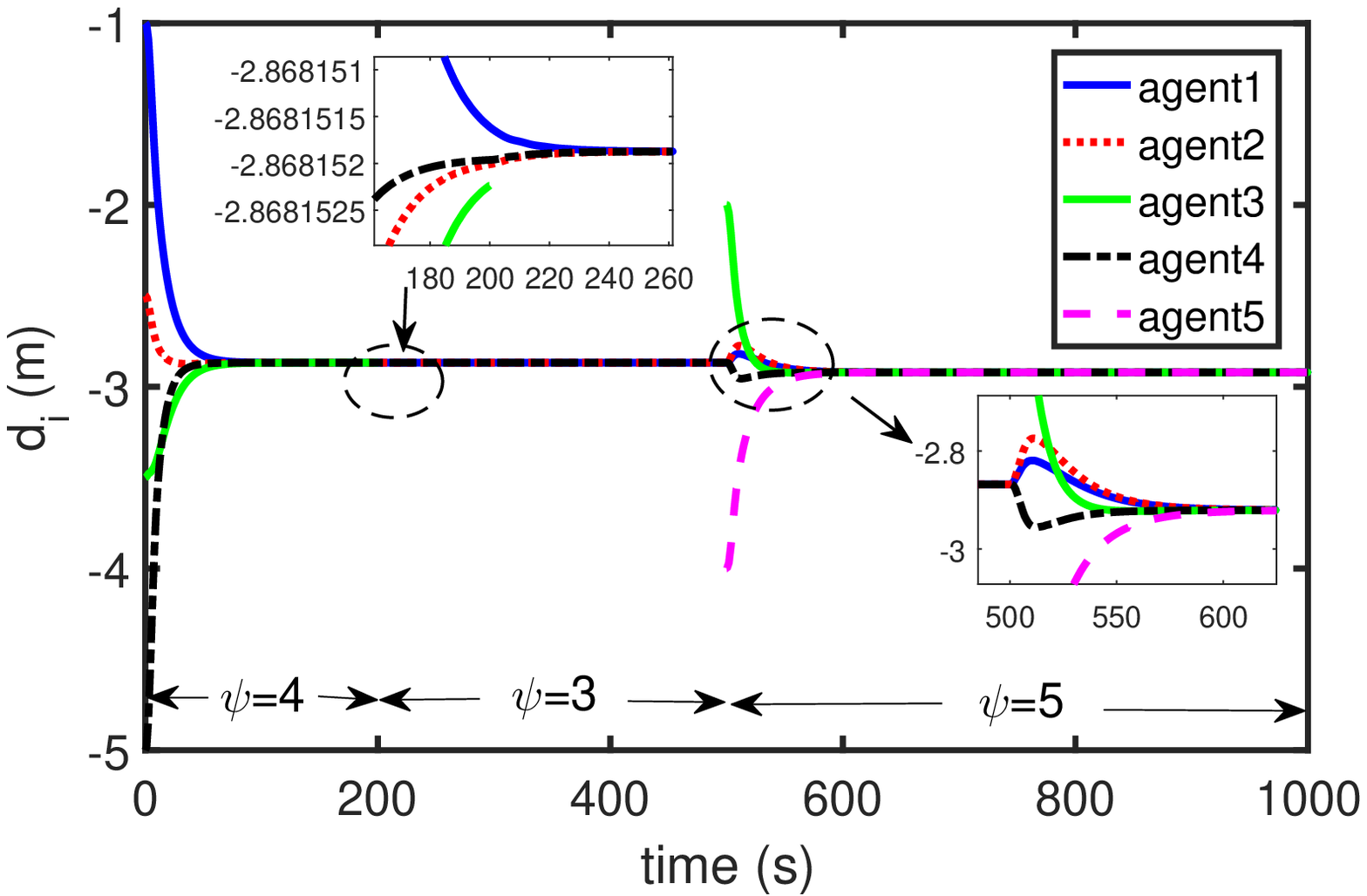}}
	\caption{Response of the CL MAS that begins its operation with four UUVs. Later, one UUV is removed from the CL MAS at the 200th~s and two UUVs are added to the CL MAS at the 500th~s. Also, $v_0$=0.3150~m/s and the communication network topologies of the CL MAS are switched at 1~s.}
	\label{fig:k}
\end{figure}
\begin{figure}[H]
	\centering
	\subfigure[\textbf{case 4}: pitch angular velocity response \label{fig:40}]{\includegraphics[width=3.5in, height=1.65in]{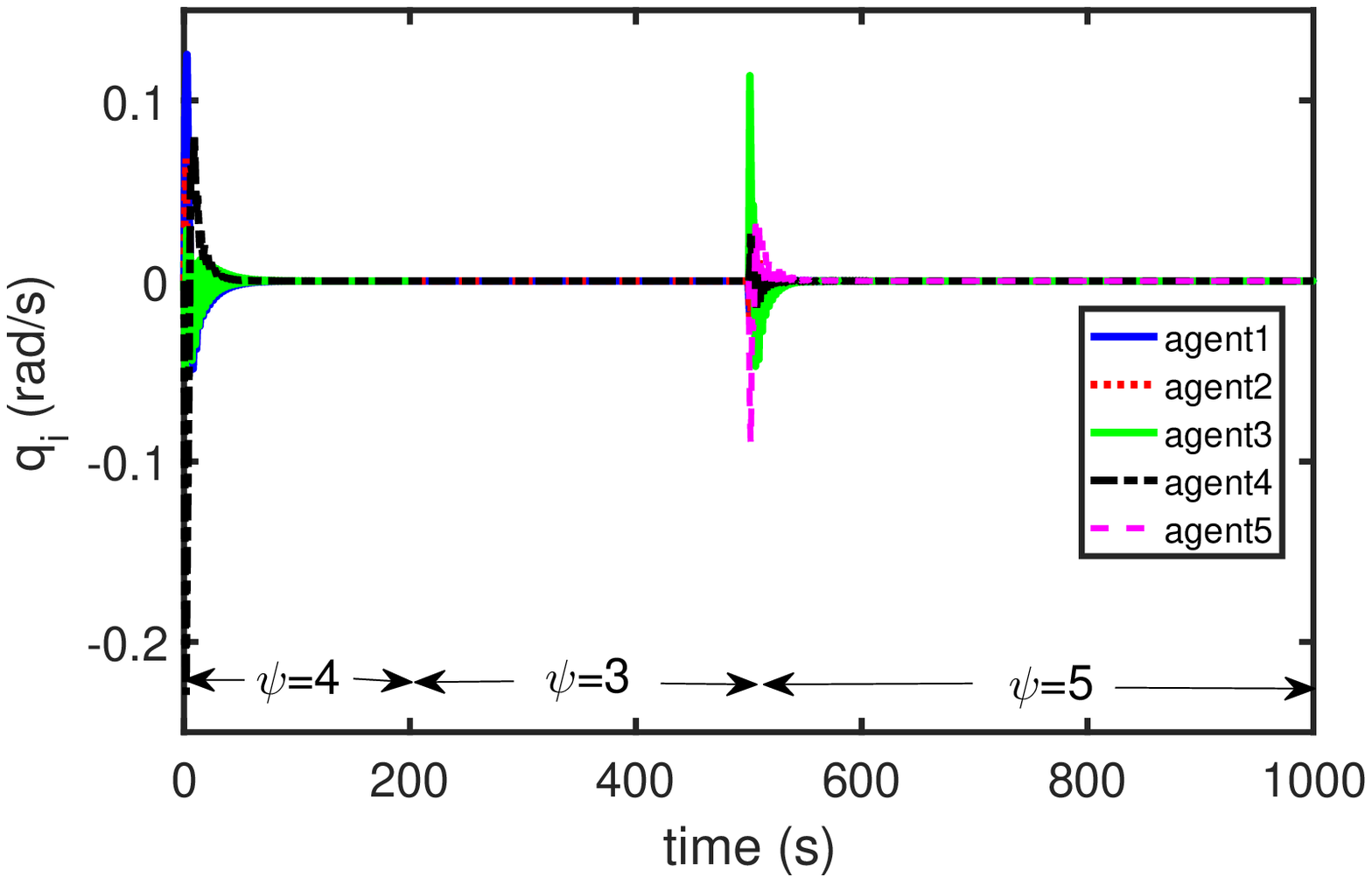}}
	\subfigure[\textbf{case 4}: pitch angle  response  \label{fig:41}]{\includegraphics[width=3.5in, height=1.65in]{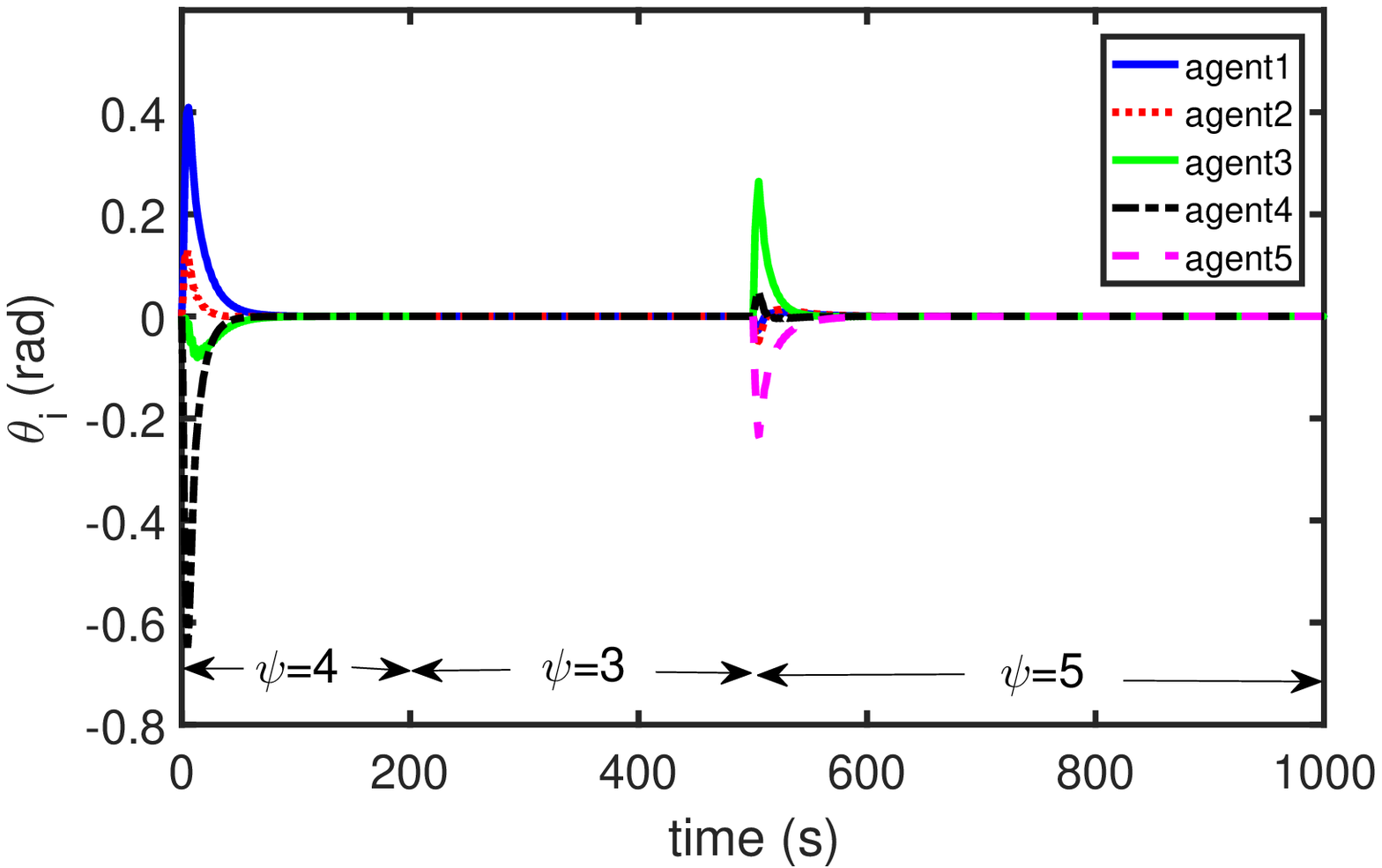}}
			\end{figure}
\begin{figure}[H]
	\subfigure[\textbf{case 4}: depth  response  \label{fig:42}]{\includegraphics[width=3.5in, height=1.65in]{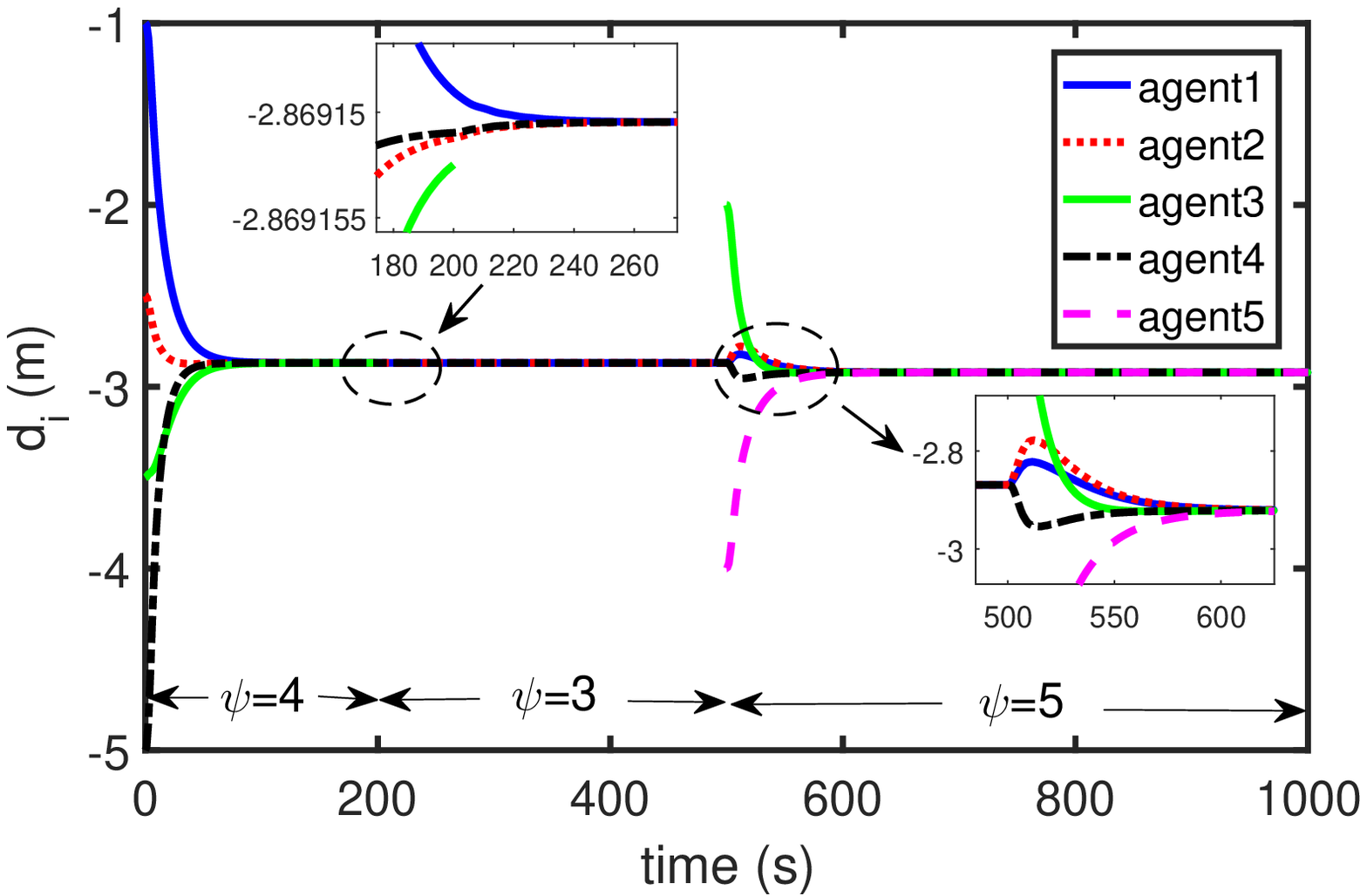}}
	\caption{Response of the CL MAS that begins its operation with four UUVs. Later, one UUV is removed from the CL MAS at the 200th~s and two UUVs are added to the CL MAS at the 500th~s. Also, $v_0$=0.2850~m/s and the communication network topologies of the CL MAS are switched at 1~s.}
	\label{fig:l}
\end{figure}
\begin{figure}[H]
	\centering
	\subfigure[\textbf{case 4}: pitch angular velocity response \label{fig:43}]{\includegraphics[width=3.5in, height=1.65in]{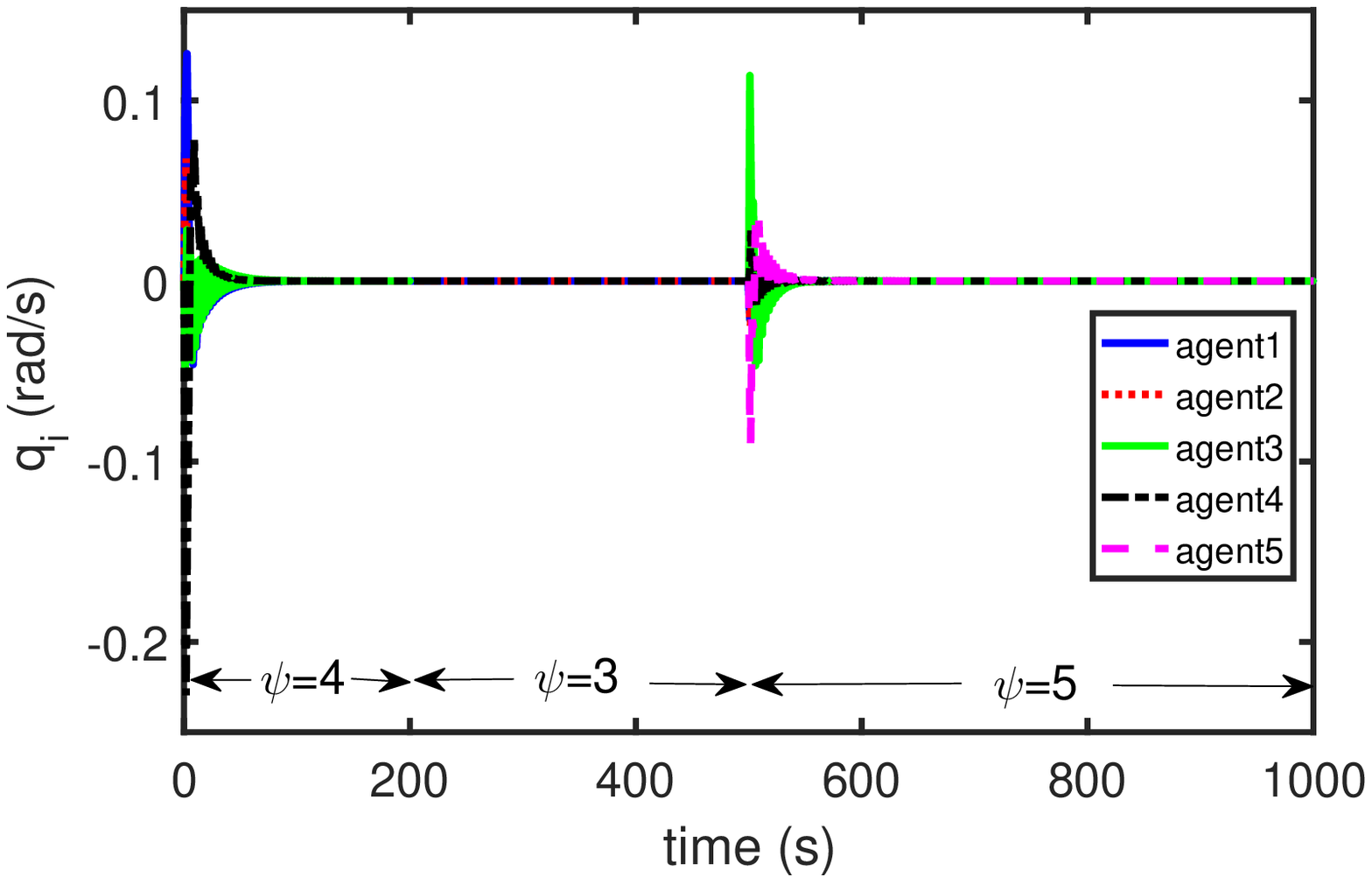}}
	\subfigure[\textbf{case 4}: pitch angle  response  \label{fig:44}]{\includegraphics[width=3.5in, height=1.65in]{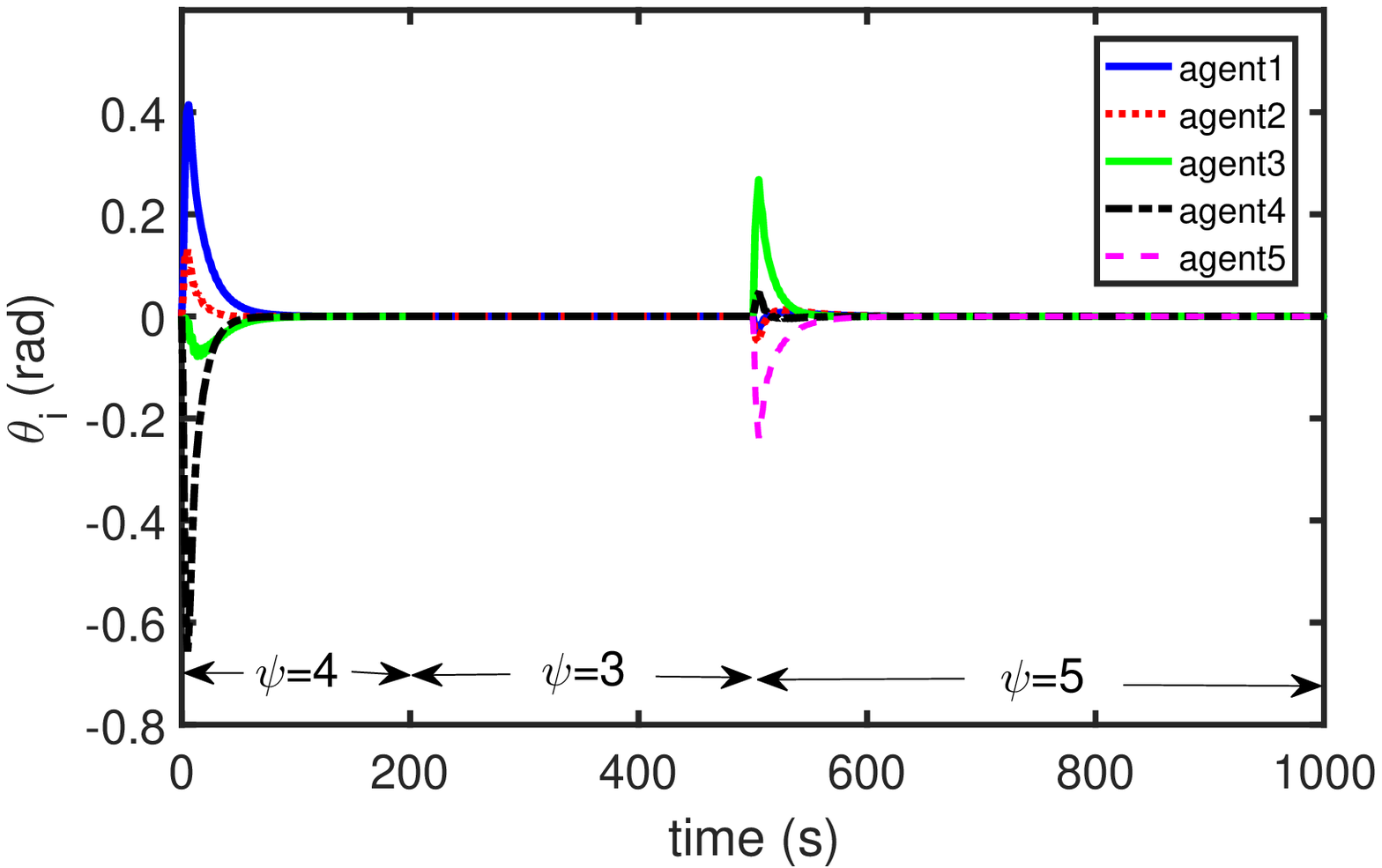}}
	\subfigure[\textbf{case 4}: depth response  \label{fig:45}]{\includegraphics[width=3.5in, height=1.65in]{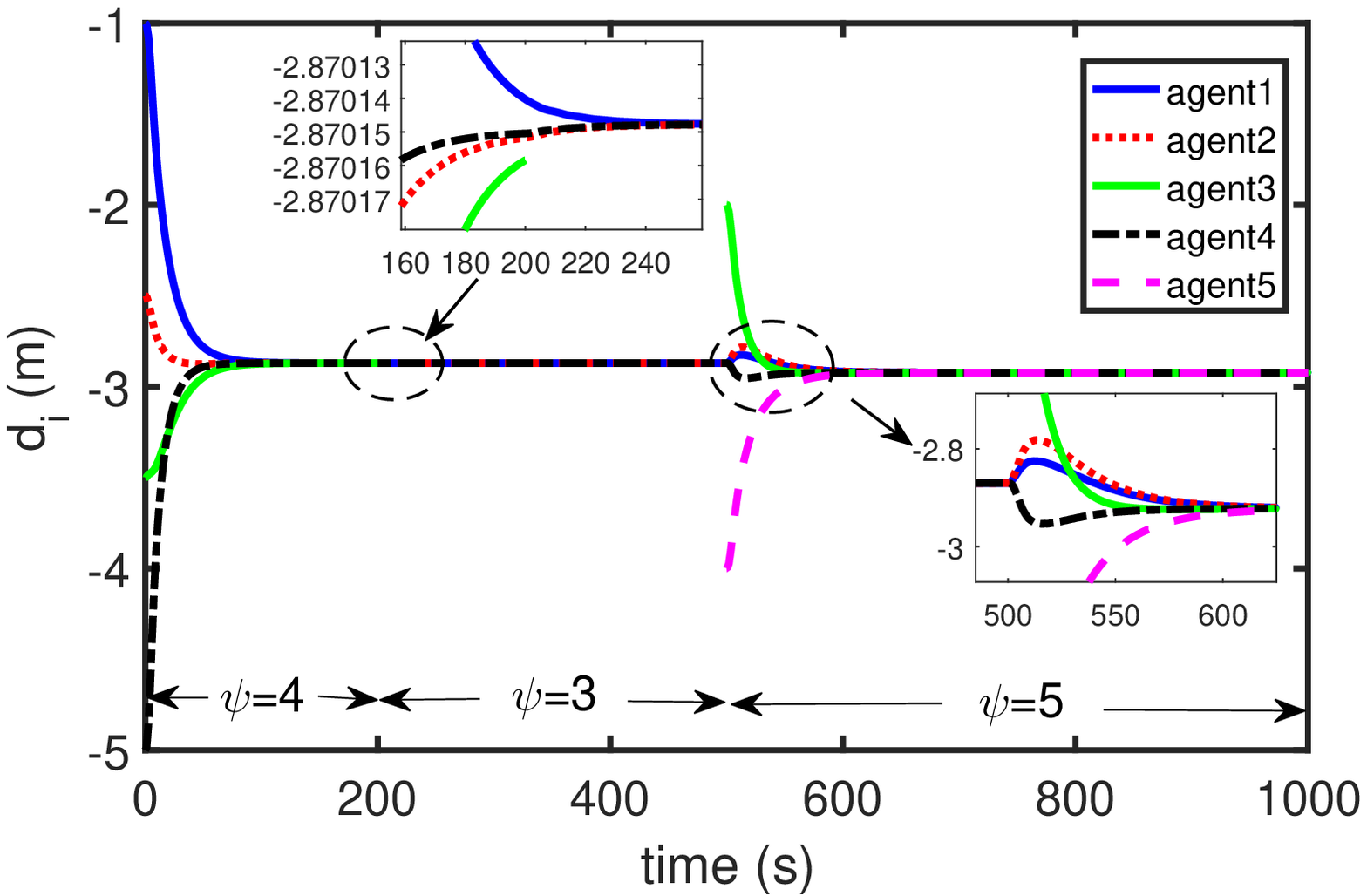}}
	\caption{Response of the CL MAS that begins its operation with four UUVs. Later, one UUV is removed from the CL MAS at the 200th~s and two UUVs are added to the CL MAS at the 500th~s. Also, $v_0$=0.2550~m/s and the communication network topologies of the CL MAS are switched at 1~s.}
	\label{fig:m}
\end{figure}
\begin{figure}[H]
	\centering
	\subfigure[\textbf{case 4}: pitch angular velocity response \label{fig:46}]{\includegraphics[width=3.5in, height=1.65in]{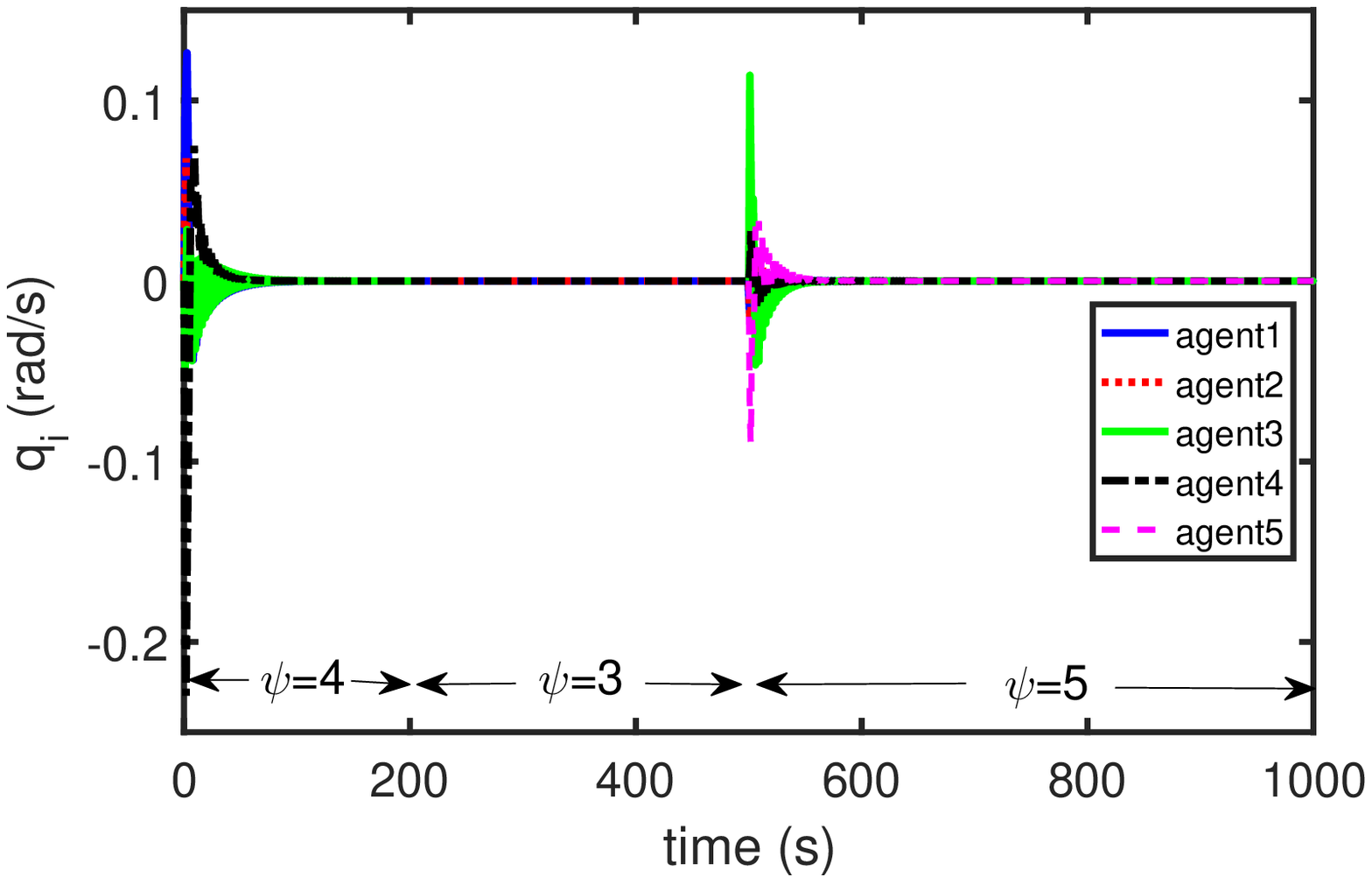}}
		\end{figure}
\begin{figure}[H]
\centering
	\subfigure[\textbf{case 4}: pitch angle  response  \label{fig:47}]{\includegraphics[width=3.5in, height=1.65in]{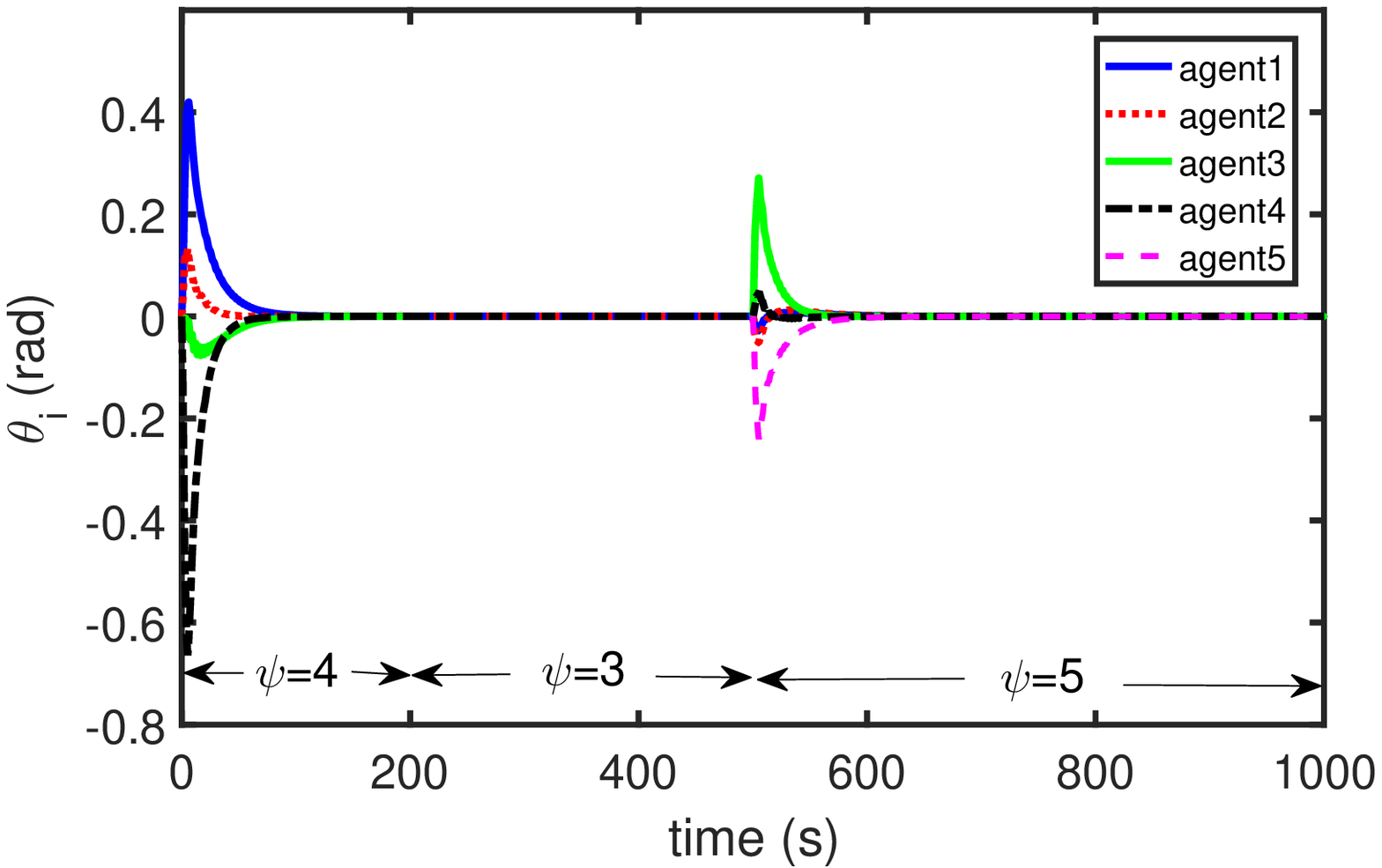}}
	\subfigure[\textbf{case 4}: depth  response  \label{fig:48}]{\includegraphics[width=3.5in, height=1.65in]{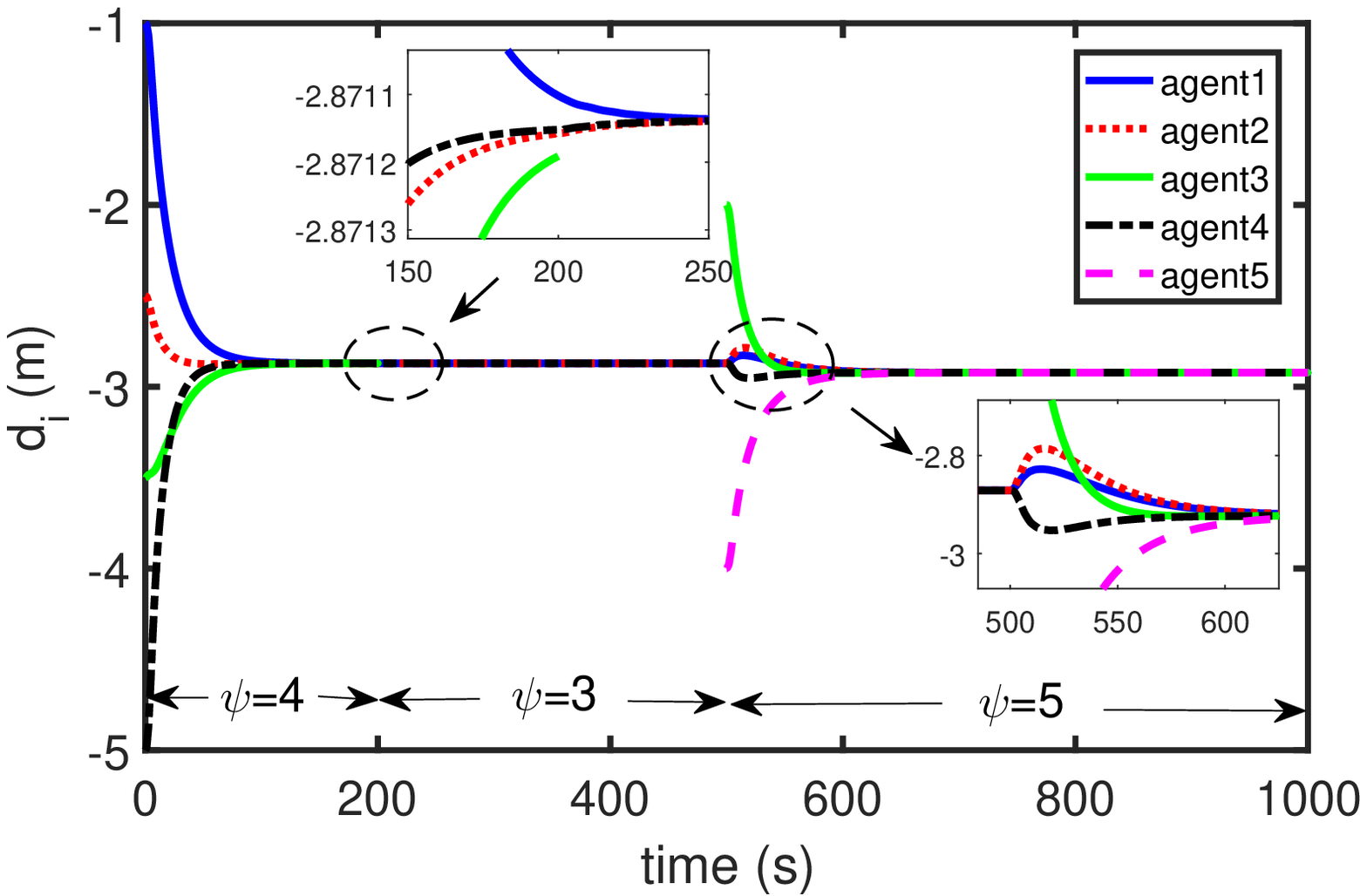}}
	\caption{Response of the CL MAS that begins its operation with four UUVs. Later, one UUV is removed from the CL MAS at the 200th~s and two UUVs are added to the CL MAS at the 500th~s. Also, $v_0$=0.2250~m/s and the communication network topologies of the CL MAS are switched at 1~s.}
	\label{fig:n}
\end{figure}

\noindent  In all the four cases, desired communication network topologies are switched at 1~s. The simulation results of all the four cases are shown in Figs. \ref{fig:a}-\ref{fig:n}. The state trajectories  shown in these figures indicate that the consensus of   MAS with 4, 3, and 5 UUVs is accomplished even with parametric uncertainties  and switching topologies.









	\section{Conclusion}\label{CL}
	
	Based on  $\nu$-gap metric-based simultaneous stabilization method, a RAIDD consensus protocol is developed for the consensus of MAS with \textit{attrition} and \textit{inclusion} of LTI higher-order uncertain homogeneous agents and switching topologies.   The sufficient condition for the existence of the RAIDD consensus protocol developed is easily testable for  MAS with a varying number of agents and switching topologies. 
	Moreover, the tractability of this condition is successfully demonstrated by generating a feasible RAIDD consensus protocol for the MAS with 4, 3, and 5 UUVs and switching topologies. The effectiveness of this protocol is validated through four cases of numerical simulations of the closed-loop system comprising the RAIDD protocol and the MAS with 4, 3, and 5 UUVs. The state trajectories of agents indicate that the consensus of   MAS with 4, 3, and 5 UUVs is achieved even with model uncertainties and switching topologies.

	
\vspace*{-1cm}
\begin{IEEEbiography}[{\includegraphics[width=1in,height=1.25in,clip,keepaspectratio]{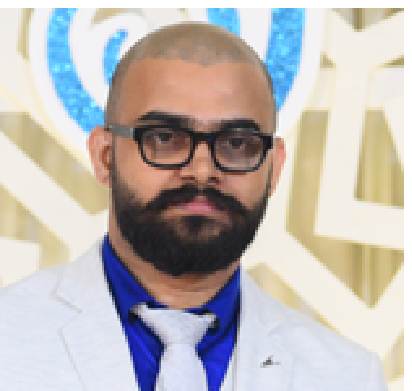}}]{Jinraj V Pushpangathan}
	received his PhD in Aerospace Engineering from the Indian Institute of Science (IISc), India, in 2018. Currently, he works as an aerospace consultant. He previously worked for Yottec Systems LLP in Bengaluru, India, as an Aero and Control specialist. He was also an international lecturer at the department of Electrical Automation Engineering, Institut Teknologi Sepuluh Nopember (ITS), Surabaya, Indonesia. From Jan. 2020 to Dec. 2020, he was a research fellow in the Artificial Intelligence Research Lab (AIRL) of the Aerospace Department, IISc. He worked as a lecturer in the electrical and electronics engineering department, Federal Institute of Science and Technology, India, from April 2007 to July 2009. His research interests include robust control, simultaneous stabilization and estimation, aerial robots, guidance and control of unmanned systems, flight dynamics and control, and reinforcement learning.
\end{IEEEbiography}
\begin{IEEEbiography}[{\includegraphics[width=1in,height=1.25in,clip,keepaspectratio]{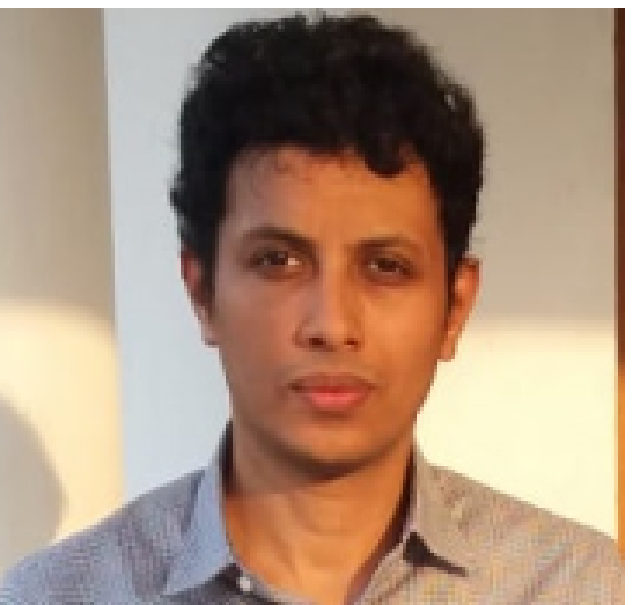}}]{K. Harikumar}
	received the Ph.D. degree in
	aerospace engineering from the Indian Institute of
	Science (IISc), Bengaluru, India, in 2015.
	He is working as an Assistant Professor with
	the Robotics Research Centre, International Institute
	of Information Technology, Hyderabad, India since March 2020.
	He was a Research Fellow with the Singapore
	University of Technology and Design, Singapore,
	from July 2019 to February 2020. He had worked
	as a Research Fellow with Nanyang Technological
	University, Singapore, from July 2016 to June 2019.
	He was a Research Assistant with the Faculty of Science and Technology,
	University of Macau, from January 2016 to March 2016. He had worked
	as an Assistant Systems Engineer with Tata Consultancy Services, from
	September 2008 to December 2009. His research interests are applications
	of control theory to unmanned systems and flight dynamics.
\end{IEEEbiography}
\vspace*{-1.1cm}
\begin{IEEEbiography}[{\includegraphics[width=1in,height=1.25in,clip,keepaspectratio]{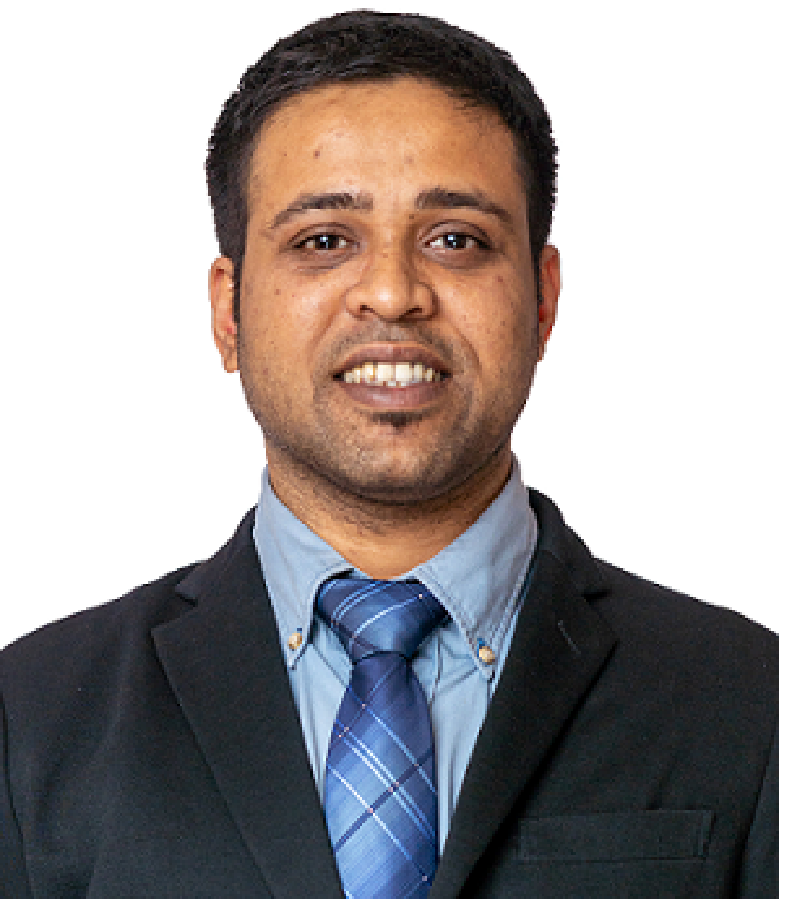}}]{Rajdeep Dutta}
	received the Bachelor of Engineering degree in Electrical Engineering, the Master of Science degree in Aerospace Engineering, and the Doctor of Philosophy degree in Electrical and Computer Engineering from Bengal Engineering and Science University (BESU) at Shibpur, Indian Institute of Science (IISc) at Bangalore, and University of Texas at San Antonio (UTSA), respectively. He is currently a research Scientist at the department of Machine Intellection, Institute for Infocom Research, Agency for Science Technology and Research (A*STAR), Singapore. Previously, he worked as a software scientist at Parabole LLC and as a postdoctoral research fellow at the department of Civil Engineering, IISc, Bangalore, India. He has earned research expertise in Dynamics and Control while working at the Unmanned Systems Laboratory, UTSA during his PhD, and gained experience in Machine Learning and Stochastic Dynamics in post-graduation career pursuit. His research interest lies in the arena of applied mathematics, dynamics and control, nonlinear systems, mathematical modelling, optimization, machine learning, and flying robotics. He is a member of Eta Kappa Nu.
\end{IEEEbiography}
\vspace*{-1.1cm}
\begin{IEEEbiography}[{\includegraphics[width=1in,height=1.25in,clip,keepaspectratio]{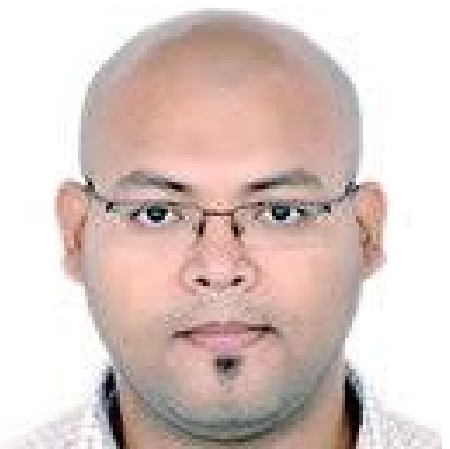}}]{Rajarshi Bardhan}
	received the Bachelor of Engineering degree in Electronics and Instrumentation Engineering, the Master of Technology degree in Electrical Engineering, and the Doctor of Philosophy degree in from the University of Burdwan, Bengal Engineering and Science University (BESU) at Shibpur, and the Department of Aerospace Engineering of Indian Institute of Science (IISc), Bangalore, respectively. He is currently a research scientist at the department of Singapore Institute of Manufacturing Technology, Agency for Science Technology and Research (A*STAR), Singapore. His research interests encompass applied mathematics, dynamics and control, nonlinear systems, optimization, and flying robotics.
\end{IEEEbiography}
\vspace*{-1.1cm}
\begin{IEEEbiography}[{\includegraphics[width=1in,height=1.25in,clip,keepaspectratio]{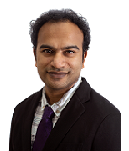}}]{J. Senthilnath}
	is a Research Scientist in Institute for Infocomm Research at Agency for Science, Technology and Research (A*STAR), Singapore. He received the PhD degree in Aerospace Engineering from the Indian Institute of Science (IISc), India. His research interests include developing new artificial intelligence (AI) techniques like deep generative models, online/continual learning, reinforcement learning and optimization in multiagent systems, accelerated materials development, failure analysis and remote sensing (Satellite/UAV). He has published over 90 referred articles in high-impact transactions/journals. For his research, he has won two best journal paper awards and three best conference paper awards. His research has contributed several AI tools for ISRO [disaster monitoring], NASA [phenology product], ST Engineering Ltd. [multiagent systems], and Airlines [predictive maintenance] - featured in Discovery Channel, 2019. He is a Senior Member of IEEE and a member of Artificial Intelligence, Analytics And Informatics (AI3), A*STAR. He was an organizing chair of the ninth IEEE SSCI'18. He is also serving as a PC co-chair of IEEE CIRS and a PC member of the 31st International Joint Conference on Artificial Intelligence (IJCAI-ECAI’22).
\end{IEEEbiography}

\end{document}